\documentclass[]{aastex631}

\usepackage{multirow}
\usepackage{graphicx} % Including figure files
\usepackage{amsmath} % Advanced maths commands
\usepackage{xcolor}
\usepackage{longtable}
\usepackage{threeparttable}
\usepackage{array}

\submitjournal{AJ}

\shorttitle{VaTEST II: Validation of 11 Exoplanets}
\shortauthors{P. Mistry et al.}
\begin{document}

\title{VaTEST II: Statistical Validation of 11 TESS-Detected Exoplanets Orbiting K-type Stars}

\correspondingauthor{Priyashkumar Mistry}
\email{priyashmistry10@gmail.com}

\author[0000-0002-4903-7950]{Priyashkumar Mistry}
\affiliation{Department of Physics, Sardar Vallabhbhai National Institute of Technology, Surat-395007, Gujarat, India}

\author[0000-0001-9365-6137]{Kamlesh Pathak}
\affiliation{Department of Physics, Sardar Vallabhbhai National Institute of Technology, Surat-395007, Gujarat, India}

\author[0000-0002-8823-8835]{Aniket Prasad}
\affiliation{Department of Physics, National Institute of Technology Agartala 799046, Tripura, India}

\author[0000-0003-3559-0840]{Georgios Lekkas}
\affiliation{Department of Physics, University of Ioannina, Ioannina, 45110, Greece}

\author[0000-0001-9026-8622]{Surendra Bhattarai}
\affiliation{Department of Physics, Indian Institute of Science Education and Research Kolkata, Mohanpur-741246, West Bengal, India}

\author[0000-0003-4845-7141]{Sarvesh Gharat}
\affiliation{Centre for Machine Intelligence and Data Science, Indian Institute of Technology Bombay}

\author[0000-0002-1918-4749]{Mousam Maity}
\affiliation{Department of Physics, Presidency University, Kolkata-700073, West Bengal, India}

\author[0000-0001-8205-0404]{Dhruv Kumar}
\affiliation{Department of Physics, National Institute of Technology Agartala 799046, Tripura, India}

%% High Angular Resolution Imaging, Stellar Parameter and Ground Based Photometry Owners

\author[0000-0001-6588-9574]{Karen A.\ Collins}
\affiliation{Center for Astrophysics \textbar \ Harvard \& Smithsonian, 60 Garden Street, Cambridge, MA 02138, USA}

\author[0000-0001-8227-1020]{Richard P.\ Schwarz}
\affiliation{Center for Astrophysics \textbar \ Harvard \& Smithsonian, 60 Garden Street, Cambridge, MA 02138, USA}

%mann@astro.umontreal.ca  (Dragonfly)
\author[0000-0002-9312-0073]{Christopher R. Mann}
\affiliation{Département de physique, Université de Montréal, 1375 Avenue Thérèse-Lavoie-Roux, Montréal, Québec, H3T 1J4, Canada}
\affiliation{Trottier Institute for Research on Exoplanets (\emph{iREx})}

\author[0000-0001-9800-6248]{Elise Furlan}
\affiliation{NASA Exoplanet Science Institute, Caltech/IPAC, Mail Code 100-22, 1200 E. California Blvd., Pasadena, CA 91125, USA}

\author[0000-0002-2532-2853]{Steve~B.~Howell}
\affiliation{NASA Ames Research Center, Moffett Field, CA 94035, USA}

\author[0000-0002-5741-3047]{David Ciardi}
\affiliation{NASA Exoplanet Science Institute, Caltech/IPAC, Mail Code 100-22, 1200 E. California Blvd., Pasadena, CA 91125, USA}

\author[0000-0001-6637-5401]{Allyson Bieryla}
\affiliation{Center for Astrophysics \textbar \ Harvard \& Smithsonian, 60 Garden Street, Cambridge, MA 02138, USA}

\author[0000-0003-0593-1560]{Elisabeth C. Matthews}
\affiliation{Max-Planck-Institut für Astronomie, Königstuhl 17, D-69117 Heidelberg, Germany}

\author[0000-0002-9329-2190]{Erica Gonzales}
\affiliation{Department of Astronomy and Astrophysics, University of California Santa Cruz, Santa Cruz, CA 95064, USA}

\author[0000-0002-0619-7639]{Carl Ziegler}
\affiliation{Department of Physics, Engineering and Astronomy, Stephen F. Austin State University, 1936 North St, Nacogdoches, TX 75962, USA}

\author[0000-0002-1835-1891]{Ian Crossfield}
\affiliation{Department of Physics \& Astronomy, University of Kansas, KS 66045, USA}

\author[0000-0002-8965-3969]{Steven Giacalone}
\affiliation{Department of Astronomy, University of California Berkeley, Berkeley, CA 94720, USA}

\author[0000-0001-5603-6895]{Thiam-Guan Tan}
\affiliation{Perth Exoplanet Survey Telescope, Perth, Western Australia}
\affiliation{Curtin Institute of Radio Astronomy, Curtin University, Bentley, Western Australia 6102}

\author[0000-0002-5674-2404]{Phil Evans}
\affiliation{El Sauce Observatory, Coquimbo Province, Chile}

\author[0000-0002-7650-3603]{Krzysztof G.\ He{\l}miniak}
\affiliation{Nicolaus Copernicus Astronomical Center, Polish Academy of Sciences, ul. Rabia\'{n}ska 8, 87-100 Toru\'{n}, Poland}

\author[0000-0003-2781-3207]{Kevin I.\ Collins}
\affiliation{George Mason University, 4400 University Drive, Fairfax, VA, 22030 USA}

%narita@g.ecc.u-tokyo.ac.jp  (MuSCAT3)
\author[0000-0001-8511-2981]{Norio Narita}
\affiliation{Komaba Institute for Science, The University of Tokyo, 3-8-1 Komaba, Meguro, Tokyo 153-8902, Japan}
\affiliation{Astrobiology Center, 2-21-1 Osawa, Mitaka, Tokyo 181-8588, Japan}
\affiliation{Instituto de Astrofisica de Canarias (IAC), 38205 La Laguna, Tenerife, Spain}

%afukui@g.ecc.u-tokyo.ac.jp   (MuSCAT3)
\author[0000-0002-4909-5763]{Akihiko Fukui}
\affiliation{Komaba Institute for Science, The University of Tokyo, 3-8-1 Komaba, Meguro, Tokyo 153-8902, Japan}
\affiliation{Instituto de Astrofisica de Canarias (IAC), 38205 La Laguna, Tenerife, Spain}

%fjpozuelos@uliege.be	
\author[0000-0003-1572-7707]{Francisco J. Pozuelos} 
\affiliation{Instituto de Astrof\'isica de Andaluc\'ia (IAA-CSIC), Glorieta de la Astronom\'ia s/n, 18008 Granada, Spain}

\author[0000-0001-8189-0233]{Courtney Dressing}
\affiliation{Department of Astronomy, University of California Berkeley, Berkeley, CA 94720, USA}

\author[0000-0002-0345-2147]{Abderahmane Soubkiou}
\affiliation{Oukaimeden Observatory, High Energy Physics and Astrophysics Laboratory, Cadi Ayyad University, Marrakech, Morocco}

\author[0000-0001-6285-9847]{Zouhair Benkhaldoun}
\affiliation{Oukaimeden Observatory, High Energy Physics and Astrophysics Laboratory, Cadi Ayyad University, Marrakech, Morocco}

\author[0000-0001-5347-7062]{Joshua E. Schlieder}
\affiliation{NASA Goddard Space Flight Center, 8800 Greenbelt Road, Greenbelt, MD 20771, USA}

\author[0000-0002-3503-3617]{Olga Suarez}
\affiliation{Universit\'e C\^ote d'Azur, Observatoire de la C\^ote d'Azur, CNRS, Laboratoire Lagrange, Bd de l'Observatoire, CS 34229, 06304 Nice cedex 4, France}

\author[0000-0003-1464-9276]{Khalid Barkaoui}
\affiliation{Astrobiology Research Unit, Université de Liège, 19C Allée du 6 Août, 4000 Liège, Belgium}
\affiliation{Department of Earth, Atmospheric and Planetary Science, Massachusetts Institute of Technology, 77 Massachusetts Avenue,\\ Cambridge, MA 02139, USA}                                 
\affiliation{Instituto  de  Astrofísica  de  Canarias  (IAC),  Calle  Vía  Láctea  s/n, 38200, La Laguna, Tenerife, Spain}

\author[0000-0003-0987-1593]{Enric Palle}
\affiliation{Instituto de Astrof\'\i sica de Canarias (IAC), 38205 La Laguna, Tenerife, Spain}
\affiliation{Departamento de Astrof\'\i sica, Universidad de La Laguna (ULL), 38206, La Laguna, Tenerife, Spain}

\author[0000-0001-9087-1245]{Felipe Murgas}
\affiliation{Instituto de Astrof\'\i sica de Canarias (IAC), 38205 La Laguna, Tenerife, Spain}
\affiliation{Departamento de Astrof\'\i sica, Universidad de La Laguna (ULL), 38206, La Laguna, Tenerife, Spain}

\author{Gregor Srdoc}
\affiliation{Kotizarovci Observatory, Sarsoni 90, 51216 Viskovo, Croatia}

\author[0000-0003-2228-7914]{Maria V. Goliguzova}
\affiliation{Sternberg Astronomical Institute, M.V. Lomonosov Moscow State University, 13, Universitetskii pr., 119234, Moscow, Russia}

\author[0000-0003-0647-6133]{Ivan A. Strakhov}
\affiliation{Sternberg Astronomical Institute, M.V. Lomonosov Moscow State University, 13, Universitetskii pr., 119234, Moscow, Russia}

\author[0000-0003-2519-6161]{Crystal Gnilka}
\affiliation{NASA Ames Research Center, Moffett Field, CA 94035, USA}

\author[0000-0002-9903-9911]{Kathryn Lester}
\affiliation{NASA Ames Research Center, Moffett Field, CA 94035, USA}

\author[0000-0001-7746-5795]{Colin Littlefield}
\affiliation{NASA Ames Research Center, Moffett Field, CA 94035, USA}
\affiliation{Bay Area Environmental Research Institute, Moffett Field, CA 94035, USA}

\author[0000-0003-1038-9702]{Nic Scott}
\affiliation{NASA Ames Research Center, Moffett Field, CA 94035, USA}

\author[0000-0001-7233-7508]{Rachel Matson}
\affiliation{U.S. Naval Observatory, Washington, D.C. 20392, USA}

\author[0000-0003-1462-7739]{Michaël Gillon}
\affiliation{Astrobiology Research Unit, Université de Liège, 19C Allée du 6 Août, 4000 Liège, Belgium}

\author[0000-0001-8923-488X]{Emmanuel Jehin}
\affiliation{Space Sciences, Technologies and Astrophysics Research (STAR) Institute, Université de Liège, 19C Allée du 6 Août, 4000 Liège, Belgium}

\author{Mathilde Timmermans}
\affiliation{Astrobiology Research Unit, Université de Liège, 19C Allée du 6 Août, 4000 Liège, Belgium}

\author[0000-0003-3986-0297]{Mourad Ghachoui}
\affiliation{Astrobiology Research Unit, Université de Liège, 19C Allée du 6 Août, 4000 Liège, Belgium}
\affiliation{Oukaimeden Observatory, High Energy Physics and Astrophysics Laboratory, Cadi Ayyad University, Marrakech, Morocco}

\author[0000-0002-0856-4527]{Lyu Abe}
\affiliation{Universit\'e C\^ote d'Azur, Observatoire de la C\^ote d'Azur, CNRS, Laboratoire Lagrange, Bd de l'Observatoire, CS 34229, 06304 Nice cedex 4, France}

\author[0000-0002-4278-1437]{Philippe Bendjoya}
\affiliation{Universit\'e C\^ote d'Azur, Observatoire de la C\^ote d'Azur, CNRS, Laboratoire Lagrange, Bd de l'Observatoire, CS 34229, 06304 Nice cedex 4, France}

\author[0000-0002-7188-8428]{Tristan Guillot}
\affiliation{Universit\'e C\^ote d'Azur, Observatoire de la C\^ote d'Azur, CNRS, Laboratoire Lagrange, Bd de l'Observatoire, CS 34229, 06304 Nice cedex 4, France}

\author[0000-0002-5510-8751]{Amaury H.M.J. Triaud}
\affiliation{School of Physics \& Astronomy, University of Birmingham, Edgbaston, Birmingham B15 2TT, United Kingdom}

%% Note that the \and command from previous versions of AASTeX is now
%% depreciated in this version as it is no longer necessary. AASTeX 
%% automatically takes care of all commas and "and"s between authors names.

%% AASTeX 6.31 has the new \collaboration and \nocollaboration commands to
%% provide the collaboration status of a group of authors. These commands 
%% can be used either before or after the list of corresponding authors. The
%% argument for \collaboration is the collaboration identifier. Authors are
%% encouraged to surround collaboration identifiers with ()s. The 
%% \nocollaboration command takes no argument and exists to indicate that
%% the nearby authors are not part of surrounding collaborations.

%% Mark off the abstract in the ``abstract'' environment. 
\begin{abstract}
NASA's Transiting Exoplanet Survey Satellite (TESS) is an all-sky survey mission designed to find transiting exoplanets orbiting nearby bright stars. It has identified more than 329 transiting exoplanets, and almost 6,000 candidates remain unvalidated. In this manuscript, we discuss the findings from the ongoing VaTEST (Validation of Transiting Exoplanets using Statistical Tools) project, which aims to validate new exoplanets for further characterization. We validated 11 new exoplanets by examining the light curves of 24 candidates using the \texttt{LATTE} and \texttt{TESS-Plot} tools and computing the False Positive Probabilities using the statistical validation tool \texttt{TRICERATOPS}. These include planets suitable for atmospheric characterization using transmission spectroscopy (TOI-2194b), emission spectroscopy (TOI-3082b and TOI-5704b) and for both transmission and emission spectroscopy (TOI-672b, TOI-1694b, and TOI-2443b); One super-Earth (TOI-2194b) orbiting a bright (V = 8.42 mag), metal-poor ([Fe/H] = -0.3720 $\pm$ 0.1) star; one short-period Neptune-like planet (TOI-5704) in the Hot Neptune Desert. In total, we validated 1 super-Earth, 7 sub-Neptunes, 1 Neptune-like, and 2 sub-Saturn or super-Neptune-like exoplanets. Additionally, we identify five likely planet candidates (TOI-323, TOI-1180, TOI-2200, TOI-2408 and TOI-3913) which can be further studied to establish their planetary nature.
\end{abstract}

%% Keywords should appear after the \end{abstract} command. 
%% The AAS Journals now uses Unified Astronomy Thesaurus concepts:
%% https://astrothesaurus.org
%% You will be asked to selected these concepts during the submission process
%% but this old "keyword" functionality is maintained in case authors want
%% to include these concepts in their preprints.

%% From the front matter, we move on to the body of the paper.
%% Sections are demarcated by \section and \subsection, respectively.
%% Observe the use of the LaTeX \label
%% command after the \subsection to give a symbolic KEY to the
%% subsection for cross-referencing in a \ref command.
%% You can use LaTeX's \ref and \label commands to keep track of
%% cross-references to sections, equations, tables, and figures.
%% That way, if you change the order of any elements, LaTeX will
%% automatically renumber them.
%%
%% We recommend that authors also use the natbib \citep
%% and \citet commands to identify citations.  The citations are
%% tied to the reference list via symbolic KEYs. The KEY corresponds
%% to the KEY in the \bibitem in the reference list below. 

\section{Introduction}
The Transiting Exoplanet Survey Satellite (TESS) \citep{2015JATIS...1a4003R} mission is an all-sky survey to discover exoplanets in nearby regions. It was launched on April 18, 2018 aboard a SpaceX Falcon 9 rocket. During its two-year primary mission, the TESS spacecraft concentrated on nearby G, K, and M type stars with apparent magnitudes $<$ 12. An area 400 times greater than the one covered by the Kepler campaign was to be surveyed, including the 1,000 nearest dwarf stars in the entire sky. The survey was divided into 26 viewing zones called sectors, each of which was 24${\degr}$ $\times$ 96${\degr}$. The spacecraft had spent two 13.7 days orbiting each sector, mapping the southern hemisphere in its first year of operation and the northern hemisphere in its second year. TESS's primary mission (cycles 1 and 2, sectors 1–26) was completed in July 2020. The first extended mission (cycles 3 and 4, sectors 27–55) ended in September 2022, and it is now on its second extended mission (cycle 5, sectors 56–69).

We currently have 323 confirmed TESS exoplanets and 6386 TESS candidates\footnote{\url{https://exoplanetarchive.ipac.caltech.edu/}, accessed on April 19, 2023} that need to be studied. By using the conventional method, i.e., a combination of transit and radial velocity to discover a new planet, it is very difficult to study this large number of candidates. There are so-called astrophysical false positives \citep{2003ApJ...593L.125B,2012Natur.492...48C}, such as eclipsing binaries, blended eclipsing binaries, and planet-sized stars in binary systems, that can generate a transit-like signal. Many tools have been developed based on transit photometry to calculate their likelihood and probability of being planets or false positives. To rule out false positives, \texttt{BLENDER} \citep{2005ApJ...619..558T} was the first approach based on $\chi^2$ statistics of eclipsing binaries and blended eclipsing binaries. \citet{2013PASP..125..889B} has presented various tests to rule out the possibility of blended eclipsing binaries. These methods include Photometric Centroid Shift, Difference Imaging and Pixel Correlation Images. In the first method the centroid shift is detected on the pixels correlated with transit signal and that shift is then used to estimate the location of the transit source. The second method uses the difference image of in- and out-of-transit pixel image to locate the transit source and the last method computes the degree to which the transit signal over time appears in each pixel. These methods make an assumption that the transit signal is solely created by the pixels under investigation (i.e., the mean flux from the TESS aperture mask pixels) and that there are no other sources of flux variation. However, when this assumption is violated, these techniques may introduce systematic errors. The nature of these errors may differ among the methods utilized. Thus, the presence of inconsistencies in the outcomes obtained from these techniques may indicate the existence of systematic error. \texttt{VESPA} \citep{2015ascl.soft03011M} was another approach that used the MCMC sampling routine to fit the Kepler light curve and produced a false-positive probability based on the fit. Both \texttt{VESPA} and \texttt{BLENDER} can include high-contrast imaging in their analysis. The framework was widely used to statistically validate exoplanets from Kepler as well as TESS. The another robust model \texttt{PASTIS} \citep{2014MNRAS.441..983D} which can take transit photometry data as well as high precision radial velocity measurements to validate the planet. Alternatively, \texttt{TRICERATOPS} \citep{2020ascl.soft02004G,2021AJ....161...24G} was specifically developed to take advantage of the unique features and requirements of the TESS mission. With a lower resolution than previous such missions, there may be a greater necessity to account for multiple star systems and scenarios like diluted transits. Such approaches can be used to validate new exoplanets in bulk without having radial velocity measurements. For our project, we made use of \texttt{TRICERATOPS} as a validation tool to calculate the False Positive Probability (FPP) of selected candidates.

The Validation of Transiting Exoplanets using Statistical Tools (VaTEST) project\footnote{\url{https://sites.google.com/view/project-vatest/home}} has its primary goal to validate multiple exoplanets with the use of various statistical validation approaches. In our first paper, we discovered our first planet, TOI-181b \citep{2023MNRAS.tmp..527M}, by utilising a similar approach. For the future, we have separated candidates based on their spectral types (temperatures) and will study them each individually in order to find out their planetary nature. However, for this manuscript, we will validate the exoplanets orbiting K-type (temperature range 3700–5200 K, \citet{1981A&AS...46..193H,2010A&A...524A..98W}) stars. Here we validated a significant number of exoplanets from the candidates observed by TESS. 

This paper is structured as follows: In section \ref{Selection of Candidates} we discuss our methodology to select the most promising candidates for the validation process, and in section \ref{Follow-up Observation} we present the high-resolution imaging and ground-based photometic observations. The algorithm and procedure for using the statistical validation tool \texttt{TRICERATOPS} is covered in Section \ref{Statistical Validation}. In section \ref{Validated Planets} we presented the main features of newly validated systems. Finally, section \ref{Likely Planets and Not Validated Candidates} describes candidates that failed the validation criteria (not validated candidates) and some likely planets that can be followed up further to validate.

\section{Selection of Candidates}
\label{Selection of Candidates}
In this manuscript, we study planets orbiting K spectral type stars. There were multiple restrictions made while selecting the targets for our study, such as:
\begin{itemize}
    \item Reported orbital period $< 20$ days
    \item Planetary radii $< 8 \text{R}_E$
    \item Removed targets with the dispositions (from ExoFOP\footnote{\url{https://exofop.ipac.caltech.edu/tess}}) CP (Confirmed Planet), KP (Known Planet), FP (False Positive) and EB (Eclipsing Binary)
\end{itemize}

As we based our validations on a combination of high-resolution imaging, ground-based photometry, and transit photometry data, it is crucial to ensure that the light curve contains a maximum number of transits to confirm the exoplanetary nature of a signal. The minimum number of transits required to confirm the existence of an exoplanet is typically at least three,  so we make sure that all of the considered targets had at least 3 transits, either in a single sector or in a combination of different sectors. For this reason we choose targets showing $< 20$ days orbital period. The major reason behind having radii $< 8 \text{R}_E$ is the statistical validation tool called \texttt{TRICERATOPS} \citep{2021AJ....161...24G}. \texttt{TRICERATOPS} under-predicts the false positive probability for planetary candidates having radii $\geq 8 \text{R}_E$. 

A total of 343 candidates from the Exoplanet Follow-up Observing Program (ExoFOP) website database are considered in this study. To identify possible binary stars, use of Renormalized Unit Weight Error (RUWE)  score \citep{LL:LL-124} from Gaia EDR3 is done. Targets with an RUWE score of $\geq 1.4$ or null \citep{LL:LL-124} are eliminated. Additionally, targets with stellar companions, lacking SPOC pipeline data, or without available stellar parameters were also excluded from this study. A visual inspection of the remaining targets was performed to eliminate any signals that were consistent with star variability, eclipsing binaries, or instrumental systematic effects. Finally, the use of \texttt{Juliet} modeling \citep{2018ascl.soft12016E} is done on the remaining set of targets to identify eclipsing binaries based on the shape (V-shaped) and characteristics of modelled transit light curves. Through this initial screening process, a total of 24 significant objects were identified for further examination of their planetary nature. Stellar parameters for these selected targets are shown in Table \ref{tab:stellar_params}. We have taken stellar parameters from ExoFOP website, which were derived using one of the three methods. First is stellar spectra collected using Fred Lawrence Whipple Observatory (FLWO). FLWO spectra were obtained at the Fred Lawrence Whipple Observatory using the Tillinghast Reflector Echelle Spectrograph (TRES; \citet{furesz}) on the 1.5m Tillinghast Reflector telescope. Stellar parameters were derived using the Stellar Parameter Classification (SPC) tool as outlined in \citet{buchhave2012,buchhave2014}. Second is Nordic Optical Telescope's high-resolution FIbre-fed Echelle Spectrograph (NOT-FIES) \citet{2014AN....335...41T}. FIES is a cross-dispersed high-resolution echelle spectrograph with a maximum spectral resolution of R = 67000. The entire spectral range 370-830 nm is covered without gaps in a single, fixed setting. And the third is \texttt{ExoFASTV2} tool \citep{2019arXiv190709480E}. For some of the targets parameters were not available from either of the method, in such cases we used the values from TESS Input Catalog \citep{2018AJ....156..102S} Stellar Parameters (TIC, version 8.2, \citet{2019AJ....158..138S}). In particular all the radii and masses are taken the from TIC Stellar Parameters.

Before conducting a thorough and computationally expensive probabilistic analysis for transit signals, it is important to check the origin of the detected signal preliminarily. We employed the open source package Lightcurve Analysis Tool for Transiting Exoplanet (\texttt{LATTE}, \citet{2022ascl.soft05006E}) which runs multiple diagnostic tests providing an approximate indication that the signal may be originating from the target star rather than any nearby sources. We discuss our interpretation of these tests in section \ref{sec:Diagnostic results from LATTE}. However, it should be noted that this preliminary test provides only approximate information and does not provide a high level of confidence regarding the origin of the signal. Further analysis and verification are required. The results of the LATTE tests for all of the considered candidates have been uploaded to a publicly available GitHub repository\footnote{\url{https://github.com/priyashmistry/VaTEST-II-Output-Files.git}} for further examination. After reviewing the results of LATTE analysis we have concluded that transit signal and star are approximately co-related. To confirm the planetary nature of the given signal we have used statistical validation tool \texttt{TRICERATOPS}. Methodology and results are discussed in sections \ref{sec:Validation with TRICERATOPS} and \ref{Validated Planets}.

\begin{longrotatetable}
\begin{deluxetable*}{cccccccccccc}
\tablecaption{Stellar parameters of the candidate systems. \label{tab:stellar_params}}
\tabletypesize{\footnotesize}
\tablehead{
\colhead{No.} & 
\colhead{TOI ID} & 
\colhead{TIC ID} &
\colhead{T$_eff$} &
\colhead{R$_s$} & 
\colhead{M$_s$} &
\colhead{logg} & 
\colhead{[Fe/H]$^{a}$} &
\colhead{V} &
\colhead{TESS} &
\colhead{RUWE} &
\colhead{Source$^{b}$} \\
\colhead{} & 
\colhead{} & 
\colhead{} &
\colhead{[K]} & 
\colhead{[$R_{\sun}$]} &
\colhead{[$M_{\sun}$]} &
\colhead{} &
\colhead{[dex]} &
\colhead{[mag]} & 
\colhead{[mag]} & 
\colhead{} & 
\colhead{}
}
\startdata
1 &  TOI 139  & 62483237 &  4570  $\pm$  50  & 0.7007 $\pm$ 0.0575 & 0.6900 $\pm$ 0.0852 & 4.705 $\pm$ 0.100 & -0.238 $\pm$ 0.080 & 10.55 & 9.36 & 0.8818 & TRES, TIC\\
2 &  TOI 323  & 251852984 &  4558  $\pm$  122  & 0.7756 $\pm$ 0.1000 & 0.7370 $\pm$ 0.0455 & 4.613 $\pm$ 0.027 & -0.040 $\pm$ 0.325 & 14.35 & 13.35 & 1.0132 & ExoFASTv2, TIC\\
3 &  TOI 493  & 19025965 &  4402  $\pm$  100  & 0.8119 $\pm$ 0.0661 & 0.6480 $\pm$ 0.0829 & 4.689 $\pm$ 0.100 & -0.181 $\pm$ 0.080 & 12.55 & 11.45 & 1.1079 & TRES, TIC\\
4 &  TOI 672  & 151825527 &  3765  $\pm$  65  & 0.5441 $\pm$ 0.0163 & 0.5399 $\pm$ 0.0204 & 4.699 $\pm$ 0.010 & -0.710 $\pm$ 0.625 & 13.58 & 11.67 & 1.1587 & TIC\\
5 &  TOI 815  & 102840239 &  4954  $\pm$  107  & 0.7594 $\pm$ 0.0426 & 0.8200 $\pm$ 0.0943 & 4.591 $\pm$ 0.081 & 0.037 $\pm$ 0.039 & 10.22 & 9.36 & 1.0129 & TIC\\
6 &  TOI 913  & 407126408 &  4969  $\pm$  129  & 0.7325 $\pm$ 0.0488 & 0.8200 $\pm$ 0.0973 & 4.622 $\pm$ 0.089 & -0.133 $\pm$ 0.100 & 10.45 & 9.62 & 1.0136 & TIC\\
7 &  TOI 1179  & 148914726 &  4998  $\pm$  50  & 0.7770 $\pm$ 0.0110 & 0.8050 $\pm$ 0.4450 & 4.462 $\pm$ 0.100 & -0.084 $\pm$ 0.080 & 10.88 & 10.13 & 1.0336 & FIES, ExoFASTv2\\
8 &  TOI 1180  & 158002130 &  4900  $\pm$  50  & 0.7272 $\pm$ 0.0518 & 0.7500 $\pm$ 0.0934 & 4.723 $\pm$ 0.100 & -0.024 $\pm$ 0.080 & 11.02 & 10.11 & 0.9415 & TRES, TIC\\
9 &  TOI 1694  & 396740648 &  5135  $\pm$  50  & 0.8183 $\pm$ 0.0477 & 0.8450 $\pm$ 0.1089 & 4.658 $\pm$ 0.100 & 0.060 $\pm$ 0.080 & 11.45 & 10.74 & 1.3827 & TRES, TIC\\
10 &  TOI 1732  & 470987100 &  3876  $\pm$  157  & 0.6326 $\pm$ 0.0187 & 0.6139 $\pm$ 0.0203 & 4.624 $\pm$ 0.011 & 0.291 $\pm$ 0.100 & 12.89 & 11.33 & 1.3104 & TIC\\ 
11 &  TOI 2194  & 271478281 &  4756  $\pm$  50  & 0.6909 $\pm$ 0.0492 & 0.7400 $\pm$ 0.0854 & 4.698 $\pm$ 0.100 & -0.372 $\pm$ 0.100 & 8.42 & 7.43 & 0.9936 & TRES, TIC \\
12 &  TOI 2200  & 142105158 &  5070  $\pm$  117  & 0.8262 $\pm$ 0.0487 & 0.8500 $\pm$ 0.1038 & 4.533 $\pm$ 0.084 & 0.273 $\pm$ 0.100 & 13.09 & 12.32 & 1.0810 & TIC\\
13 &  TOI 2408  & 67630845 &  4935  $\pm$  132  & 0.7485 $\pm$ 0.0510 & 0.8100 $\pm$ 0.0969 & 4.598 $\pm$ 0.092 & -   & 12.78 & 11.92 & 1.0093 & TIC\\
14 &  TOI 2443  & 318753380 &  4357  $\pm$  100  & 0.7321 $\pm$ 0.0713 & 0.6600 $\pm$ 0.0789 & 4.709 $\pm$ 0.100 & -0.439 $\pm$ 0.080 & 9.51 & 8.30 & 1.2536 & TRES, TIC\\
15 &  TOI 2459  & 192790476 &  4195  $\pm$  124  & 0.6751 $\pm$ 0.0630 & 0.6600 $\pm$ 0.0763 & 4.599 $\pm$ 0.107 & -   & 10.77 & 9.40 & 1.1358 & TIC\\
16 &  TOI 3082  & 428699140 &  4263  $\pm$  100  & 0.6847 $\pm$ 0.0613 & 0.6640 $\pm$ 0.0798 & 4.625 $\pm$ 0.100 & 0.170 $\pm$ 0.080 & 12.93 & 11.77 & 1.2058 & TRES, TIC\\
17 &  TOI 3568  & 160390955 &  4890  $\pm$  50  & 0.7858 $\pm$ 0.0517 & 0.7920 $\pm$ 0.0942 & 4.540 $\pm$ 0.100 & 0.002 $\pm$ 0.080 & 12.88 & 12.07 & 0.9779 & TRES, TIC\\
18 &  TOI 3896  & 445837596 &  5043  $\pm$  50  & 0.7478 $\pm$ 0.0431 & 0.8600 $\pm$ 0.1032 & 4.419 $\pm$ 0.100 & -0.279 $\pm$ 0.080 & 12.43 & 11.68 & 0.9606 & TRES, TIC\\
19 &  TOI 3913  & 155898758 &  4180  $\pm$  123  & 0.8257 $\pm$ 0.0767 & 0.6540 $\pm$ 0.0813 & 4.420 $\pm$ 0.110 & 0.137 $\pm$ 0.110 & 13.71 & 12.58 & 1.0625 & TIC\\
20 &  TOI 4090  & 289373041 &  4740  $\pm$  124  & 0.8194 $\pm$ 0.0567 & 0.7600 $\pm$ 0.0909 & 4.492 $\pm$ 0.095 & 0.760 $\pm$ 0.091 & 13.40 & 12.55 & 1.0152 & TIC\\
21 &  TOI 4308  & 144193715 &  5243  $\pm$  126  & 0.7934 $\pm$ 0.0465 & 0.9000 $\pm$ 0.1133 & 4.593 $\pm$ 0.087 & -   & 11.25 & 10.34 & 0.8720 & TIC\\
22 &  TOI 5584  & 29169215 &  4372  $\pm$  100  & 0.7451 $\pm$ 0.0684 & 0.6400 $\pm$ 0.0789 & 4.725 $\pm$ 0.100 & 0.128 $\pm$ 0.065 & 15.83 & 10.73 & 1.0809 & TRES, TIC\\
23 &  TOI 5704  & 148673433 &  4590  $\pm$  126  & 0.7575 $\pm$ 0.0593 & 0.7300 $\pm$ 0.0846 & 4.543 $\pm$ 0.095 & 0.428 $\pm$ 0.100 & 11.53 & 10.61 & 1.2287 & TIC\\
24 &  TOI 5803  & 466382581 &  5134  $\pm$  121  & 0.7625 $\pm$ 0.0451 & 0.8700 $\pm$ 0.1032 & 4.613 $\pm$ 0.085 & -   & 10.66 & 9.94 & 0.9834 & TIC\\
\enddata
\tablenotetext{a}{TRES derives a [m/H], not [Fe/H], In other words it derive a mixture of metals and not just Fe, which is an important distinction to note when comparing the metallicity of different objects.}
\tablenotetext{b}{In the source column, wherever TIC is included, the radius and mass are taken from TICv8.2, while the rest of the parameters are taken from the other sources listed.}
\end{deluxetable*}
\end{longrotatetable}

\subsection{Results of diagnostic tests}
\label{sec:Diagnostic results from LATTE}

The \texttt{LATTE} program performs multiple tests on a chosen TESS target by considering all the transit events observed by the TESS to be passed through diagnosis. Taking inspiration from SPOC pipeline data validation, LATTE performs multiple checks for background flux, in-out transit flux, pixel-level light curves, and centroid correlation/positions for each transit. The checks at the level of each transit event being tested individually help us to rule out any possible False alarms (FA), or account for any instrumental error affecting transits.

\begin{figure*}
\begin{center}
\includegraphics[scale=0.50]{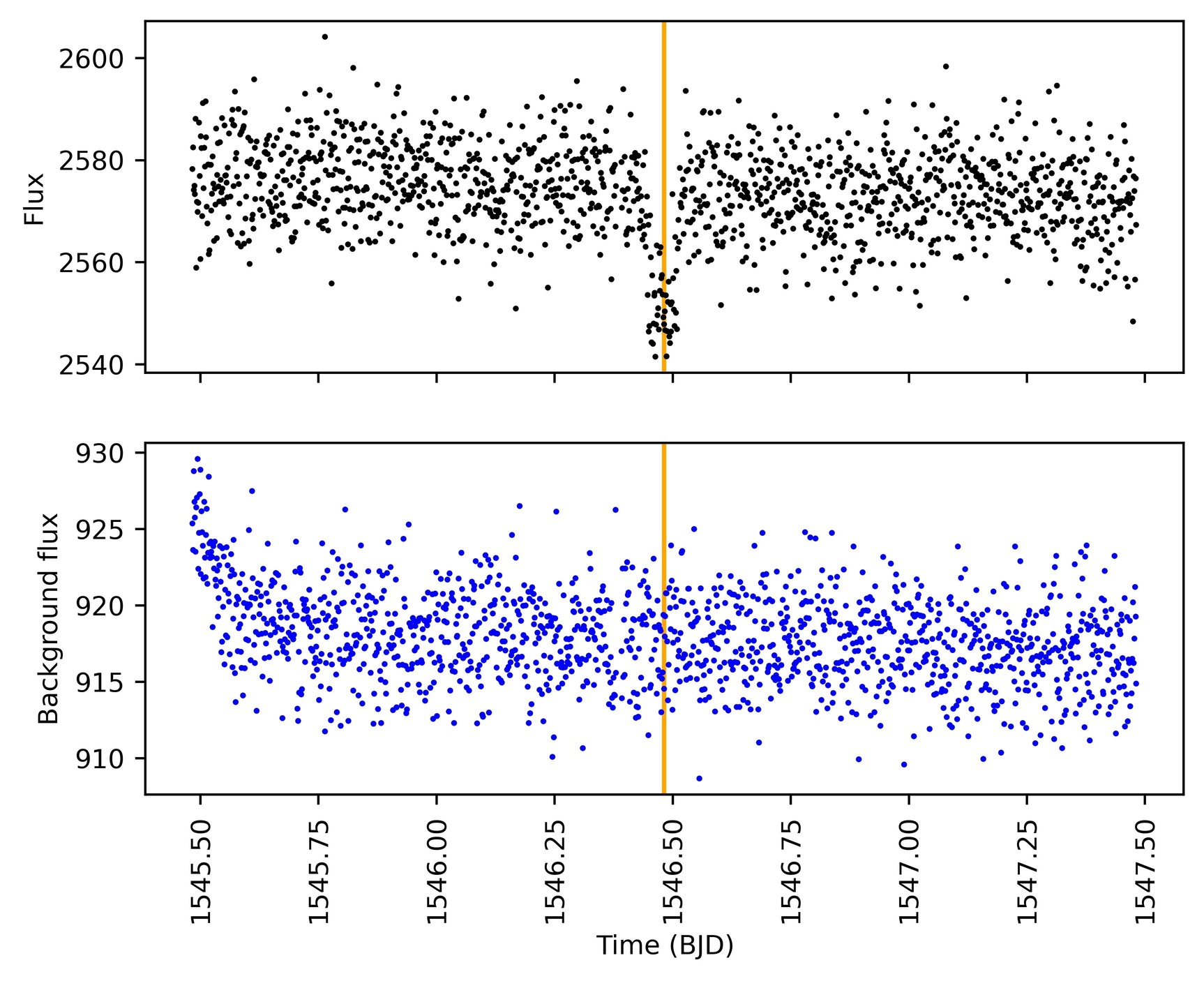}
\includegraphics[scale=0.50]{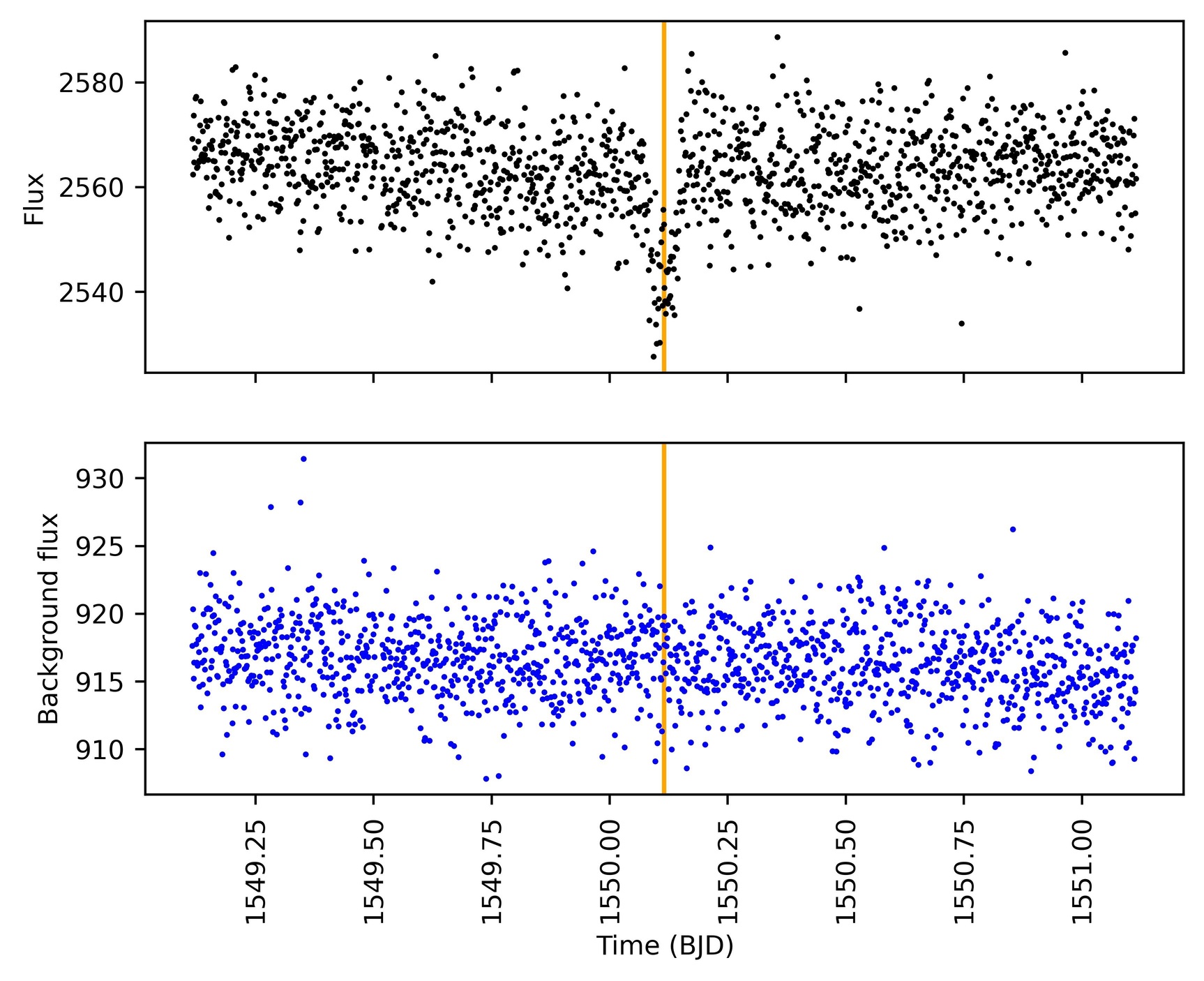}
\end{center}
\caption{TOI-672 Sector-09 Background flux at the time of first two transits. No spikes are observed at the time of transit.}
\label{fig:bgf}
\end{figure*}

We utilized comparative plots of background flux and overall flux to rule out the possibility of any background object such as a solar system object or asteroid in line-of-sight mimicking a transit-like event. Each time TESS attains the perigee of its eccentric orbit around the Earth enhanced scattered light in the telescope optics can cause the background flux to sharply increase. We therefore look at the background plots to ensure that there are no spikes at the time of the transit-like events. Figure \ref{fig:bgf} simultaneously shows the transit event and the background flux for TOI-672 (Sector-09). There is no obvious change in the background flux that is correlated with the transit signals. Also, no correlations were seen in any other TOI light curves.

\begin{figure*}
\begin{center}
\includegraphics[scale=0.3]{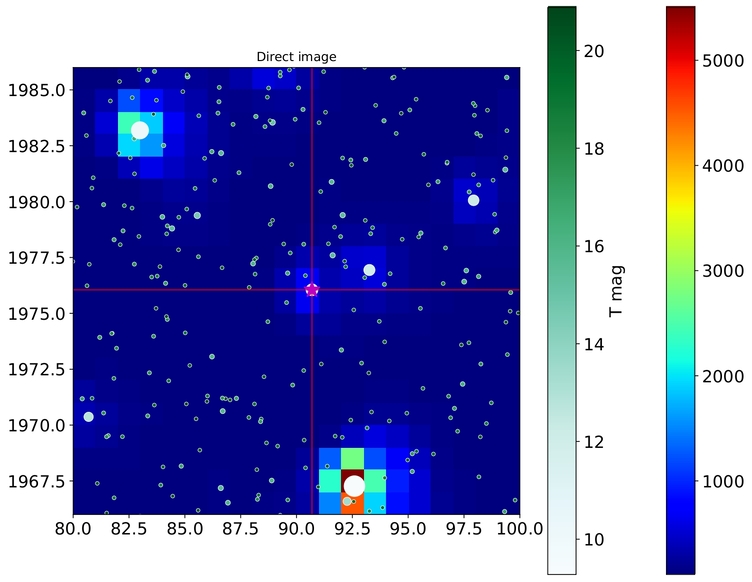}
\includegraphics[scale=0.3]{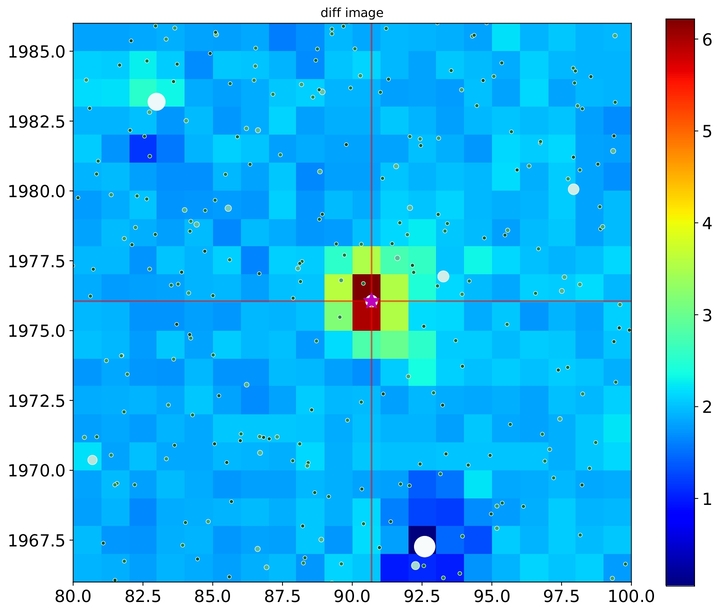}
\end{center}
\caption{Difference between in and out transit flux for TOI-672. Left: Direct Image, Right: Difference Image. It can be understood by observing these images that variation in the brightness and target are approximately co-related.}
\label{fig:io}
\end{figure*}

Another diagnostic test we have used as a filter in this work is generating in- and out-of-transit flux comparison plots. We have used \texttt{TESS-Plot}\footnote{\url{https://github.com/mkunimoto/TESS-plots}} package to generate the plots. By analyzing the difference images, we were able to determine whether the observed transit-like signal was related to the target star or if it was occurring due to a background source such as an eclipsing binary or a nearby transiting planet. If the change in brightness occurred on the target star pixel, then it was indicative that the signal may be related to the target star. However, if the change in brightness occurred elsewhere on the image, it was evident that the transit-like signal was occurring due to an off-target source. Figure \ref{fig:io} displays the difference image for our target, TOI-672. As can be observed from the figure, the change in brightness is on our selected target (indicated by the star symbol, other dots represent the nearby stars). For all the TOIs there exists an approximate correlation between the transit signal and the target star.

\begin{figure*}
\begin{center}
\includegraphics[scale=0.40]{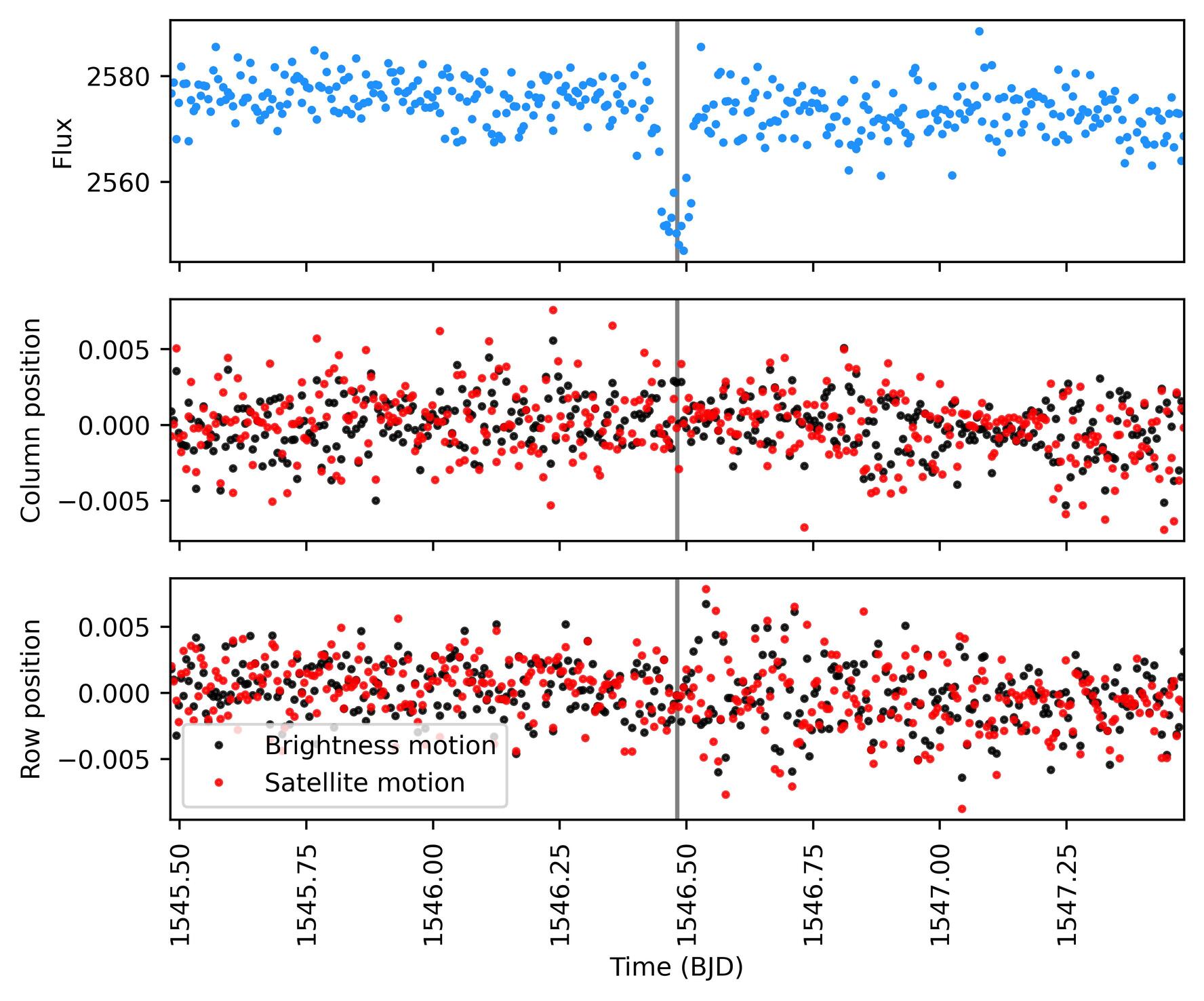}
\includegraphics[scale=0.40]{TOI_672_S_09_CPT_0.jpeg}
\end{center}
\caption{Centroid raw and column position at transit events.}
\label{fig:centroid}
\end{figure*}

We also considered the location of the row and column centroid at each transit event as shown in Figure \ref{fig:centroid} centroid plot for TOI-672. The centroid is the point in aperture where the average amount of light from the stars fall. Since a false positive scenario like blended binary could result visible change in the position of centroid, We created diagnostic plots to track the target's row and column positions throughout each transit. However, no significant differences were found upon visual inspection. It is to note that the targets with almost similar magnitude within the aperture might still show correlations in centroid positions and hence, despite the apparent lack of visual shifts it is imperative to account for nearby neighbours while performing validation.

Hence, the insights from these diagnostic tests helped increase our confidence to follow up with \texttt{TRICERATOPS} for a deeper analysis. The results of these tests for other candidates have been uploaded in GitHub. By observing those plots, it is inferred that the source of the transit-like signal is approximately related to the target pixel. Furthermore, we performed ground-based follow-up observations, high-resolution imaging, and false positive probability calculation using \texttt{TRICERATOPS} to validate the planetary nature of the given transit signal.

\section{Follow-up Observations}
\label{Follow-up Observation}

\subsection{High-resolution Imaging}
By utilizing adaptive optics and speckle imaging techniques, we captured high-contrast images of our Target Objects of Interest (TOIs). The observations were conducted by the members of TFOP Sub Group 3 (SG3) and are presented in Table \ref{tab:cc_data}, illustrated in Figure \ref{fig:cc_images}, and elaborated below.

\subsubsection{Gemini-N/'Alopeke, Gemini-N/NIRI \& Gemini-S/Zorro}
Speckle interferometric observations for TOI-139, TOI-323, TOI-493, TOI-672, TOI-815, TOI-913, TOI-1179, TOI-1694, TOI-1732, and TOI-2443 were performed by 'Alopeke, and Zorro installed at the calibration ports at Gemini North and South, respectively (e.g., see \citet{2009AJ....137.5057H,2021FrASS...8..138S}). The full set of observations taken in 562 nm ($\Delta \lambda$ = 54 nm) and 832 nm ($\Delta \lambda$ = 40 nm) was then combined in Fourier space to produce their power spectrum and autocorrelation functions. The data reduction pipeline produces final data products that include 5$\sigma$ contrast curves and reconstructed images \citep{2011AJ....142...19H}. Contrast curves are presented in Figure \ref{fig:cc_images}. No secondary sources were detected within the reconstructed images.

\subsubsection{Keck/NIRC2}
High-resolution imaging observations for TOI-139, TOI-493, and TOI-1694 were made on UT 2018 September 18, 2019 March 25, and 2020 September 09 respectively, using NIRC2 \citep{2020yCat..18730065S} which is situated on Keck II's left Nasmyth Platform \citep{2000PASP..112..315W}, behind the AO bench. By injecting simulated sources in 45$^{\circ}$ azimuthal incremenrs at discrete separations that were integer multiples of the Full Width at Half Maxima (FWHM) of the central source, we measured the sensitivity of the companions \citep{2017AJ....153...71F,2021FrASS...8...63S}. To determine the contrast sensitivity, the flux of each simulated source was raised until aperture photometry provided a detection of 5$\sigma$. Averaging all of the limits at that separation resulted in the final contrast sensitivity as a function of separation. Observations for TOI-139 were made in the BrGamma ($\lambda_0$ = 2.168; $\Delta \lambda$ = 0.033 $\mu$m) and Jcont ($\lambda_0$ = 1.213; $\Delta \lambda$ = 0.019 $\mu$m) filters, TOI-1694 was observed using Ks ($\lambda_0$ = 2.146; $\Delta \lambda$ = 0.311 $\mu$m) filter and TOI-493 was observed using BrGamma filter. The Keck AO observations revealed no additional stellar companions to within a resolution of $\approx 0.05\arcsec$ FWHM.

\subsubsection{Palomar/PHARO}
PHARO \citep{2001PASP..113..105H} is a near-infrared camera made to work with the 200-inch Hale telescope at Palomar Observatory and the Palomar Adaptive Optics system. Detector has 1024 $\times$ 1024 Rockwell HAWAII HgCdTe pixel array with wavelength sensitivity of 1 - 1.25 microns. It has diffraction-limited angular resolutions of 0.063$\arcsec$ and 0.111$\arcsec$ for J and K band imaging, respectively. Its large-format detector has a field of view of 25$\arcsec$ to 40$\arcsec$. AO images for TOI-1732, TOI-2443, TOI-3568, TOI-3896, TOI-3913, and TOI-4090 were collected in BrGamma ($\lambda_0$ = 2.166; $\Delta \lambda$ = 0.02 $\mu$m) using PHARO instrument. Estimated contrasts at different separation are presented in Table \ref{tab:cc_data}, No secondary sources were detected within the reconstructed images.

\subsubsection{Shane/ShARCS}
\begin{figure*}
    \centering
    \includegraphics[scale = 0.45]{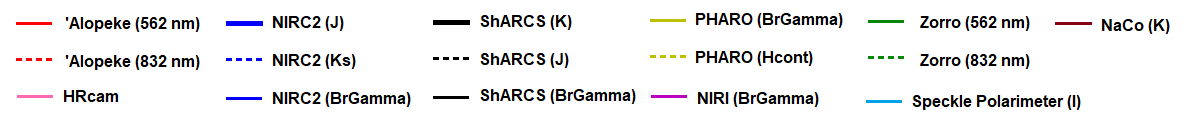}\\
    \includegraphics[scale = 0.32]{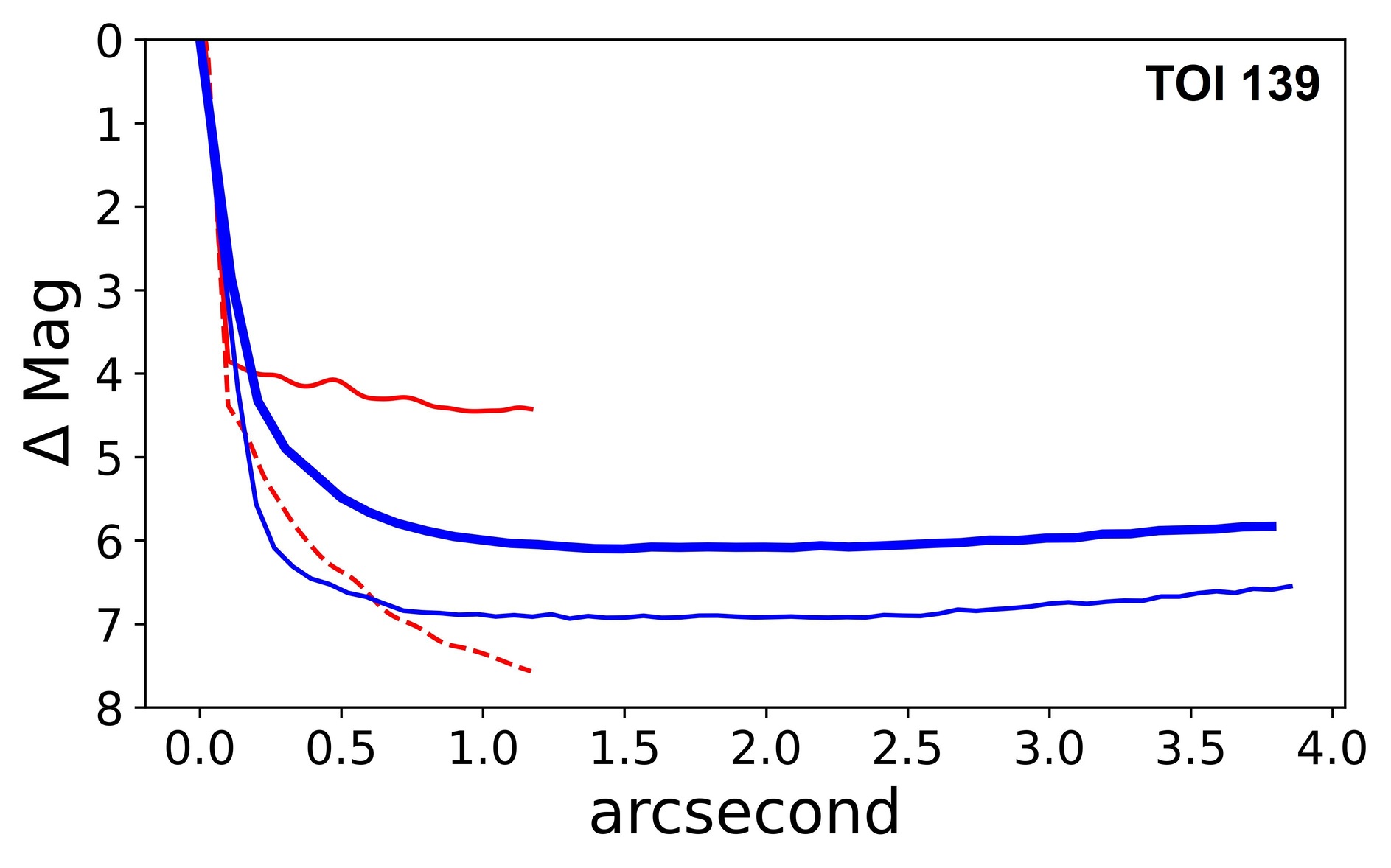}
    \includegraphics[scale = 0.32]{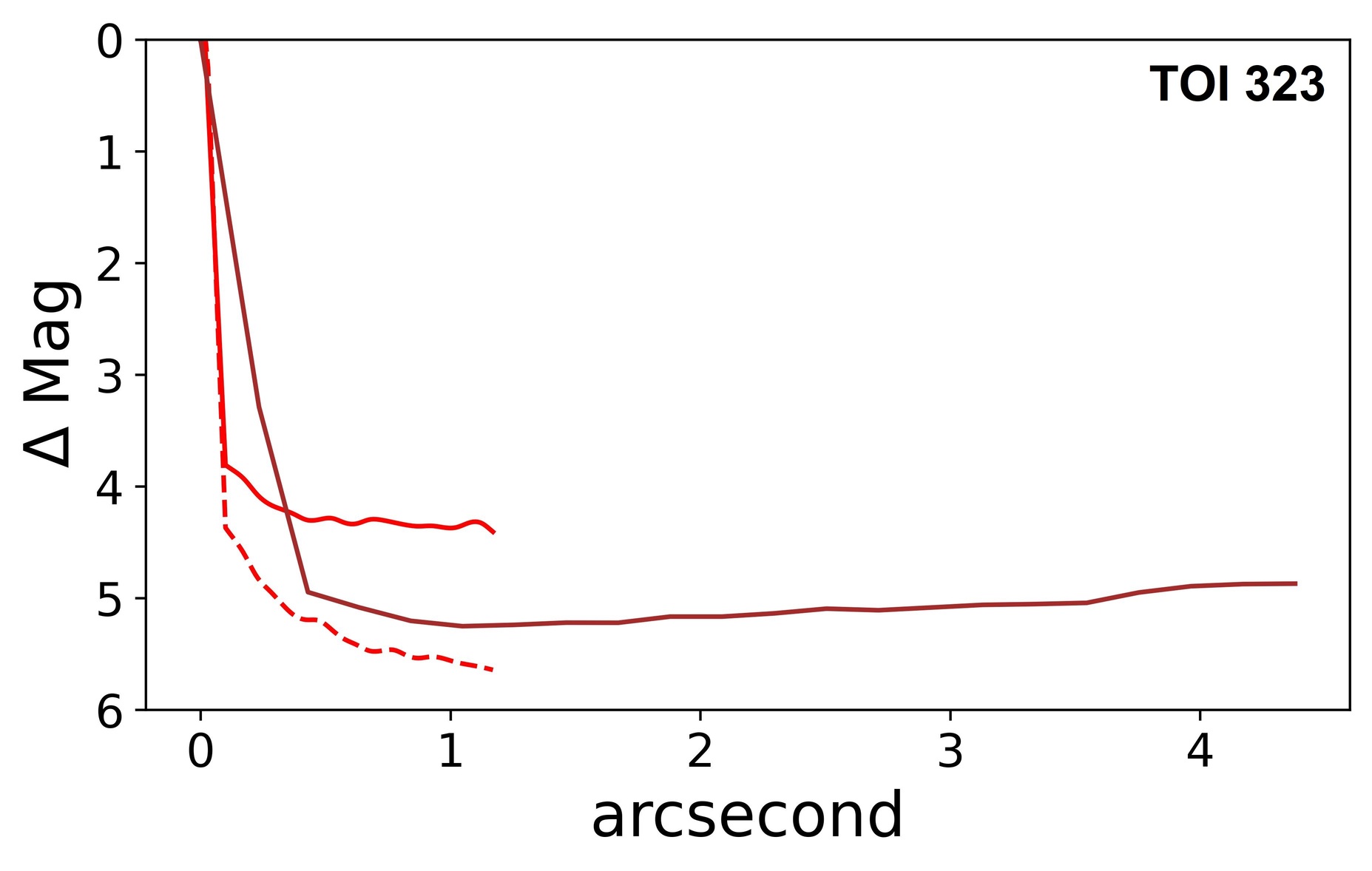}
    \includegraphics[scale = 0.32]{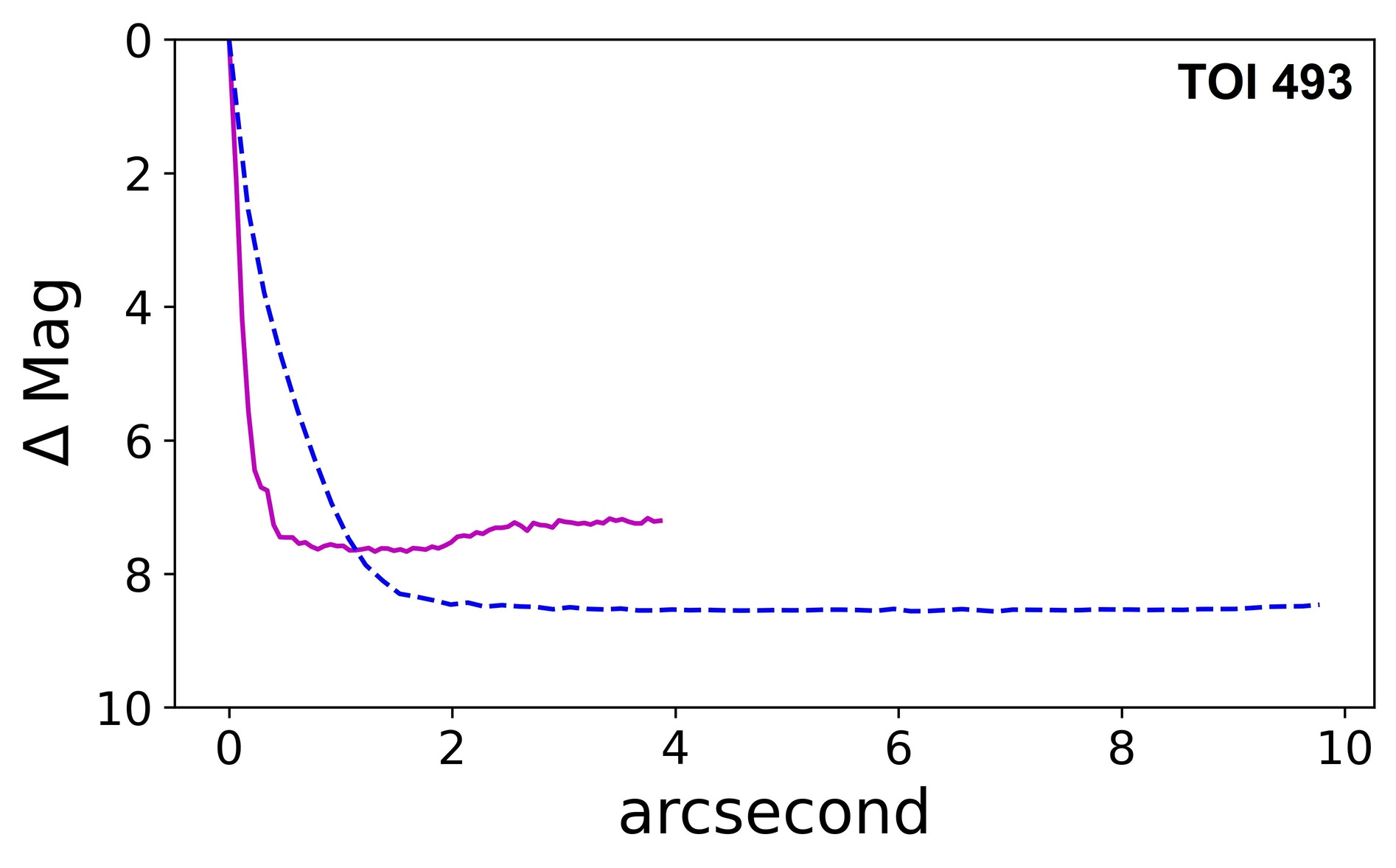}
    \includegraphics[scale = 0.32]{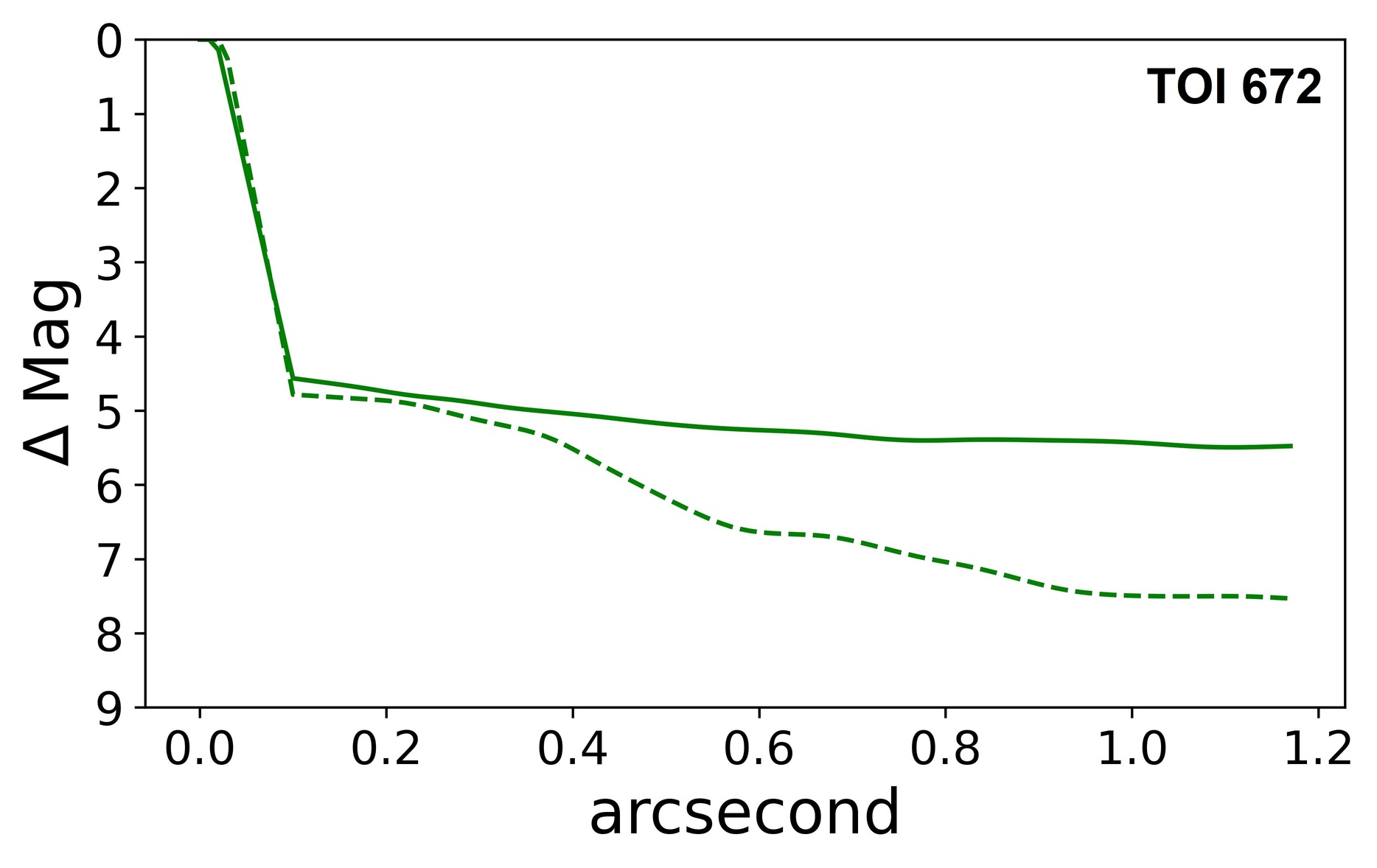}
    \includegraphics[scale = 0.32]{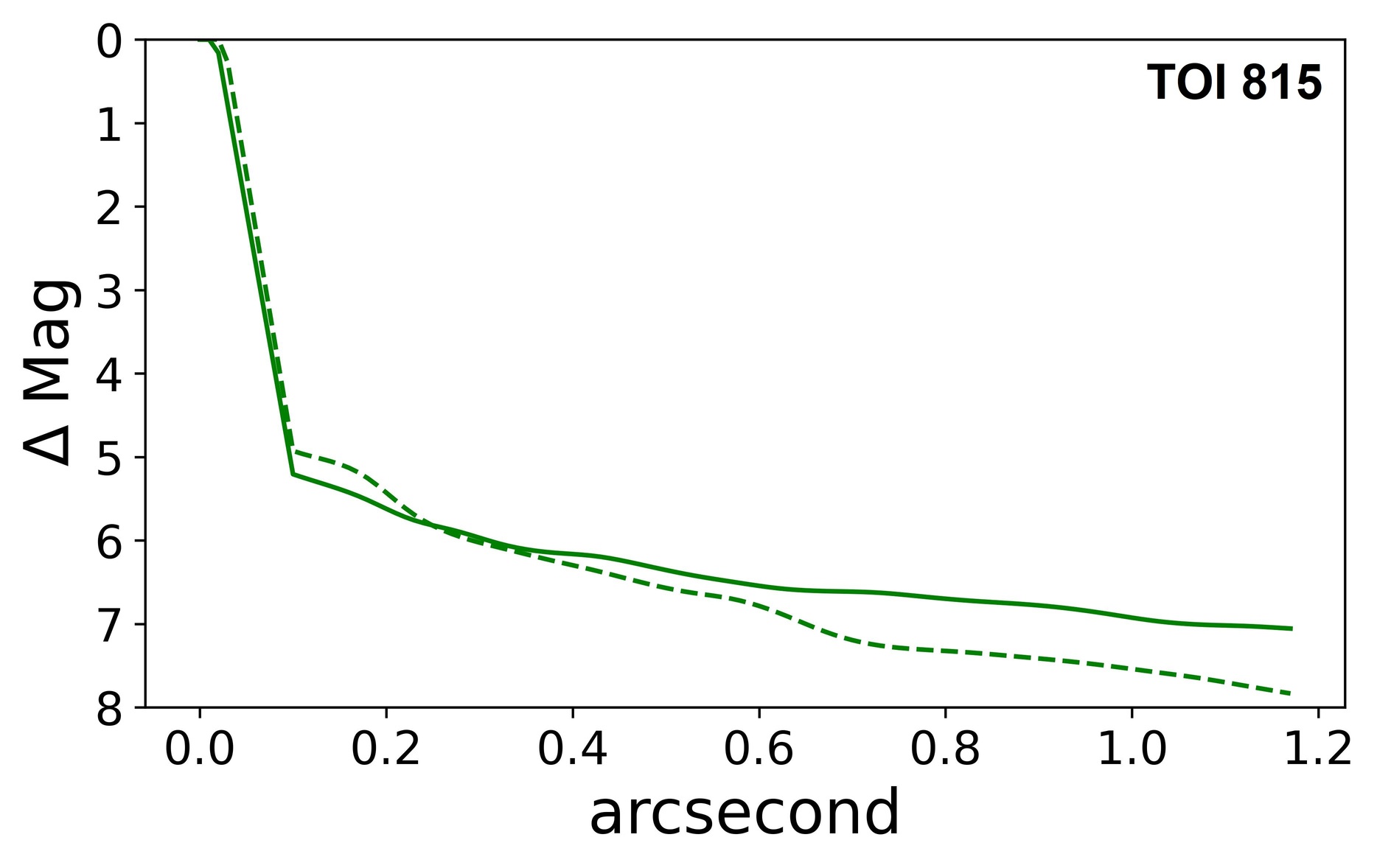}
    \includegraphics[scale = 0.32]{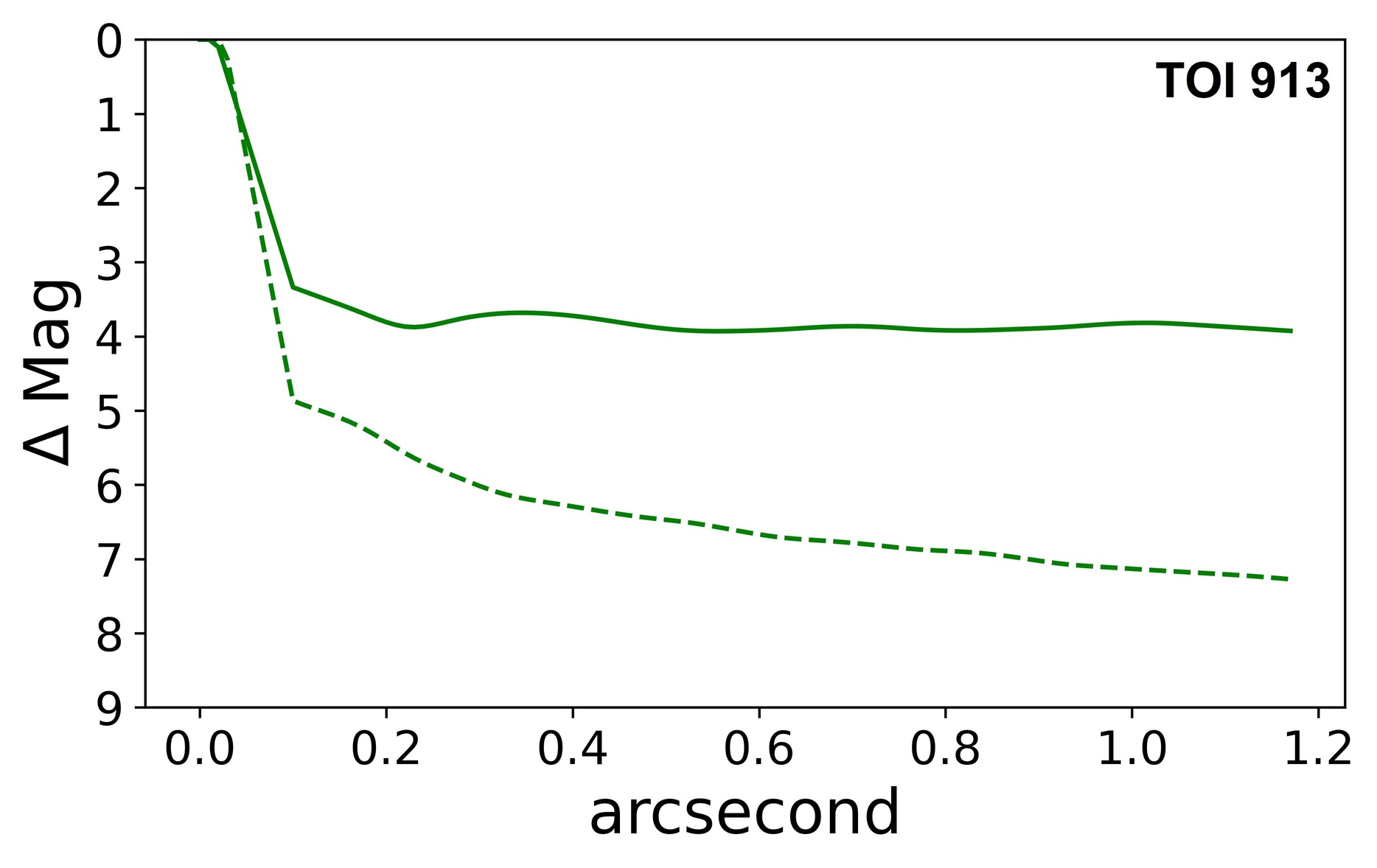}
    \includegraphics[scale = 0.32]{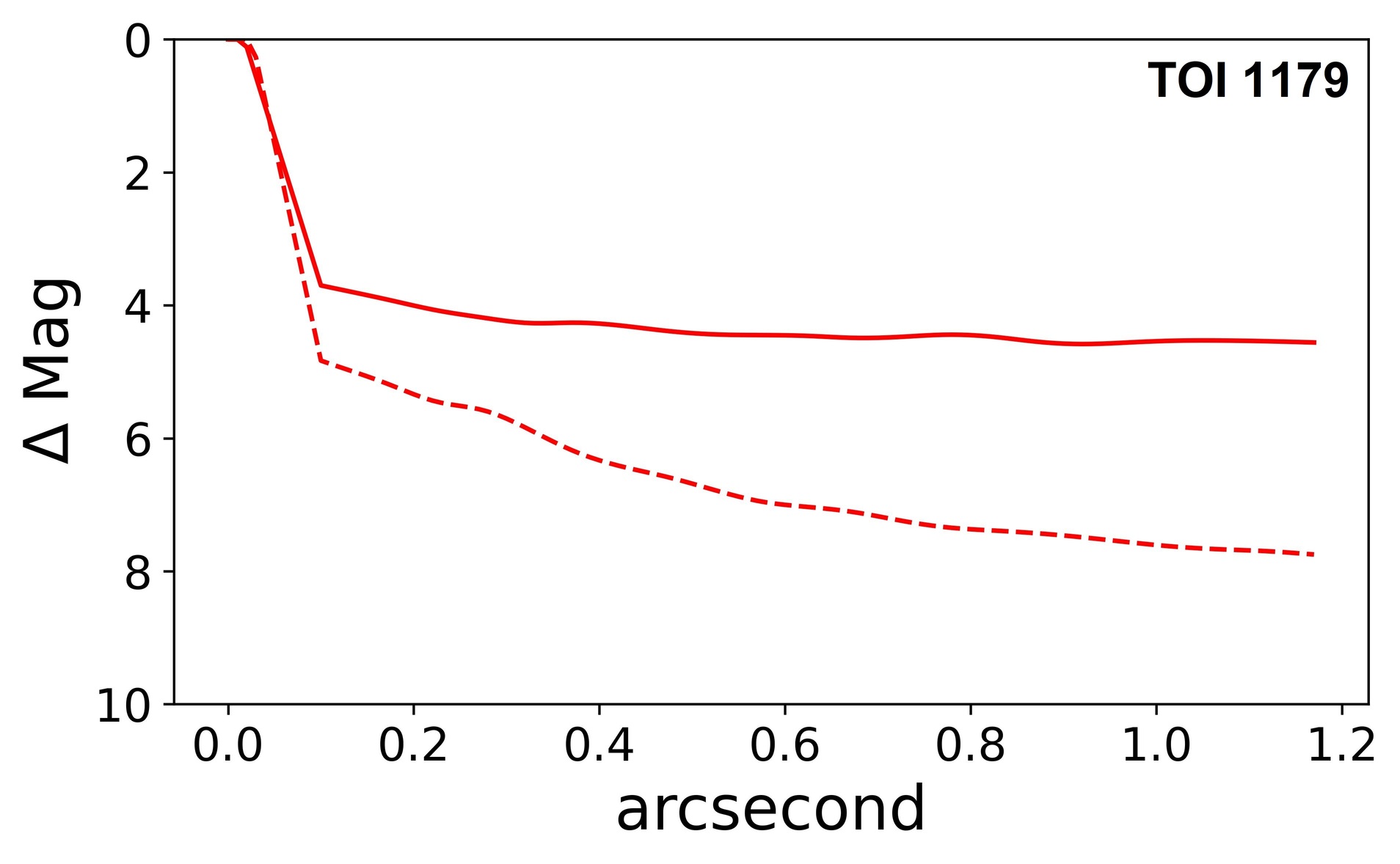}
    \includegraphics[scale = 0.32]{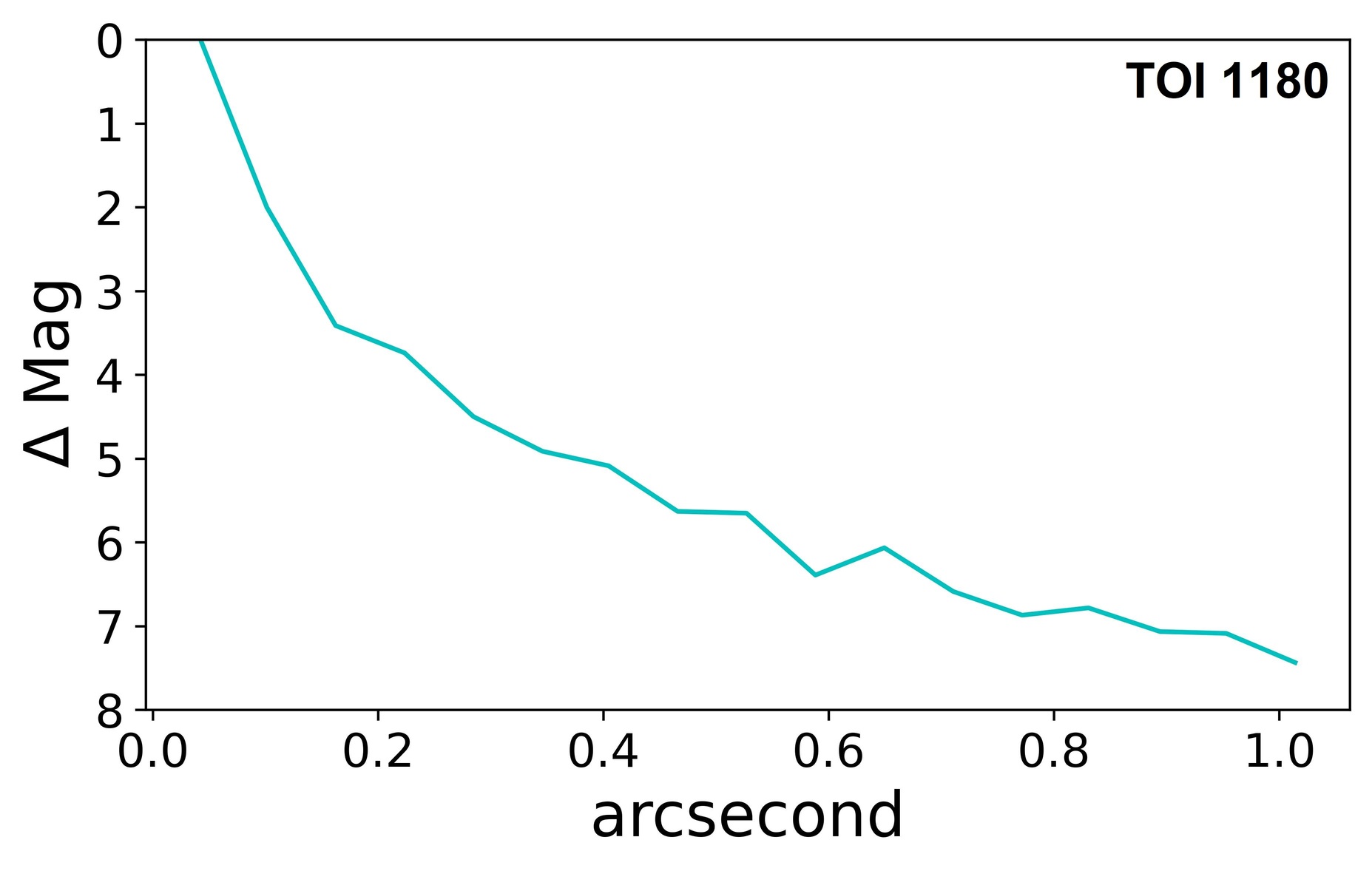}
    \includegraphics[scale = 0.32]{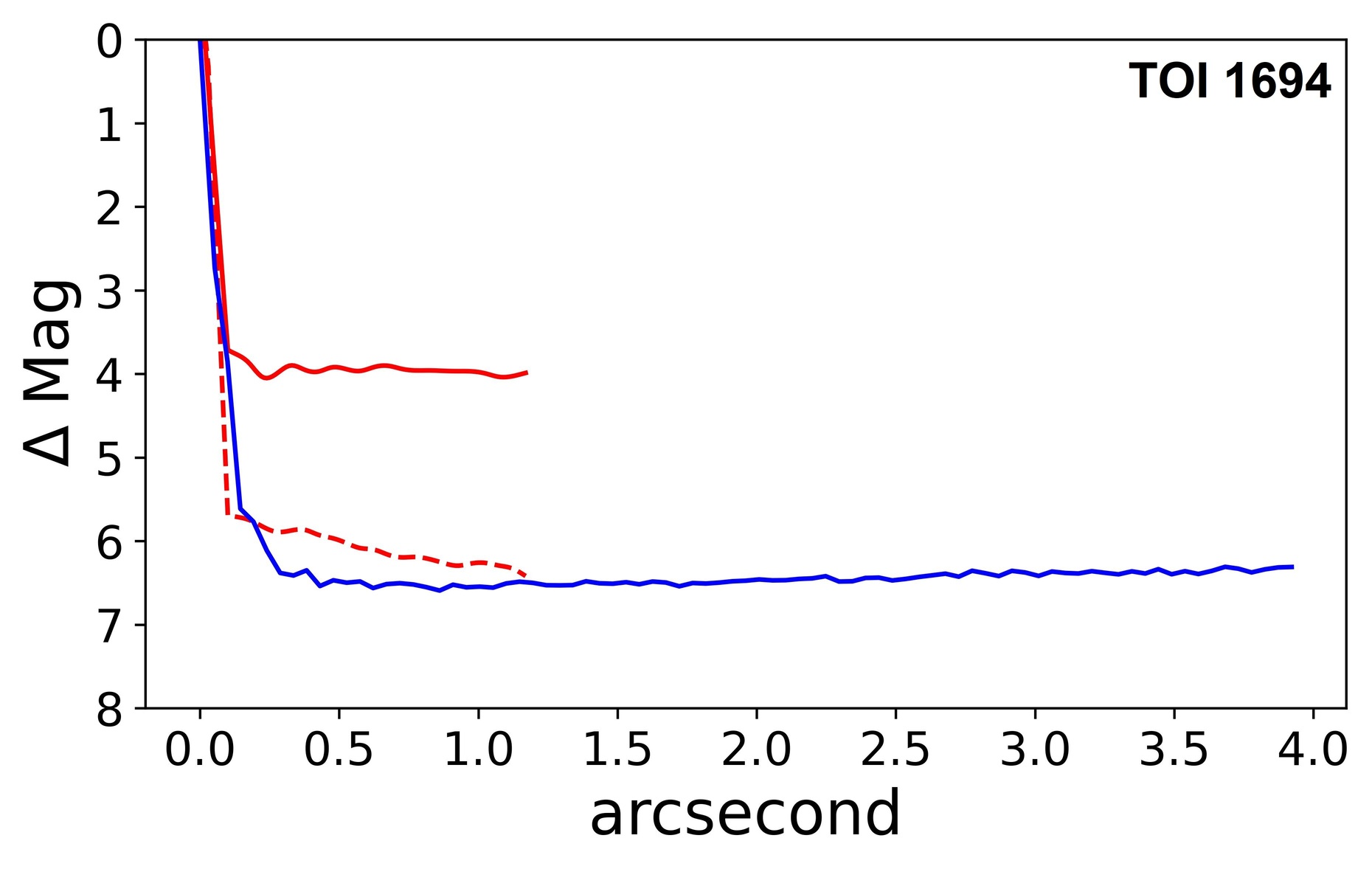}
    \includegraphics[scale = 0.32]{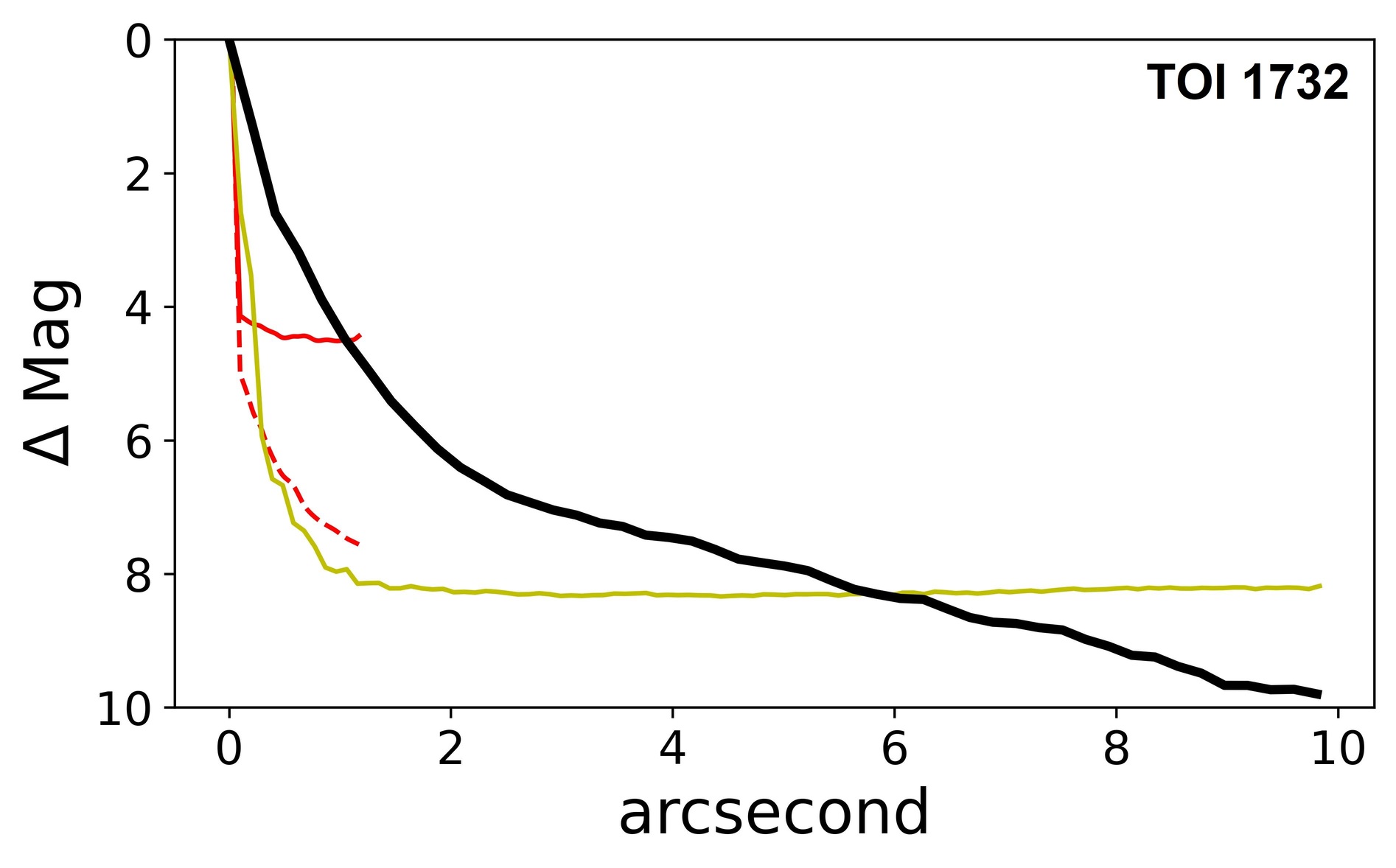}
    \includegraphics[scale = 0.32]{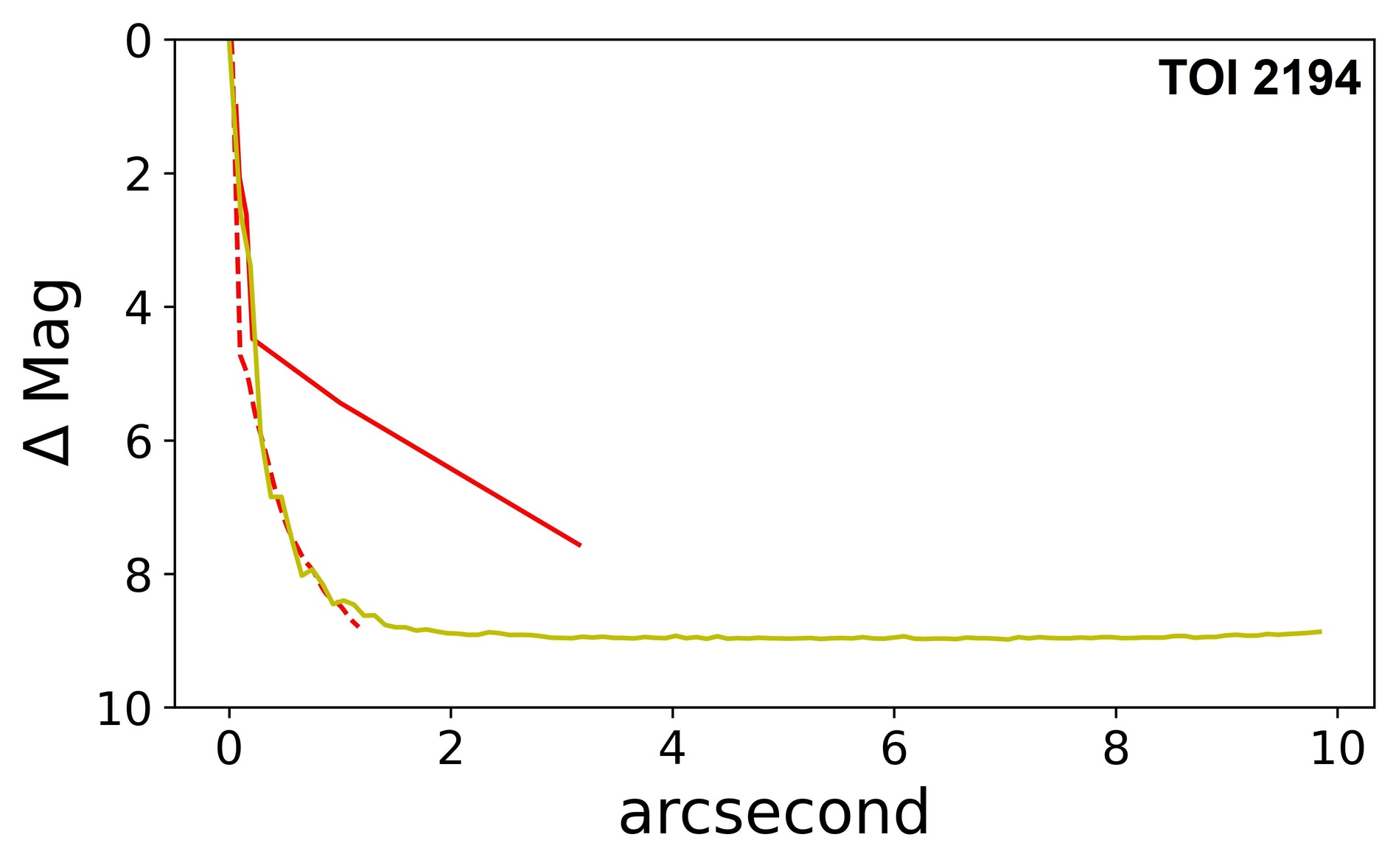}
    \includegraphics[scale = 0.32]{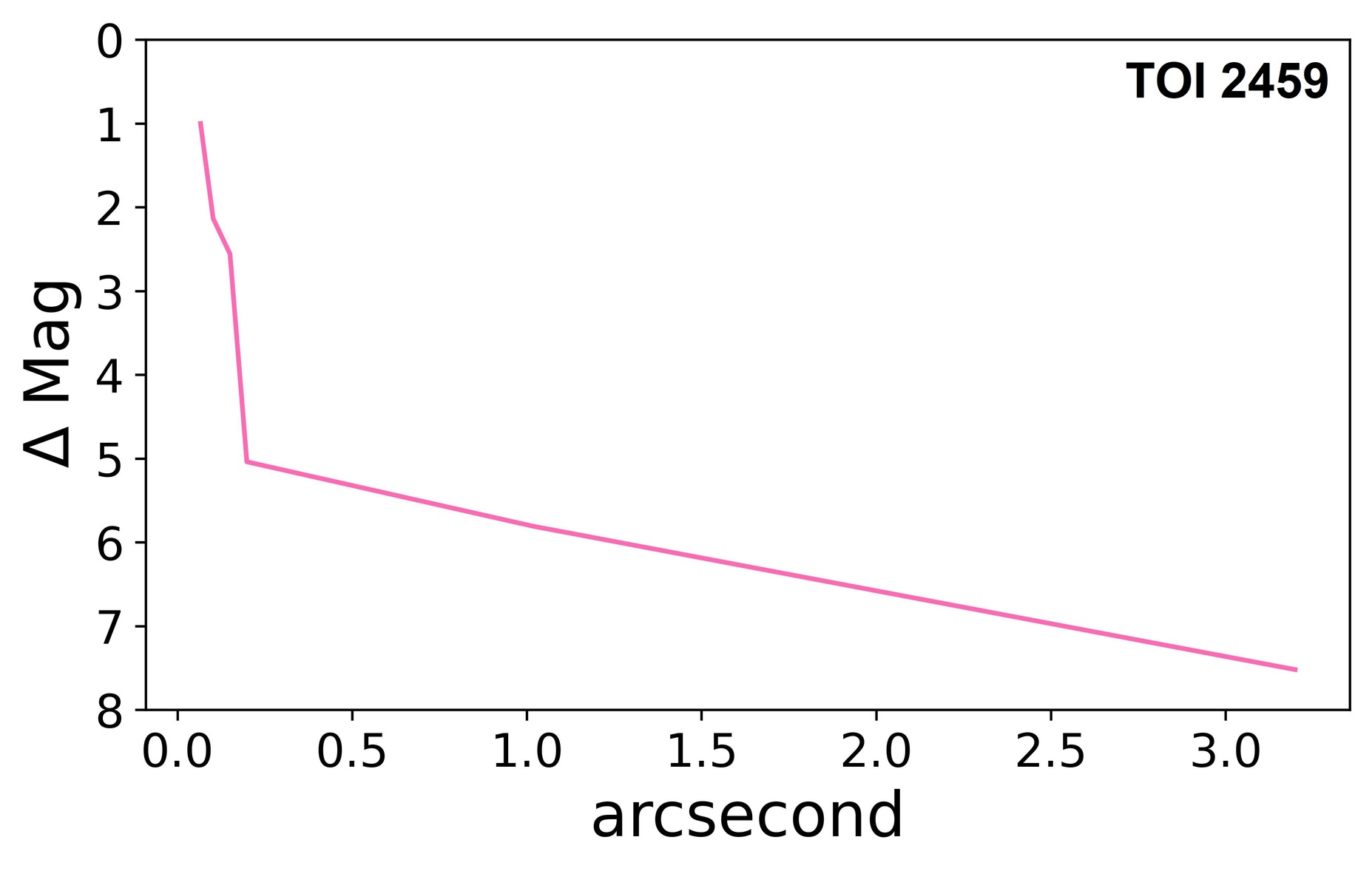}
    \includegraphics[scale = 0.32]{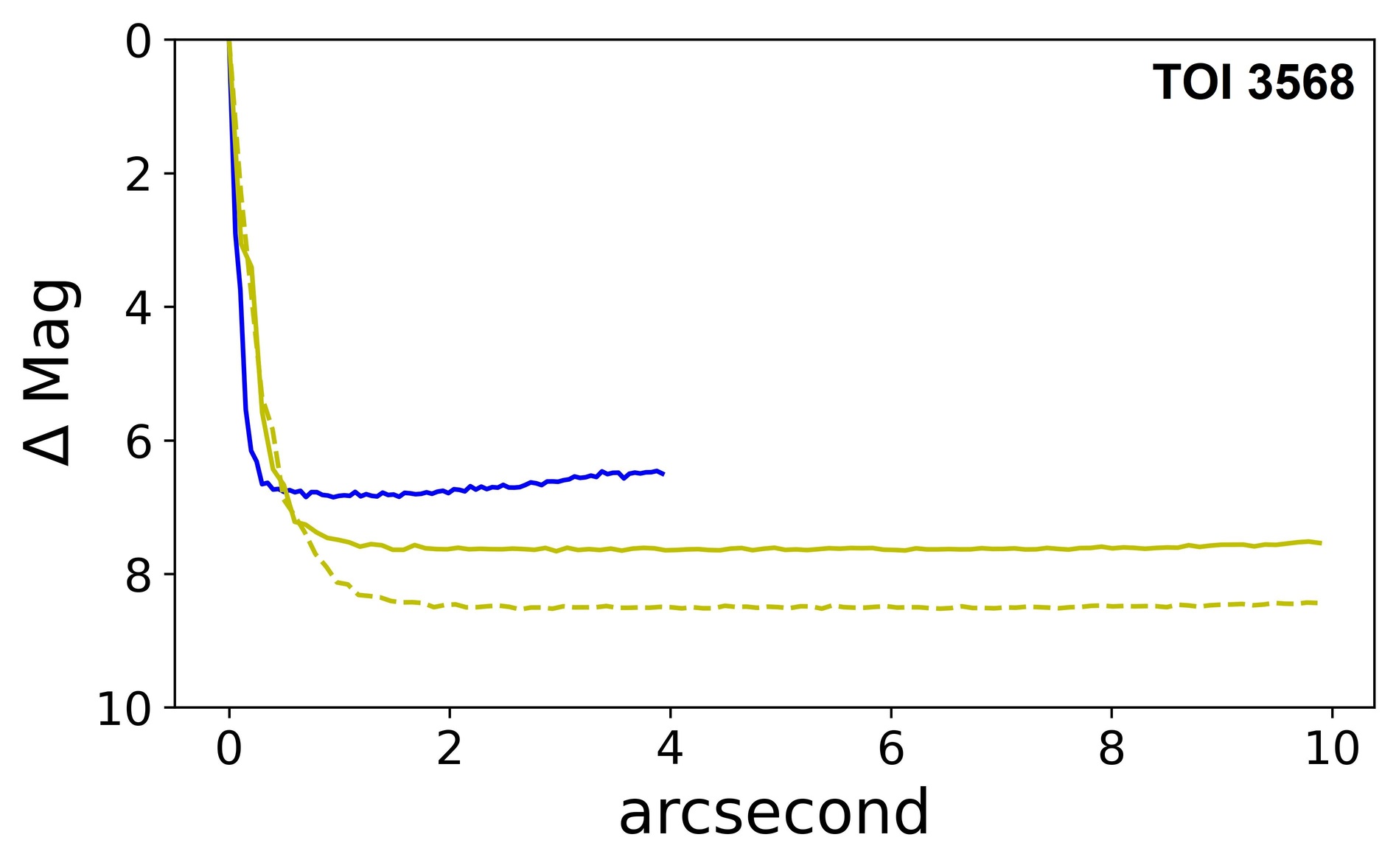}
    \includegraphics[scale = 0.32]{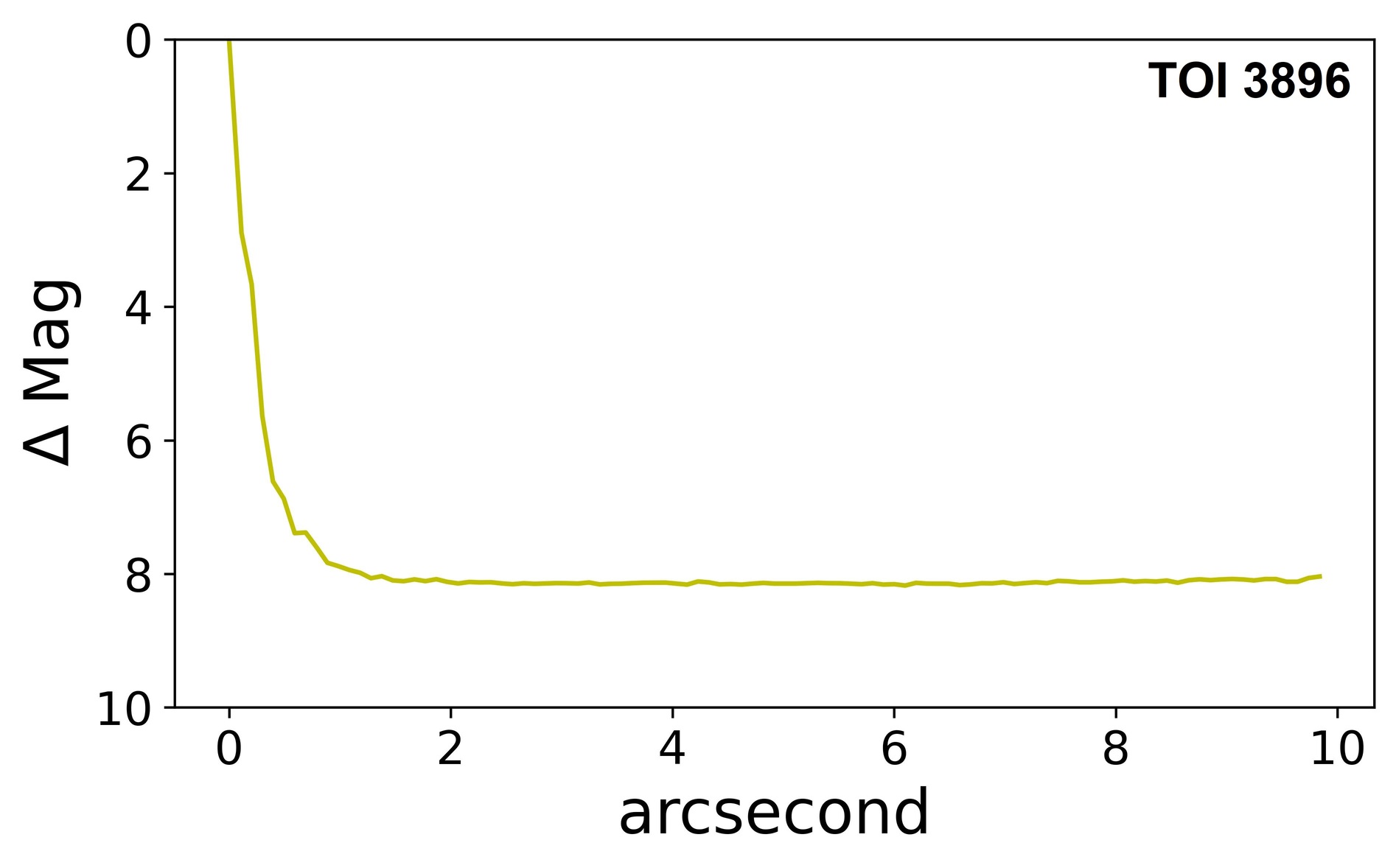}
    \includegraphics[scale = 0.32]{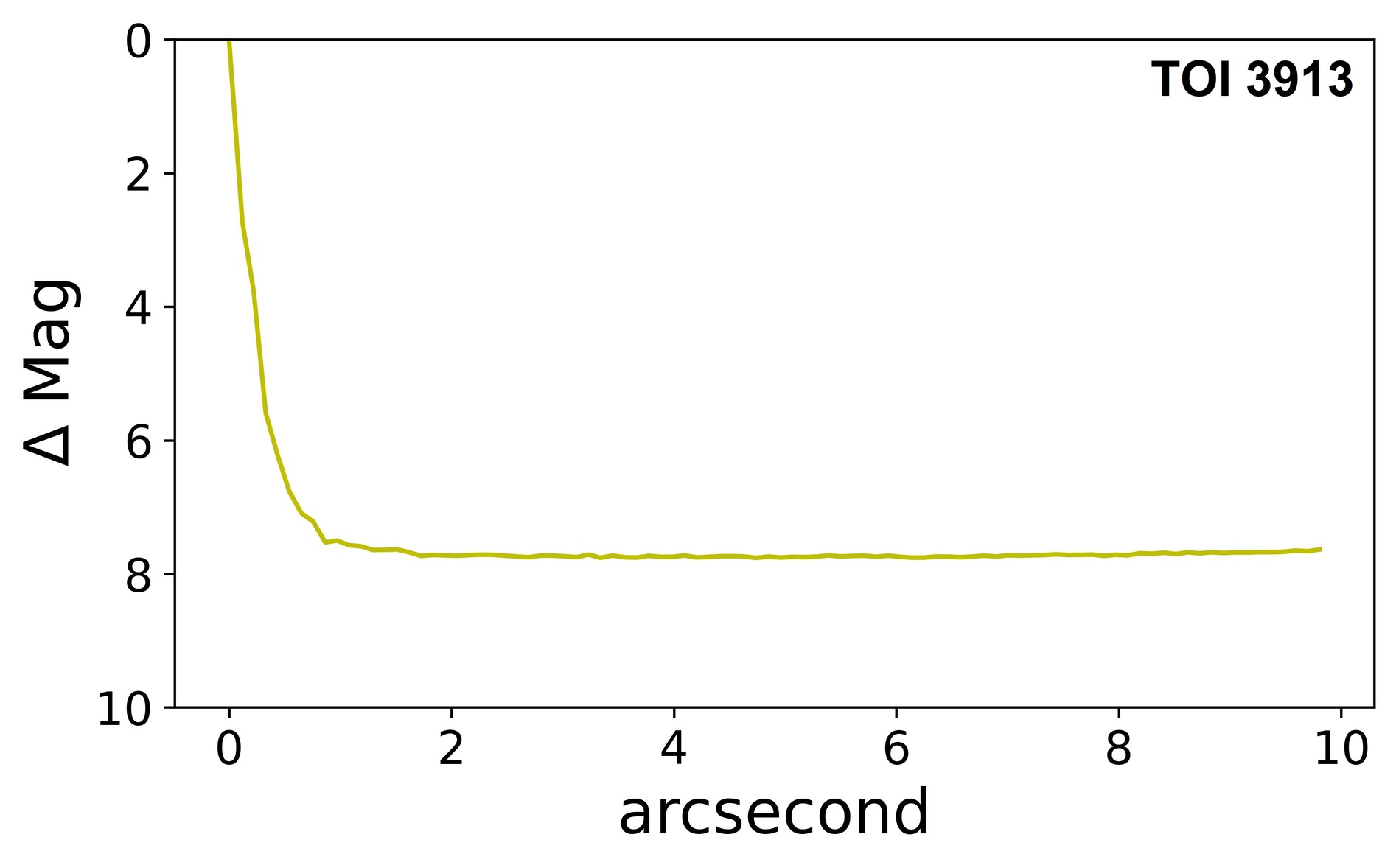}
    \includegraphics[scale = 0.32]{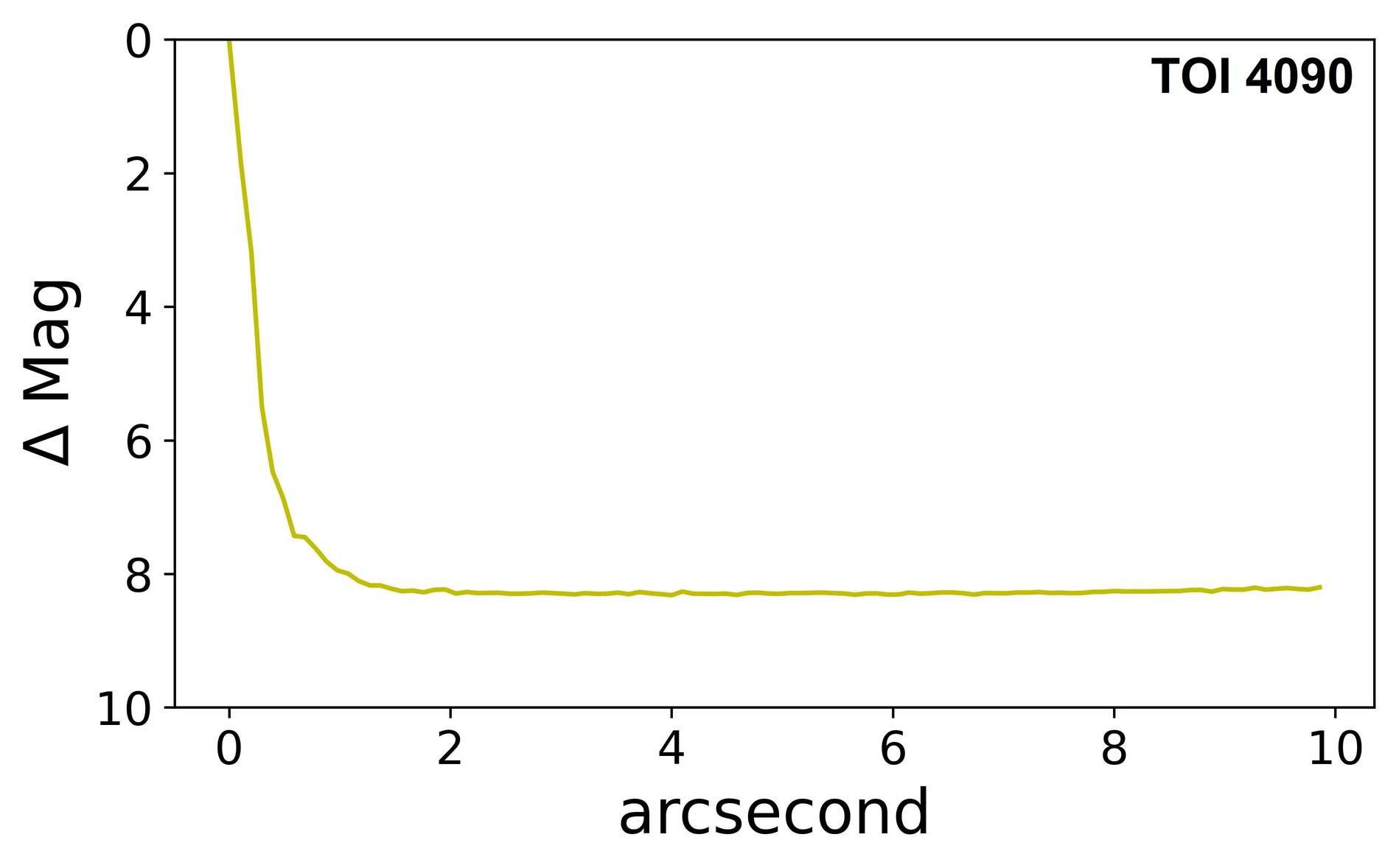}
    \includegraphics[scale = 0.32]{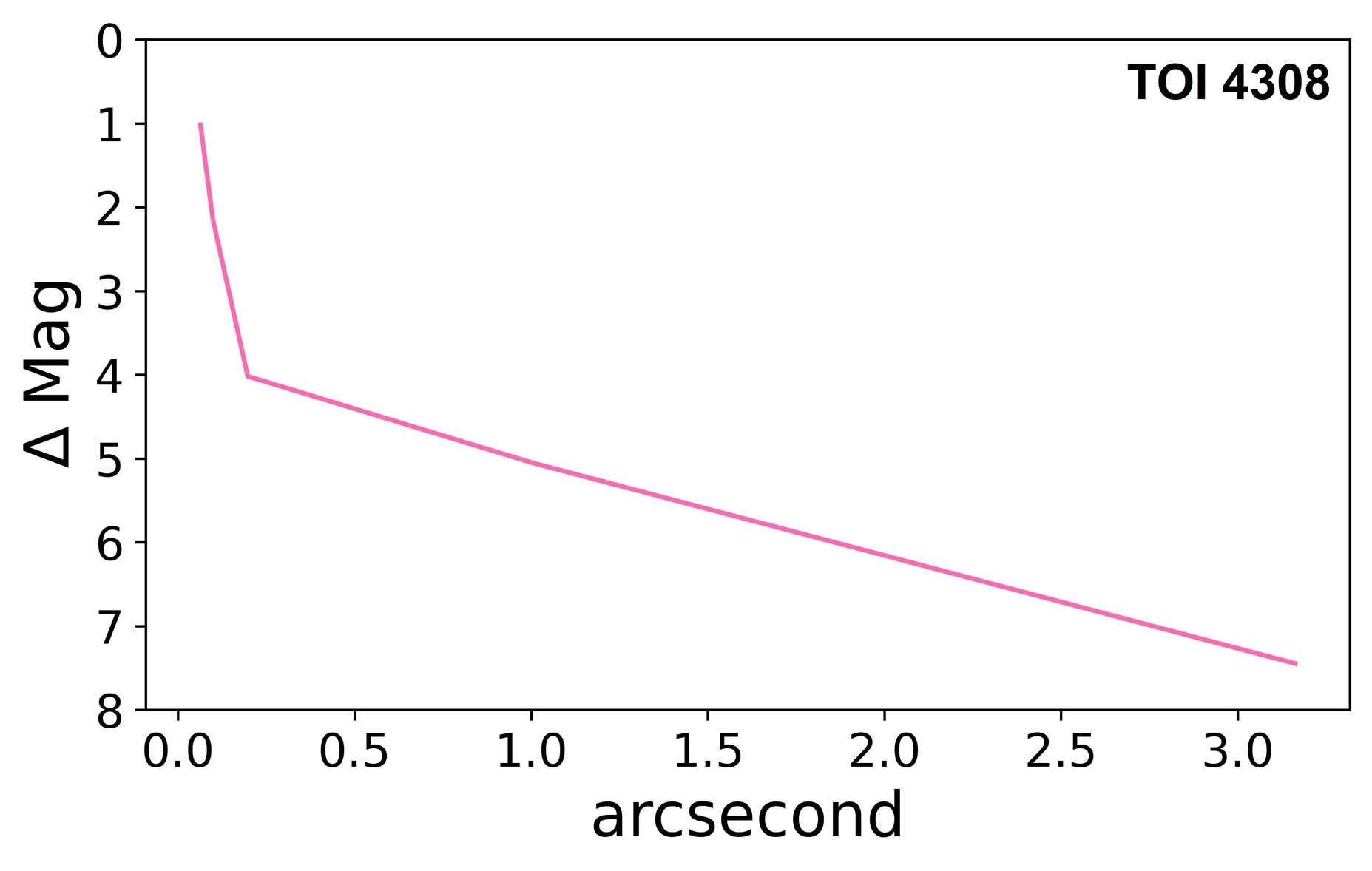}
    \includegraphics[scale = 0.32]{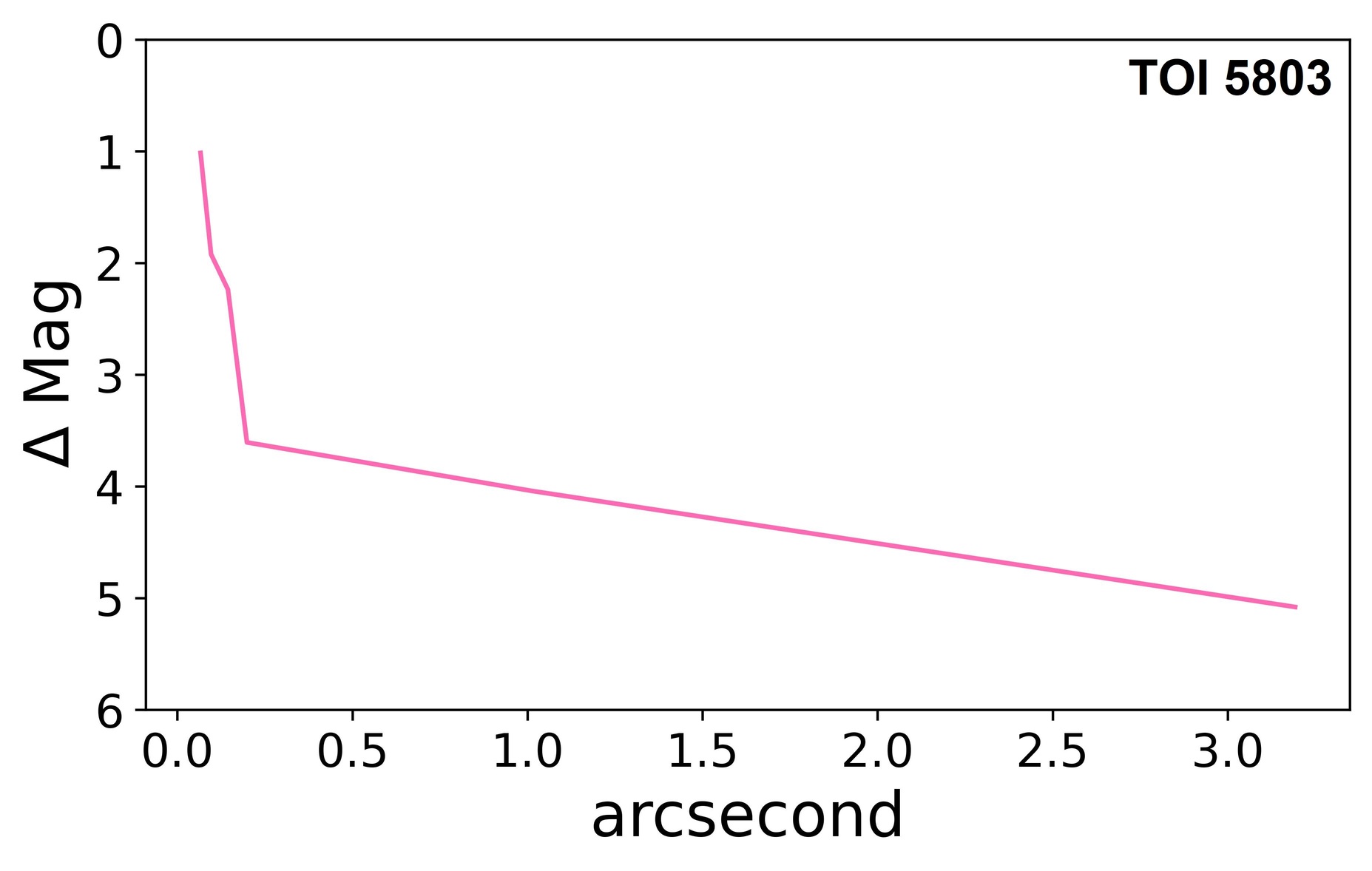}
    \caption{Contrast curves extracted from the high-resolution follow-up observations, which allows us to rule out companions at a given
separation above a certain $\Delta$ Magnitude.}
    \label{fig:cc_images}
\end{figure*}

We observed TIC 470987100 (TOI-1732) on UT 2020 December 01 using the ShARCS camera on the Shane 3-meter telescope at Lick Observatory \citep{2012SPIE.8447E..3GK, 2014SPIE.9148E..05G, 2014SPIE.9148E..3AM}. Observations were taken with the Shane adaptive optics system in natural guide star mode in order to search for nearby, unresolved stellar companions. We collected sequences of observations using $K_S$ filter ($\lambda_0 = 2.150$ $\mu$m, $\Delta \lambda = 0.320$ $\mu$m). We reduced the data using the publicly available \texttt{SImMER} pipeline \citep{2020AJ....160..287S, 2022PASP..134l4501S}.\footnote{\url{https://github.com/arjunsavel/SImMER}} Our reduced images and corresponding contrast curves are shown in Figure \ref{fig:cc_images}. Our observations achieve contrasts of 4.5 (Br$\gamma$) and 2.7 ($J$) at 1$\arcsec$. We find no nearby stellar companions within our detection limits.

\subsubsection{SOAR/HRCam}
A high-resolution camera (HRCam) that can observe the $9.9\arcsec \times 7.5\arcsec$ field of the sky has a 658 $\times$ 496 pixel array, with each pixel able to collect light from a 15 mas region \citep{2010AJ....139..743T}. This instrument was used to collect the speckle imaging observations for TOI-2194, TOI-2459, TOI-4308 and TOI-5803. It is a fast imager designed to work at the SOAR telescope, which uses a CCD detector with internal electro-multiplication (EMCCD). These observations and their related analyses are outlined in
\citep{2020AJ....159...19Z, 2021AJ....162..192Z}. We suggest the reader to those papers for more information.

\subsubsection{VLT/NaCo}
Observations for TOI-323 was performed using NaCo instrument with K filter. NaCo is the Paranal Observatory's instrument, which is a combination of NAOS \citep{2000SPIE.4007...72R} (Nasmyth Adaptive Optics System) and CONICA \citep{1998SPIE.3354..606L} (Near-Infrared Imager and Spectrograph) installed on the Very Large Telescope (VLT). It is able to compensate for the atmospheric variabilities and provides a diffraction-limited resolution for observing wavelengths ranging from 1 to 5 microns. It can collect imaging data with broad and narrow band filters, a field of view of 14$\arcsec$-56$\arcsec$, and a pixel scale of 13-54 mas per pixel.

\subsubsection{SAI/Speckle Polarimeter}
We observed TOI-1180 on 2020 December 02 UT with the Speckle Polarimeter \citep{2017AstL...43..344S} on the 2.5m telescope at the Caucasian Observatory of Sternberg Astronomical Institute (SAI) of Lomonosov Moscow State University. SPP uses Electron Multiplying CCD Andor iXon 897 as a detector. The atmospheric dispersion compensator allowed observation of this relatively faint target through the wide-band $I_c$ filter. The power spectrum was estimated from 4000 frames with 30 ms exposure. The detector has a pixel scale of $20.6$ mas pixel$^{-1}$, and the angular resolution was 89 mas. We did not detect any stellar companions brighter than $\Delta I_C=4$ and $7.2$ at $\rho=0\farcs25$ and $1\farcs0$, respectively, where $\rho$ is the separation between the source and the potential companion.

\startlongtable
\begin{deluxetable*}{cccccccccc}
\tablecaption{Details of High-resolution Imaging data. \label{tab:cc_data}}
\tablehead{
\colhead{TOI} & 
\colhead{Telescope} & 
\colhead{Instrument} &
\colhead{Filter} &
\colhead{Image Type} & 
\multicolumn{5}{c}{Contrast $\Delta$mag} \\
\cline{6-10}
\colhead{} & 
\colhead{} & 
\colhead{} &
\colhead{} & 
\colhead{} &
\colhead{0$\arcsec$.1} &
\colhead{0$\arcsec$.5} &
\colhead{1$\arcsec$.0} &
\colhead{1$\arcsec$.5} & 
\colhead{2$\arcsec$.0}
}
\startdata
139 & Gemini-N (8m) & 'Alopeke & 562 nm & Speckle & 3.849 & 4.104 & 4.450 & --- & --- \\
    & Gemini-N (8m) & 'Alopeke & 832 nm & Speckle & 4.383 & 6.372 & 7.354 & --- & --- \\
    & Keck-2 (10m) & NIRC2 & BrGamma & AO & 4.201 & 6.628 & 6.991 & 6.923 & 6.917 \\
    & Keck-2 (10m) & NIRC2 & Jcont & AO & 2.860 & 5.489 & 6.035 & 6.101 & 6.085 \\
\hline
323 & Gemini-N (8m) & 'Alopeke & 562 nm & Speckle & 3.807 & 4.285 & 4.372 & --- & --- \\
    & Gemini-N (8m) & 'Alopeke & 832 nm & Speckle & 4.369 & 5.234 & 5.558 & --- & --- \\
    & VLT (8m) & NaCo & Ks & AO & 1.405 & 4.993 & 5.240 & 5.219 & 5.165 \\
\hline
493 & Gemini-N (8m) & NIRI & BrGamma & AO & 1.491 & 4.929 & 6.946 & --- & --- \\
    & Keck-2 (10m) & NIRC2 & BrGamma & AO & 3.515 & 7.454 & 7.581 & 7.646 & 7.528 \\
\hline
672 & Gemini-S (8m) & Zorro & 562 nm & Speckle & 4.565 & 5.180 & 5.429 & --- & --- \\
    & Gemini-S (8m) & Zorro & 832 nm & Speckle & 4.784 & 6.182 & 7.494 & --- & --- \\
\hline
815 & Gemini-S (8m) & Zorro & 562 nm & Speckle & 5.206 & 6.354 & 6.925 & --- & --- \\
    & Gemini-S (8m) & Zorro & 832 nm & Speckle & 4.927 & 6.569 & 7.539 & --- & --- \\
\hline
913 & Gemini-S (8m) & Zorro & 562 nm & Speckle & 3.338 & 3.896 & 3.819 & --- & --- \\
    & Gemini-S (8m) & Zorro & 832 nm & Speckle & 4.868 & 6.472 & 7.132 & --- & --- \\
\hline
1179 & Gemini-N (8m) & 'Alopeke & 562 nm & Speckle & 3.701 & 4.419 & 4.541 & --- & --- \\
     & Gemini-N (8m) & 'Alopeke & 832 nm & Speckle & 4.832 & 6.683 & 7.606 & --- & --- \\
\hline
1180 & SAI (2.5m) & Speckle Polarimeter & I & Speckle & 1.960 & 5.643 & 7.355 & --- & --- \\
\hline
1694 & Gemini-N (8m) & 'Alopeke & 562 nm & Speckle & 3.712 & 3.923 & 3.977 & --- & --- \\
     & Gemini-N (8m) & 'Alopeke & 832 nm & Speckle & 5.691 & 5.989 & 6.258 & --- & --- \\
     & Keck2 (10m) & NIRC2 & Ks & AO & 3.854 & 6.482 & 6.545 & 6.504 & 6.459 \\
\hline
1732 & Palomar (5m) & PHARO & BrGamma & AO & 2.592 & 6.768 & 8.216 & 8.275 & 8.250 \\
     & Gemini-N (8m) & 'Alopeke & 562 nm & Speckle & 4.134 & 4.467 & 4.509 & --- & --- \\
     & Gemini-N (8m) & 'Alopeke & 832 nm & Speckle & 5.008 & 6.554 & 7.399 & --- & --- \\
     & Shane (3m) & ShARCS & K & AO & 0.613 & 2.836 & 4.481 & 5.482 & 6.408\\
\hline
2194 & SOAR (4.1m) & HRCam & I & Speckle & 2.118 & 4.831 & 5.437 & 5.931 & 6.423 \\
\hline
2443 & Gemini-N (8m) & 'Alopeke & 562 nm & Speckle & 5.219 & 5.729 & 5.952 & --- & --- \\
     & Gemini-N (8m) & 'Alopeke & 832 nm & Speckle & 5.389 & 6.943 & 8.223 & --- & --- \\
     & Palomar (5m) & PHARO & BrGamma & AO & 2.351 & 6.697 & 7.910 & 8.753 & 9.146\\
\hline
2459 & SOAR (4.1m) & HRCam & I & Speckle & 2.134 & 5.321 & 5.807 & 6.186 & 6.578 \\
\hline
3568 & Keck-2 (10m) & NIRC2 & Ks & AO & 3.746 & 6.773 & 6.833 & 6.812 & 6.791 \\
     & Palomar (5m) & PHARO & BrGamma & AO & 2.809 & 6.670 & 7.491 & 7.639 & 7.627 \\
     & Palomar (5m) & PHARO & Hcont & AO & 2.273 & 6.898 & 8.135 & 8.414 & 8.462 \\
\hline 
3896 & Palomar (5m) & PHARO & BrGamma & AO & 2.585 & 6.908 & 7.893 & 8.101 & 8.128 \\
\hline
3913 & Palomar (5m) & PHARO & BrGamma & AO & 2.289 & 6.551 & 7.520 & 7.635 & 7.726 \\
\hline
4090 & Palomar (5m) & PHARO & BrGamma & AO & 1.724 & 6.929 & 7.958 & 8.235 & 8.263 \\
\hline
4308 & SOAR (4.1m) & HRCam & I & Speckle & 2.143 & 4.405 & 5.053 & 5.602 & 6.156 \\
\hline
5803 & SOAR (4.1m) & HRCam & I & Speckle & 1.921 & 3.766 & 4.039 & 4.272 & 4.510 \\
\hline
\enddata
\end{deluxetable*}

\subsection{Light Curve Follow-up Observations}
The \textit{TESS} pixel scale is $\sim 21\arcsec$ pixel$^{-1}$ and photometric apertures typically extend out to roughly 1 arcminute, generally causing multiple stars to blend in the \textit{TESS} aperture. To rule out a nearby eclipsing binary (NEB) or shallower nearby planet candidate (NPC) blend as the potential source of a \textit{TESS} detection and attempt to detect the signal on-target, we observed our target stars and the nearby fields as part of the \textit{TESS} Follow-up Observing Program\footnote{\url{https://tess.mit.edu/followup}} Sub Group 1 \citep[TFOP;][]{collins:2019}. In some cases, we also observed in multiple bands across the optical spectrum to check for wavelength dependent transit depth differences, which can be suggestive of a planet candidate false positive. We used the {\tt TESS Transit Finder}, which is a customized version of the {\tt Tapir} software package \citep{Jensen:2013}, to schedule our transit observations.

All of our light curve follow-up observations are summarized in Table \ref{tab:transitnebfollowup} and all light curve data are available on the {\tt EXOFOP-TESS} website\footnote{\href{https://exofop.ipac.caltech.edu/tess}{https://exofop.ipac.caltech.edu/tess}}. We also provide a short summary of each light curve result and an overall final photometric follow-up disposition in Table \ref{tab:transitnebfollowup}. We assign four light curve follow-up dispositions (PC, CPC, VPC, VPC+) to indicate differing levels of confidence that a \textit{TESS} detection is on-target, as described below.

The planet candidate (PC) disposition indicates that we either have no light curve follow-up observations, or the light curve observations are unable to confirm that the \textit{TESS}-detected event is on-target relative to Gaia DR3 and TIC version 8 stars.

The cleared planet candidate (CPC) disposition indicates that we have confined the \textit{TESS}-detection to occur on the target star relative to all Gaia DR3 and TIC version 8 stars. Using ground-based photometry, we check all stars out to $2\farcm5$ from the target star that are bright enough, assuming a 100\% eclipse, in \textit{TESS}-band to produce the \textit{TESS}-detected depth at mid-transit. To account for possible delta-magnitude differences between \textit{TESS}-band and the follow-up band, and to account for \textit{TESS}-band magnitude errors, we included an extra 0.5 magnitudes fainter in \textit{TESS}-band. For these cases, the transit depth is generally too shallow to detect on-target in ground-based follow-up observations, so we often saturate the target star on the detector to enable a complete search of all necessary fainter nearby stars. Since the \textit{TESS} point-spread-function has full-width-half-maximum of $\sim40\arcsec$, and the irregularly shaped SPOC photometric apertures and circular QLP photometric apertures generally extend to $\sim1\arcmin$ from the target star, we check for events in stars out to $2\farcm5$ from the target star. For a star to be considered ``cleared'' of an NEB signal, we require its light curve to have an flat model residual RMS value to be at least a factor of 3 smaller than the eclipse depth required to produce the \textit{TESS} detection in the star. We ensure that the predicted ephemeris uncertainty is covered by at least $\pm3\sigma$ relative to the most precise SPOC or QLP ephemeris available at the time of publication. Finally, we check all nearby star light curves by eye to ensure that no obvious eclipse-like event is present. By process of elimination, we conclude that when all necessary nearby stars are ``cleared'' of NEBs, the transit is indeed occurring on-target, or in a star so close to the target star, that it was not detected by Gaia DR3 and is not in TIC version 8.

The verified planet candidate (VPC) disposition indicates that we have confirmed using ground-based follow-up light curve photometry that the \textit{TESS}-detected event is occurring on-target. This is accomplished using follow-up photometric apertures that are small enough to exclude most or all of the flux from the nearest Gaia DR3 and/or TIC version 8 star that is bright enough to be capable of producing the \textit{TESS} signal.

The verified planet candidate plus (VPC+) disposition is the same as VPC, except we have measured transit depths in the target star follow-up photometric apertures across several optical bands. We elevate the disposition to VPC+ if no strong ($>3\sigma$) transit depth difference is detected across the bands.

\startlongtable
\setlength{\tabcolsep}{3pt}
\begin{deluxetable}{>{\centering}p{0.12\linewidth} c >{\centering}p{0.18\linewidth} c c p{0.32\linewidth} c}
\tablecaption{Ground-based light curve observations. \label{tab:transitnebfollowup}}
\tablehead{
\colhead{Observatory} & 
\colhead{Ap (m)} & 
\colhead{Location} &
\colhead{UTC Date} &
\colhead{Filter} & 
\colhead{Result} &
\colhead{Disp.$^{a}$}
}
\startdata
TOI-139.01 & & & & & & \\
\cline{1-1}
SLR2$^{b}$-SAAO & 0.5 & Sutherland, S. Africa & 2018-10-31 & V & possible NEB at 73$\arcsec$ (TIC 62482371) & CPC\\
LCOGT$^{c}$-CTIO & 1.0 & Cerro Tololo, Chile & 2019-07-24 & Y$^{d}$ & cleared all 5 NEB check stars, including the 73$\arcsec$ star & \\
[1.5mm]
TOI-672.01 & & & & & & \\
\cline{1-1}
LCOGT-CTIO & 0.4 & Cerro Tololo, Chile & 2019-05-12 & $i'$ & $\sim9$ ppt transit in $4\arcsec$ target aperture & \\
Evans & 0.36 & El Sauce, Chile & 2019-05-12 & $\rm R_c$ & $\sim8$ ppt egress in $7\arcsec$ target aperture & \\
PEST$^{e}$ & 0.3 & Perth, Australia & 2019-05-26 & $R_c$ & $\sim8$ ppt transit in $7\arcsec$ target aperture & VPC+\\
TRAPPIST-S & 0.6 & La Silla, Chile & 2019-06-09 & $z'$ & $\sim8$ ppt transit in $5\arcsec$ target aperture & \\
Evans & 0.36 & El Sauce, Chile & 2020-02-01 & B & $\sim8$ ppt transit in $6\arcsec$ target aperture & \\
LCOGT-SSO & 1.0 & Siding Spring, Australia & 2020-03-19 & $g'$ & $\sim8$ ppt transit in $4\arcsec$ target aperture & \\
[1.5mm]
TOI-913.01 & & & & & & \\
\cline{1-1} 
LCOGT-CTIO & 1.0 & Cerro Tololo, Chile & 2020-03-05 & $z_s$$^{g}$ & tentative $\sim1$ ppt transit in 5$\arcsec$ target aper. & VPC\\
LCOGT-SAAO & 1.0 & Sutherland, S. Africa & 2020-05-21 & $z_s$ & $\sim1.1$ ppt transit in 5$\arcsec$ target aperture & \\
LCOGT-CTIO & 1.0 & Cerro Tololo, Chile & 2021-05-12 & $z_s$ &  tentative $\sim1$ ppt transit in 5$\arcsec$ target aper. & \\
ASTEP$^{f}$ & 0.4 & Dome C, Antarctica & 2022-09-12 & $R_c$ & tentative $\sim1-2$ ppt ingress in contaminated 11$\arcsec$ target aper. & \\
[1.5mm]
TOI-1694.01 & & & & & & \\
\cline{1-1}
Catania Obs. & 0.91 & Catania, Italy & 2020-02-16 & R & $\sim4$ ppt transit in 3$\arcsec$ target aperture & \\
Grand-Pra Obs. & 0.4 & Valais Sion, Switzerland & 2020-12-10 & $g'$ & $\sim4$ ppt egress in 6$\arcsec$ target aperture & VPC+ \\
Dragonfly & 1.0 & Mayhill, NM, USA & 2021-01-02 & $g'$, $r'$ & $\sim5$ ppt egress in $<26\arcsec$ target apertures & \\
[1.5mm] 
TOI-2194.01 & & & & & & \\
\cline{1-1}
LCOGT-SAAO & 1.0 & Sutherland, S. Africa & 2021-07-09 & Y & all 7 NEB check stars cleared & CPC\\
LCOGT-CTIO & 1.0 & Cerro Tololo, Chile & 2022-05-28 & $z_s$ & tentative 700 ppm transit in $9\arcsec$ target aper. & \\
[1.5mm]
TOI-2443.01 & & & & & & \\
\cline{1-1} 
LCOGT-SSO & 1.0 & Siding Spring, Australia & 2021-09-19 & Y & cleared all 3 NEB check stars & CPC \\
[1.5mm]
TOI-2459.01 & & & & & & \\
\cline{1-1}  
LCOGT-CTIO & 1.0 & Cerro Tololo, Chile & 2021-02-09 & $z_s$ & $\sim2$ ppt ingress in 5$\arcsec$ target aperture & VPC\\
PEST & 0.3 & Perth, Australia & 2021-11-03 & $g'$ & $\sim3$ ppt transit in 8$\arcsec$ target aperture & \\
[1.5mm] 
TOI-3082.01 & & & & & & \\
\cline{1-1}
TRAPPIST-S & 0.6 & La Silla, Chile & 2022-04-20 & I+$z'$ & $\sim2.5$ ppt transit in $5\arcsec$ target aperture & \multirow{4}{*}{VPC+}\\
LCOGT-McD & 1.0 & McDonald Obs, TX, USA & 2022-05-16 & $i'$ & $\sim2.5$ ppt transit in $4\arcsec$ target aperture & \\
LCOGT-CTIO & 1.0 & Cerro Tololo, Chile & 2022-05-16 & $i'$ & $\sim2.5$ ppt transit in $4\arcsec$ target aperture & \\
TCS-MuSCAT2$^{h}$ & 1.52 & Canaries, Spain & 2022-05-21 & $g'$, $r'$, $i'$, $z_s$ & $\sim2.5$ ppt transits in $11\arcsec$ target apertures (transit depths $1.5\sigma$ consistent across bands) & \\
[1.5mm]
TOI-4308.01 & & & & & &\\
\cline{1-1}
 - & - & - & - & - & No follow-up available & PC \\
[1.5mm]
TOI-5704.01 & & & & & & \\
\cline{1-1}
LCOGT MuSCAT3 & 2.0 & Haleakala, Hawaii & 2023-01-24 & $g'$, $r'$, $i'$, $z_s$ & tentative $\sim1.5$ ppt event in  7$\arcsec$ target apertures that are contaminated with 1.5$\arcsec$ neighbor TIC 900281091 ($\Delta$T = 5.42) & PC \\ 
[1.5mm]
TOI-5803.01 & & & & & & \\
\cline{1-1}
 - & - & - & - & - & No follow-up available & PC \\
[1.5mm]
\enddata
\tablenotetext{a}{The overall follow-up disposition. CPC $=$ cleared of NEBs, VPC $=$ on-target relative to Gaia DR3 stars, VPC+ $=$ achromatic on-target relative to Gaia DR3 stars. See the text for full disposition definitions.}
\tablenotetext{b}{{\it Solaris} network of telescopes of the Nicolaus Copernicus Astronomical Center of the Polish Academy of Sciences.}
\tablenotetext{c}{Las Cumbres Observatory Global Telescope \citep[LCOGT;][]{Brown:2013} 0.4\,m, 1.0\,m, 2.0\,m network nodes at Cerro Tololo Inter-American Observatory (CTIO), South Africa Astronomical Observatory (SAAO), Siding Spring Observatory (SSO), McDonald Observatory (McD), and MuSCAT3 \citep{Narita:2020} on Faulkes Telescope North at Haleakala Observatory. Images calibrated by {\tt BANZAI} pipeline \citep{McCully:2018} and photometry extracted using {\tt AstroImageJ} \citep{Collins:2017}.}
\tablenotetext{d} {Pan-STARRS Y band ($\lambda_{\rm c} = 10040$\,\AA, ${\rm Width} =1120$\,\AA)}
\tablenotetext{e}{Perth Exoplanet Survey Telescope. Images calibrated and photometry extracted using {\tt C-Munipack}\footnote{http://c-munipack.sourceforge.net}.}
\tablenotetext{f}{Antarctica Search for Transiting ExoPlanets (ASTEP; \citet{2015AN....336..638G}): 0.4m Newton Telescope installed at the Concordia station, Antarctica using a camera functioning in the R band \citep{2022SPIE12182E..2OS}. Data reduction follows \citet{2016MNRAS.463...45M}}
\tablenotetext{g} {Pan-STARRS $z$-short band ($\lambda_{\rm c} = 8700$\,\AA, ${\rm Width} =1040$\,\AA)}
\tablenotetext{h}{MuSCAT2 \citep{2019JATIS...5a5001N} 4-color multi-band simultaneous camera on the 1.52{\~{}}m Telescopio Carlos S$\backslash$'anchez (TCS). Data reduction follows \citet{Parviainen:2020}.}
\end{deluxetable}

\section{Statistical Validation}
\label{Statistical Validation}
With the advent of dedicated space missions  for finding exoplanets, the number of possible planet-like candidates has increased rapidly. This creates a potential bottleneck between finding an exoplanet candidate and confirming the discovery with multiple follow-up observations. It is expected that TESS alone would be adding ~12,000 potential exoplanets in the database over its 7-year extended mission lifetime \citep{2022AJ....163..290K}. As a result, statistical validation of exoplanets becomes a viable alternative to confirming each candidate with dedicated observations with follow-up telescopes. Furthermore, statistical validation of such likely candidates could also act as a vetting and prioritization procedure for space missions and surveys such as \texttt{JWST} \citep{2006SSRv..123..485G}, \texttt{CHEOPS} \citep{2014cosp...40E.890F} or upcoming \texttt{PLATO} \citep{2022EPSC...16..453R}. The constant improvement in knowledge of exoplanets occurrence rates and studies related to stellar populations have been utilised to derive a statistical threshold for confidently validating transit events as exoplanets.Various codes have been developed over the years with this objective such as \texttt{Pastis} \citep{2014MNRAS.441..983D}, \texttt{DAVE} \citep{2019AJ....157..124K}, \texttt{VESPA} \citep{2015ascl.soft03011M} and \texttt{TRICERATOPS} \citep{2020ascl.soft02004G}.

\texttt{VESPA} can be used if there are no known stars within the maximum radius (\texttt{maxrad}). For our selected targets there are known nearby targets within the \texttt{maxrad}, due to this reason it is not possible to use \texttt{VESPA} for the validation process. We have chosen \texttt{TRICERATOPS} which was developed recently with a focus on the specifics of the TESS mission profile adding to features of \texttt{VESPA}. It has shown positive results for validations of TESS candidates \citep{2021AJ....161...24G}. Unlike \texttt{VESPA}, \texttt{TRICERATOPS} includes known nearby stars in its analysis. Further details about this tool is provided in section \ref{sec:Validation with TRICERATOPS}.

\subsection{Validation with TRICERATOPS}
\label{sec:Validation with TRICERATOPS}
TRICERATOPS \citep{2021AJ....161...24G} is used to validate planet candidates using the Bayesian framework. The algorithm first starts searching for stars within a 2.5' radius of the target star. It determines the contamination of the flux from these stars to the TESS aperture. For the target star and other stars that seem to contribute enough to the transit signal, TRICERATOPS calculates the probability of that signal being generated by a transiting planet, an eclipsing binary, or a nearby eclipsing binary based on the measurements of marginal likelihood for each scenario. This is then combined with prior probability, based on which it calculates the final False Positive Probability (FPP) and Nearby False Positive Probability (NFPP). Mathematically it can be expressed as follows:
\begin{eqnarray}
    FPP &=& 1 - (\mathcal{P}_{TP} + \mathcal{P}_{PTP} + \mathcal{P}_{DTP})\\
    NFPP &=& \Sigma (\mathcal{P}_{NTP} + \mathcal{P}_{NEB} + \mathcal{P}_{NEBX2P})
\end{eqnarray}

Where $\mathcal{P}_j$ shows probability of each scenarios that can be found on Table 1 of \citet{2021AJ....161...24G}, (i.e., TP = No unresolved companion; transiting planet with Period around target star, PTP = Unresolved bound companion; transiting planet with Period around primary star, DTP = Unresolved background star; transiting planet with Period around target star, NTP = No unresolved companion; transiting planet with Period around nearby star, NEB = No unresolved companion; eclipsing binary with Period around nearby star and NEBX2P = No unresolved companion; eclipsing binary with 2 $\times$ Period around nearby star), which can be calculated by,
\begin{eqnarray}
    \mathcal{P}_j = \frac{p(S_j | D)}{\Sigma p(S_j | D)}
\end{eqnarray}
where $p(S_j | D) \propto p(S_j)p(D|S_j)$. $p(S_j)$ is prior probability of each scenario and $p(D | S_j)$ is Marginal likelihood or Bayesian Evidence. 

It can also use high-resolution imaging follow-up observations to constrain the area of sky around the target where an unresolved companion star can exist. To calculate FPP and NFPP using TRICERATOPS we give following input parameters and files: Orbital period in days, transit depth, data of transit photometry, cadence in days, name of filter used by high resolution imaging and data of contrast curve. We calculated FPP and NFPP for each selected target with 15 iterations and tabulated the mean and standard deviation values in table \ref{tab:FPPs}.

%\begin{longrotatetable}
\startlongtable
\begin{deluxetable*}{c c c c c c c}
\tablecaption{False Positive Probabilities of all the targets calculated using TRICERATOPS. $\mu$ represents the mean value and $\sigma$ represents the standard deviation in the values of FPP and NFPP. CC File depicts the name of the instrument (Filter Used) from which high resolution image was taken and respective filters used. TRICERATOPS takes CC file one at a time so we calculated FPP and NFPP for each CC file separately. SNR = Signal to Noise Ratio and FAP = False Alarm Probability, calculated using Transit Least Squares (TLS). \label{tab:FPPs}} 
\tablehead{
\colhead{TOI ID} &
\colhead{TIC ID} &
\colhead{SNR} & 
\colhead{FAP} &
\multicolumn{2}{c}{TRICERATOPS} &
\colhead{CC File} \\
\colhead{} & 
\colhead{} &
\colhead{} & 
\colhead{[\%]} &
\colhead{$\mu$(FPP) $\pm$ $\sigma$(FPP)} &
\colhead{$\mu$(NFPP) $\pm$ $\sigma$(NFPP)} &
\colhead{}
}     
\startdata
\multicolumn{7}{c}{\textit{\textbf{Validated Planets}}} \\
        TOI 139.01 & TIC 62483237 & \\
                   & Sector 01 & 18.6619 & 0.01 & $7.88 \times 10^{-04} \pm 4.88 \times 10^{-04}$ & 0.00 $\pm$ 0.00 & \multirow{2}{*}{'Alopeke (562 nm)} \\
                   & Sector 28 & 15.0097 & 0.01 & $ 9.03 \times 10^{-04} \pm 2.87 \times 10^{-04}$ & 0.00 $\pm$ 0.00 & \\
                   & Sector 01 & & & $ 2.25 \times 10^{-04} \pm 7.13 \times 10^{-05}$ & 0.00 $\pm$ 0.00 & \multirow{2}{*}{'Alopeke (832 nm)} \\
                   & Sector 28 & & & $ 3.05 \times 10^{-04} \pm 7.06 \times 10^{-05}$ & 0.00 $\pm$ 0.00 & \\
                   & Sector 01 & & & $ 3.12 \times 10^{-04} \pm 1.58 \times 10^{-04}$ & 0.00 $\pm$ 0.00 & \multirow{2}{*}{NIRC2 (BrGamma)} \\
                   & Sector 28 & & & $ 3.34 \times 10^{-04} \pm 2.16 \times 10^{-04}$ & 0.00 $\pm$ 0.00 & \\
                   & Sector 01 & & & $ 3.78 \times 10^{-04} \pm 1.99 \times 10^{-04}$ & 0.00 $\pm$ 0.00 & \multirow{2}{*}{NIRC2 (J)}\\
                   & Sector 28 & & & $ 8.44 \times 10^{-04} \pm 3.76 \times 10^{-04}$ & 0.00 $\pm$ 0.00 & \\
         \hline
        TOI 672.01 & 151825527  \\
                   & Sector 09 & 43.1068 & 0.01 & $ 7.34 \times 10^{-09} \pm 9.90 \times 10^{-09}$ & $ 2.52 \times 10^{-13} \pm 2.63 \times 10^{-13}$ & \multirow{3}{*}{Zorro (562 nm)} \\
                   & Sector 10 & 38.8176 & 0.01 & $ 5.56 \times 10^{-07} \pm 2.00 \times 10^{-06}$ & $ 5.14 \times 10^{-13} \pm 2.56 \times 10^{-13}$ \\
                   & Sector 36 & 41.0399 & 0.01 & $ 9.12 \times 10^{-06} \pm 3.07 \times 10^{-05}$ & $ 3.26 \times 10^{-46} \pm 1.84 \times 10^{-46}$ & \\
          & \\
                   & Sector 09 & & &  $ 2.01 \times 10^{-07} \pm 4.45 \times 10^{*07}$ & $ 1.20 \times 10^{-13} \pm 6.99 \times 10^{-14}$ & \multirow{3}{*}{Zorro (832 nm)}\\
                   & Sector 10 & & &  $ 2.42 \times 10^{-08} \pm 5.67 \times 10^{*08}$ & $ 5.48 \times 10^{-13} \pm 2.62 \times 10^{-13}$ \\
                   & Sector 36 & & &  $ 6.76 \times 10^{-08} \pm 1.99 \times 10^{*07}$ & $ 2.71 \times 10^{-46} \pm 1.68 \times 10^{-46}$ \\
         \hline
        TOI 913.01 & 407126408 \\
                   & Sector 12 & 15.6221 & 0.01 & $ 4.01 \times 10^{-04} \pm 1.34 \times 10^{-04}$ & $ 1.05 \times 10^{-26} \pm 4.40 \times 10^{-28}$ & \multirow{2}{*}{Zorro (562 nm)} \\
                   & Sector 13 & 16.5437 & 0.01 & $ 2.59 \times 10^{-03} \pm 4.95 \times 10^{-04}$ & $ 1.82 \times 10^{-71} \pm 9.15 \times 10^{-73}$ \\
          & \\
                   & Sector 12 & & & $ 1.04 \times 10^{-04} \pm 3.71 \times 10^{-05}$ & $ 1.11 \times 10^{-26} \pm 5.58 \times 10^{-28}$ & \multirow{2}{*}{Zorro (832 nm)} \\
                   & Sector 13 & & & $ 8.26 \times 10^{-04} \pm 2.04 \times 10^{-04}$ & $ 1.90 \times 10^{-71} \pm 1.14 \times 10^{-72}$ \\
         \hline
        TOI 1694.01 & 396740648 \\
                    & Sector 19 & 49.3381 & 0.01 & $ 1.76 \times 10^{-03} \pm 2.53 \times 10^{-03}$ & 0.00 $\pm$ 0.00 & \multirow{2}{*}{'Alopeke (562 nm)} \\
                    & Sector 20 & 48.6979 & 0.01 & $ 1.59 \times 10^{-03} \pm 2.52 \times 10^{-03}$ & $ 5.26 \times 10^{-108} \pm 6.93 \times 10^{-108}$ \\
          & \\
                    & Sector 19 & & & $ 2.71 \times 10^{-03} \pm 3.24 \times 10^{-03}$ & 0.00 $\pm$ 0.00 & \multirow{2}{*}{'Alopeke (832 nm)} \\
                    & Sector 20 & & & $ 2.82 \times 10^{-03} \pm 3.74 \times 10^{-03}$ & $ 6.73 \times 10^{-108} \pm 1.14 \times 10^{-107}$ \\
          & \\
                    & Sector 19 & & & $ 1.92 \times 10^{-05} \pm 4.58 \times 10^{-05}$ & 0.00 $\pm$ 0.00 & \multirow{2}{*}{NIRC2 (Ks)}\\
                    & Sector 20 & & & $ 8.93 \times 10^{-06} \pm 1.69 \times 10^{-05}$ & $ 3.51 \times 10^{-108} \pm 5.77 \times 10^{-108}$ \\
         \hline
        TOI 2194.01 & 271478281 \\
                     & Sector 27 & 29.3712 & 0.01 & $ 1.32 \times 10^{-06} \pm 3.88 \times 10^{-06}$ & 0.00 $\pm$ 0.00 & HRCam (I)\\
        \hline
        TOI 2443.01 & 318753380 \\
                     & Sector 31 & 33.3454 & 0.01 & $ 1.58 \times 10^{-17} \pm 8.24 \times 10^{-17}$ & $ 1.97 \times 10^{-19} \pm 1.00 \times 10^{-19}$ & 'Alopeke (562 nm)\\
                     & Sector 31 & & & $ 1.05 \times 10^{-13} \pm 3.95 \times 10^{-13}$ & $ 2.28 \times 10^{-19} \pm 1.21 \times 10^{-19}$ & 'Alopeke (832 nm)\\
                     & Sector 31 & & & $ 1.11 \times 10^{-11} \pm 3.55 \times 10^{-11}$ & $ 1.50 \times 10^{-19} \pm 5.81 \times 10^{-20}$ & PHARO (BrGamma)\\
        \hline
        TOI 2459.01 & 192790476 \\
                     & Sector 05 & & & $ 8.52 \times 10^{-04} \pm 7.63 \times 10^{-05}$ & $ 8.28 \times 10^{-04} \pm 6.07 \times 10^{-05}$ & \multirow{4}{*}{HRCam (I)} \\
                     & Sector 06 & & & $ 3.71 \times 10^{-04} \pm 2.03 \times 10^{-04}$ & $ 2.56 \times 10^{-04} \pm 2.04 \times 10^{-05}$ & \\
                     & Sector 32 & & & $ 2.44 \times 10^{-06} \pm 6.22 \times 10^{-06}$ & $ 1.44 \times 10^{-07} \pm 1.67 \times 10^{-08}$ & \\
                     & Sector 33 & & & $ 7.08 \times 10^{-04} \pm 6.01 \times 10^{-05}$ & $ 6.98 \times 10^{-04} \pm 4.91 \times 10^{-05}$ & \\
        \hline
        TOI 3082.01 & 428699140 \\
                     & Sector 37 & 16.8096 & 0.01 & $ 6.78 \times 10^{-03} \pm 1.06 \times 10^{-03}$ & $ 1.39 \times 10^{-27} \pm 1.06 \times 10^{-27}$ & - \\
        \hline
        TOI 4308.01 & 144193715 \\
                     & Sector 01 & 8.9739 & 0.01 & $ 5.98 \times 10^{-03} \pm 3.63 \times 10^{-04}$ & $ 1.64 \times 10^{-10} \pm 2.40 \times 10^{-11}$ & HRCam (I) \\
        \hline
        TOI 5704.01 & 148673433 \\
                     & Sector 22 & 18.0160 & 0.01 & $ 8.56 \times 10^{-03} \pm 4.97 \times 10^{-05}$ & $ 5.51 \times 10^{-04} \pm 1.49 \times 10^{-05}$ & \multirow{2}{*}{-} \\
                     & Sector 48 & 16.8362 & 0.01 & $ 6.61 \times 10^{-03} \pm 1.03 \times 10^{-03}$ & $ 5.82 \times 10^{-06} \pm 5.28 \times 10^{-07}$ \\
        \hline
        TOI 5803.01 & 466382581\\
                     & Sector 55 & 18.9821 & 0.01 & $ 9.31 \times 10^{-03} \pm 2.74 \times 10^{-03}$ & $ 4.54 \times 10^{-08} \pm 3.21 \times 10^{-09}$ & HRCam (I) \\
        \hline
         \multicolumn{7}{c}{\textit{\textbf{Likely Planets}}} \\
        TOI 323 & 251852984 \\
                    & Sector 37 & 12.8622 & 0.01 & $ 2.49 \times 10^{-01} \pm 2.30 \times 10^{-02}$ & $ 3.20 \times 10^{-11} \pm 7.36 \times 10^{-12}$ & 'Alopeke (562 nm)\\
                    & Sector 37 & & & $ 2.57 \times 10^{-01} \pm 2.75 \times 10^{-02}$ & $ 3.59 \times 10^{-11} \pm 5.88 \times 10^{-12}$ & 'Alopeke (832 nm) \\
                    & Sector 37 & & & $ 2.56 \times 10^{-01} \pm 2.09 \times 10^{-02}$ & $ 2.78 \times 10^{-11} \pm 4.54 \times 10^{-12}$ & NaCo (K) \\
        \hline
        TOI 1180 & 158002130 \\
                     & Sector 14 & 17.9456 & 0.01 & $ 7.99 \times 10^{-03} \pm 5.87 \times 10^{-04}$ & $ 7.13 \times 10^{-04} \pm 2.59 \times 10^{-05}$ & \multirow{8}{*}{Speckle Polarimeter (I)} \\
                     & Sector 19 & 19.8802 & 0.01 & $ 6.56 \times 10^{-03} \pm 1.13 \times 10^{-03}$ & $ 2.34 \times 10^{-04} \pm 1.67 \times 10^{-05}$ \\
                     & Sector 20 & 13.5258 & 0.01 & $ 1.28 \times 10^{-02} \pm 1.48 \times 10^{-03}$ & $ 4.39 \times 10^{-03} \pm 1.13 \times 10^{-04}$ \\
                     & Sector 21 & 13.9659 & 0.01 & $ 1.30 \times 10^{-02} \pm 1.14 \times 10^{-03}$ & $ 4.31 \times 10^{-03} \pm 1.15 \times 10^{-04}$ \\
                     & Sector 40 & 20.0742 & 0.01 & $ 5.07 \times 10^{-03} \pm 1.15 \times 10^{-03}$ & $ 1.71 \times 10^{-05} \pm 2.48 \times 10^{-06}$ \\
                     & Sector 41 & 11.1516 & 0.01 & $ 4.87 \times 10^{-03} \pm 1.11 \times 10^{-03}$ & $ 2.22 \times 10^{-05} \pm 3.03 \times 10^{-06}$ \\
                     & Sector 47 & 17.2377 & 0.01 & $ 1.86 \times 10^{-02} \pm 2.32 \times 10^{-03}$ & $ 9.15 \times 10^{-03} \pm 8.68 \times 10^{-04}$ \\
                     & Sector 48 & 17.6670 & 0.01 & $ 3.99 \times 10^{-02} \pm 3.78 \times 10^{-03}$ & $ 1.86 \times 10^{-02} \pm 1.40 \times 10^{-03}$ \\
        \hline
        TOI 2200 & 142105158 & \\
                     & Sector 27 & 31.8576 & 0.01 & $ 1.05 \times 10^{-01} \pm 5.56 \times 10^{-02}$ & $ 1.15 \times 10^{-19} \pm 3.67 \times 10^{-10}$ & \multirow{12}{*}{-} \\
                     & Sector 28 & 33.0523 & 0.01 & $ 1.46 \times 10^{-01} \pm 5.07 \times 10^{-02}$ & $ 2.01 \times 10^{-07} \pm 5.37 \times 10^{-08}$ \\
                     & Sector 29 & 35.2422 & 0.01 & $ 6.15 \times 10^{-02} \pm 2.88 \times 10^{-02}$ & $ 7.44 \times 10^{-12} \pm 2.17 \times 10^{-12}$ \\
                     & Sector 30 & 34.8712 & 0.01 & $ 5.01 \times 10^{-01} \pm 9.29 \times 10^{-02}$ & $ 2.95 \times 10^{-06} \pm 4.58 \times 10^{-08}$ \\
                     & Sector 31 & 33.0510 & 0.01 & $ 5.85 \times 10^{-02} \pm 2.04 \times 10^{-02}$ & $ 1.48 \times 10^{-10} \pm 4.34 \times 10^{-11}$ \\
                     & Sector 32 & 33.4764 & 0.01 & $ 3.94 \times 10^{-02} \pm 2.08 \times 10^{-02}$ & $ 2.83 \times 10^{-14} \pm 1.28 \times 10^{-14}$ \\
                     & Sector 33 & 34.5528 & 0.01 & $ 4.82 \times 10^{-01} \pm 9.15 \times 10^{-02}$ & $ 2.42 \times 10^{-05} \pm 4.74 \times 10^{-06}$ \\
                     & Sector 34 & 35.6292 & 0.01 & $ 2.95 \times 10^{-01} \pm 6.65 \times 10^{-02}$ & $ 5.68 \times 10^{-10} \pm 2.01 \times 10^{-10}$ \\
                     & Sector 36 & 33.9929 & 0.01 & $ 3.52 \times 10^{-01} \pm 9.12 \times 10^{-02}$ & $ 6.92 \times 10^{-05} \pm 9.69 \times 10^{-05}$ \\ 
                     & Sector 37 & 34.8538 & 0.01 & $ 7.23 \times 10^{-02} \pm 2.41 \times 10^{-02}$ & $ 5.37 \times 10^{-13} \pm 1.13 \times 10^{-13}$ \\
                     & Sector 38 & 34.9817 & 0.01 & $ 1.13 \times 10^{-01} \pm 4.40 \times 10^{-02}$ & $ 1.19 \times 10^{-10} \pm 3.36 \times 10^{-11}$ \\
                     & Sector 39 & 37.0436 & 0.01 & $ 4.86 \times 10^{-01} \pm 1.21 \times 10^{-01}$ & $ 1.01 \times 10^{-08} \pm 3.36 \times 10^{-09}$ \\   
        \hline
        TOI 2408 & 67630845 \\
                     & Sector 30 & 19.5685 & 0.01 & $ 1.75 \times 10^{-01} \pm 2.89 \times 10^{-02}$ & 0.00 $\pm$ 0.00 & - \\
        \hline
        TOI 3913 & 155898758 \\
                     & Sector 49 & 16.1696 & 0.01 & $ 6.16 \times 10^{-01} \pm 4.30 \times 10^{-02}$ & 0.00 $\pm$ 0.00 & \multirow{2}{*}{PHARO (BrGamma)} \\
                     & Sector 50 & 14.5601 & 0.01 & $ 5.63 \times 10^{-02} \pm 6.18 \times 10^{-03}$ & 0.00 $\pm$ 0.00 \\
        \hline
        \multicolumn{7}{c}{\textit{\textbf{Not Validated}}} \\
        TOI 493 & 19025965 \\
                    & Sector 34 & 9.4971 & 0.01 & $ 1.75 \times 10^{-02} \pm 1.62 \times 10^{-03}$ & $ 3.31 \times 10^{-03} \pm 5.99 \times 10^{-04}$ & \multirow{4}{*}{NIRI (BrGamma)} \\
                    & Sector 44 & 14.7922 & 0.01 & $ 1.41 \times 10^{-02} \pm 3.09 \times 10^{-03}$ & $ 1.06 \times 10^{-02} \pm 2.51 \times 10^{-03}$ \\
                    & Sector 45 & 12.3651 & 0.01 & $ 2.24 \times 10^{-02} \pm 3.61 \times 10^{-03}$ & $ 1.88 \times 10^{-02} \pm 3.45 \times 10^{-03}$ \\
                    & Sector 46 & 9.7146 & 0.01 & $ 2.83 \times 10^{-02} \pm 7.96 \times 10^{-03}$ & $ 1.72 \times 10^{-02} \pm 7.69 \times 10^{-03}$ \\
           & \\
                    & Sector 34 & & &  $ 1.23 \times 10^{-02} \pm 1.62 \times 10^{-03}$ & $ 3.56 \times 10^{-03} \pm 6.47 \times 10^{-04}$ & \multirow{4}{*}{NIRC2 (BrGamma)}\\
                    & Sector 44 & & &  $ 1.08 \times 10^{-02} \pm 2.07 \times 10^{-03}$ & $ 1.00 \times 10^{-02} \pm 1.85 \times 10^{-03}$ \\
                    & Sector 45 & & &  $ 1.94 \times 10^{-02} \pm 4.41 \times 10^{-03}$ & $ 1.87 \times 10^{-02} \pm 4.39 \times 10^{-03}$ \\
                    & Sector 46 & & &  $ 2.99 \times 10^{-02} \pm 9.54 \times 10^{-03}$ & $ 2.28 \times 10^{-02} \pm 9.19 \times 10^{-03}$ \\
        \hline
        TOI 815 & 102840239 \\
                    & Sector 36 & 15.7674 & 1.00 & $ 2.23 \times 10^{-03} \pm 1.32 \times 10^{-03}$ & $ 1.87 \times 10^{-03} \pm 1.32 \times 10^{-03}$ & Zorro (562 nm) \\
                    & Sector 36 & & & $ 2.13 \times 10^{-03} \pm 1.32 \times 10^{-03}$ & $ 1.77 \times 10^{-03} \pm 1.32 \times 10^{-03}$ & Zorro (832 nm) \\
        \hline
        TOI 1179 & 148914726 \\
                     & Sector 14 & 41.5323 & 0.01 & $ 9.98 \times 10^{-01} \pm 1.51 \times 10^{-03}$ & $ 2.79 \times 10^{-02} \pm 9.85 \times 10^{-03}$ & \multirow{6}{*}{'Alopeke (562 nm)} \\
                     & Sector 15 & 47.9257 & 0.01 & $ 9.87 \times 10^{-01} \pm 6.69 \times 10^{-03}$ & $ 7.11 \times 10^{-02} \pm 2.72 \times 10^{-02}$ \\
                     & Sector 21 & 49.6108 & 0.01 & $ 9.72 \times 10^{-01} \pm 1.57 \times 10^{-02}$ & $ 5.84 \times 10^{-02} \pm 2.29 \times 10^{-02}$ \\
                     & Sector 22 & 43.6678 & 0.01 & $ 2.51 \times 10^{-02} \pm 5.48 \times 10^{-03}$ & $ 4.08 \times 10^{-03} \pm 1.48 \times 10^{-03}$ \\
                     & Sector 41 & 42.6402 & 0.01 & $ 1.56 \times 10^{-01} \pm 7.08 \times 10^{-02}$ & $ 3.46 \times 10^{-02} \pm 2.20 \times 10^{-02}$ \\
                     & Sector 48 & 42.5446 & 0.01 & $ 9.82 \times 10^{-01} \pm 7.00 \times 10^{-03}$ & $ 1.20 \times 10^{-01} \pm 4.19 \times 10^{-02}$ \\
           & \\
                     & Sector 14 & & & $ 9.96 \times 10^{-01} \pm 4.87 \times 10^{-03}$ & $ 2.78 \times 10^{-01} \pm 7.95 \times 10^{-03}$ & \multirow{6}{*}{'Alopeke (832 nm)} \\
                     & Sector 15 & & & $ 9.78 \times 10^{-01} \pm 1.98 \times 10^{-02}$ & $ 9.48 \times 10^{-02} \pm 2.78 \times 10^{-02}$ \\
                    & Sector 21 & & & $ 9.82 \times 10^{-01} \pm 1.02 \times 10^{-02}$ & $ 7.31 \times 10^{-02} \pm 2.08 \times 10^{-02}$ \\
                     & Sector 22 & & & $ 2.17 \times 10^{-02} \pm 6.25 \times 10^{-03}$ & $ 4.69 \times 10^{-03} \pm 1.74 \times 10^{-03}$ \\
                    & Sector 41 & & & $ 1.32 \times 10^{-01} \pm 9.88 \times 10^{-02}$ & $ 4.59 \times 10^{-02} \pm 3.57 \times 10^{-02}$ \\
                     & Sector 48 & & & $ 9.67 \times 10^{-01} \pm 2.17 \times 10^{-02}$ & $ 1.67 \times 10^{-01} \pm 4.12 \times 10^{-02}$ \\
        \hline   
        TOI 1732 & 470987100 \\
                     & Sector 20 & 10.6478 & 0.01 & $ 8.10 \times 10^{-03} \pm 1.26 \times 10^{-04}$ & $ 7.33 \times 10^{-03} \pm 1.37 \times 10^{-04}$ & \multirow{2}{*}{PHARO (BrGamma)} \\
                     & Sector 47 & 12.6158 & 0.01 & $ 2.45 \times 10^{-04} \pm 9.89 \times 10^{-06}$ & $ 2.15 \times 10^{-04} \pm 6.56 \times 10^{-06}$ \\
           & \\
                     & Sector 20 & & &  $ 1.02 \times 10^{-02} \pm 1.53 \times 10^{-04}$ & $ 8.08 \times 10^{-03} \pm 1.46 \times 10^{-04}$ & \multirow{2}{*}{'Alopeke (562 nm)} \\
                     & Sector 47 & & &  $ 2.72 \times 10^{-04} \pm 1.46 \times 10^{-05}$ & $ 3.37 \times 10^{-04} \pm 1.34 \times 10^{-05}$ \\
           & \\
                     & Sector 20 & & &  $ 8.84 \times 10^{-03} \pm 2.44 \times 10^{-04}$ & $ 8.26 \times 10^{-03} \pm 2.43 \times 10^{-04}$ & \multirow{2}{*}{'Alopeke (832 nm)} \\
                     & Sector 47 & & &  $ 2.45 \times 10^{-04} \pm 1.01 \times 10^{-05}$ & $ 2.34 \times 10^{-04} \pm 7.39 \times 10^{-06}$ \\
           & \\
                    & Sector 20 & & &  $ 1.17 \times 10^{-02} \pm 1.64 \times 10^{-04}$ & $ 6.97 \times 10^{-03} \pm 9.31 \times 10^{-05}$ & \multirow{2}{*}{ShARCS (K)}\\
                    & Sector 47 & & &  $ 3.92 \times 10^{-04} \pm 2.86 \times 10^{-05}$ & $ 2.07 \times 10^{-04} \pm 1.06 \times 10^{-05}$ \\
        \hline      
        TOI 3568 & 160390955 \\
                     & Sector 55 & 21.0566 & 0.01 & $ 1.17 \times 10^{-02} \pm 6.66 \times 10^{-03}$ & $ 1.88 \times 10^{-03} \pm 3.82 \times 10^{-04}$ & NIRC2 (K) \\
                     & Sector 55 & & & $ 3.61 \times 10^{-02} \pm 1.07 \times 10^{-02}$ & $ 2.22 \times 10^{-03} \pm 4.26 \times 10^{-04}$ & PHARO (Hcont) \\
           & Sector 55 & & & $ 2.32 \times 10^{-03} \pm 3.36 \times 10^{-03}$ & $ 2.09 \times 10^{-03} \pm 5.63 \times 10^{-04}$ & PHARO (BrGamma)\\
        \hline
        TOI 3896 & 445837596 \\
                     & Sector 48 & 11.1953 & 0.01 & $ 1.40 \times 10^{-02} \pm 8.42 \times 10^{-04}$ & $ 1.04 \times 10^{-03} \pm 3.75 \times 10^{-05}$ & PHARO (BrGamma)\\
        \hline       
        TOI 4090 & 289373041 \\
                     & Sector 53 & 12.7536 & 0.01 & $ 3.95 \times 10^{-02} \pm 4.76 \times 10^{-03}$ & $ 2.51 \times 10^{-02} \pm 1.69 \times 10^{-03}$ & \multirow{2}{*}{PHARO (BrGamma)} \\
                     & Sector 54 & 15.2323 & 0.01 & $ 5.63 \times 10^{-02} \pm 6.18 \times 10^{-03}$ & $ 2.82 \times 10^{-02} \pm 2.79 \times 10^{-03}$ \\
           & \\
        \hline
        TOI 5584 & 29169215 \\
                    & Sector 21 & 13.8631 & 0.01 & $ 1.91 \times 10^{-02} \pm 1.33 \times 10^{-03}$ & $ 7.18 \times 10^{-03} \pm 4.19 \times 10^{-04}$ & \multirow{2}{*}{-}\\
                     & Sector 47 & 7.0555  & 0.10 & $ 1.81 \times 10^{-01} \pm 6.31 \times 10^{-03}$ & $ 1.46 \times 10^{-01} \pm 5.59 \times 10^{-03}$ \\
         \hline
\enddata
\end{deluxetable*}
%\end{longrotatetable}

\section{Validated Planets}
\label{Validated Planets}
\begin{figure*}
    \centering
    \includegraphics[scale = 0.41]{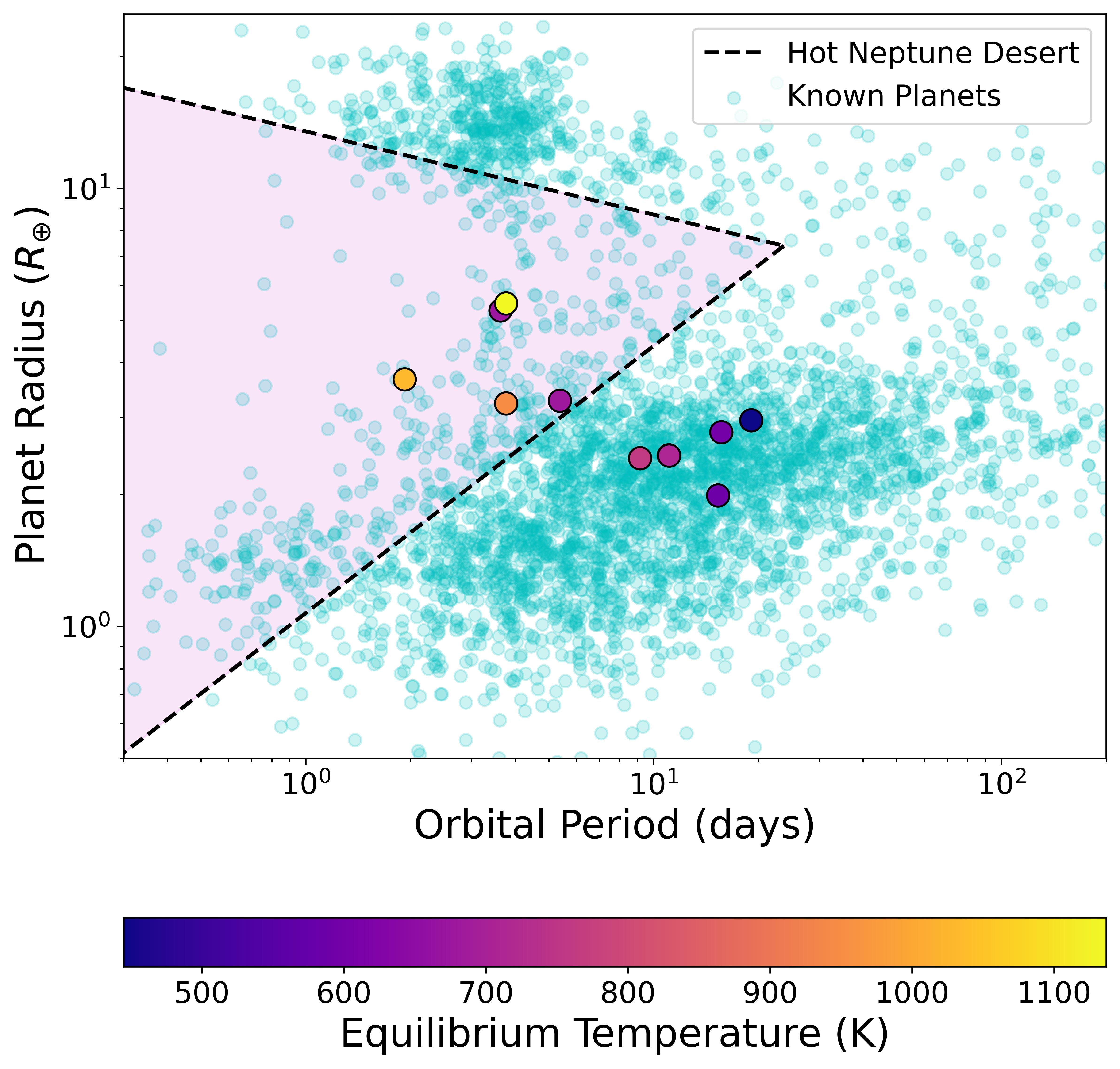}
    \includegraphics[scale = 0.41]{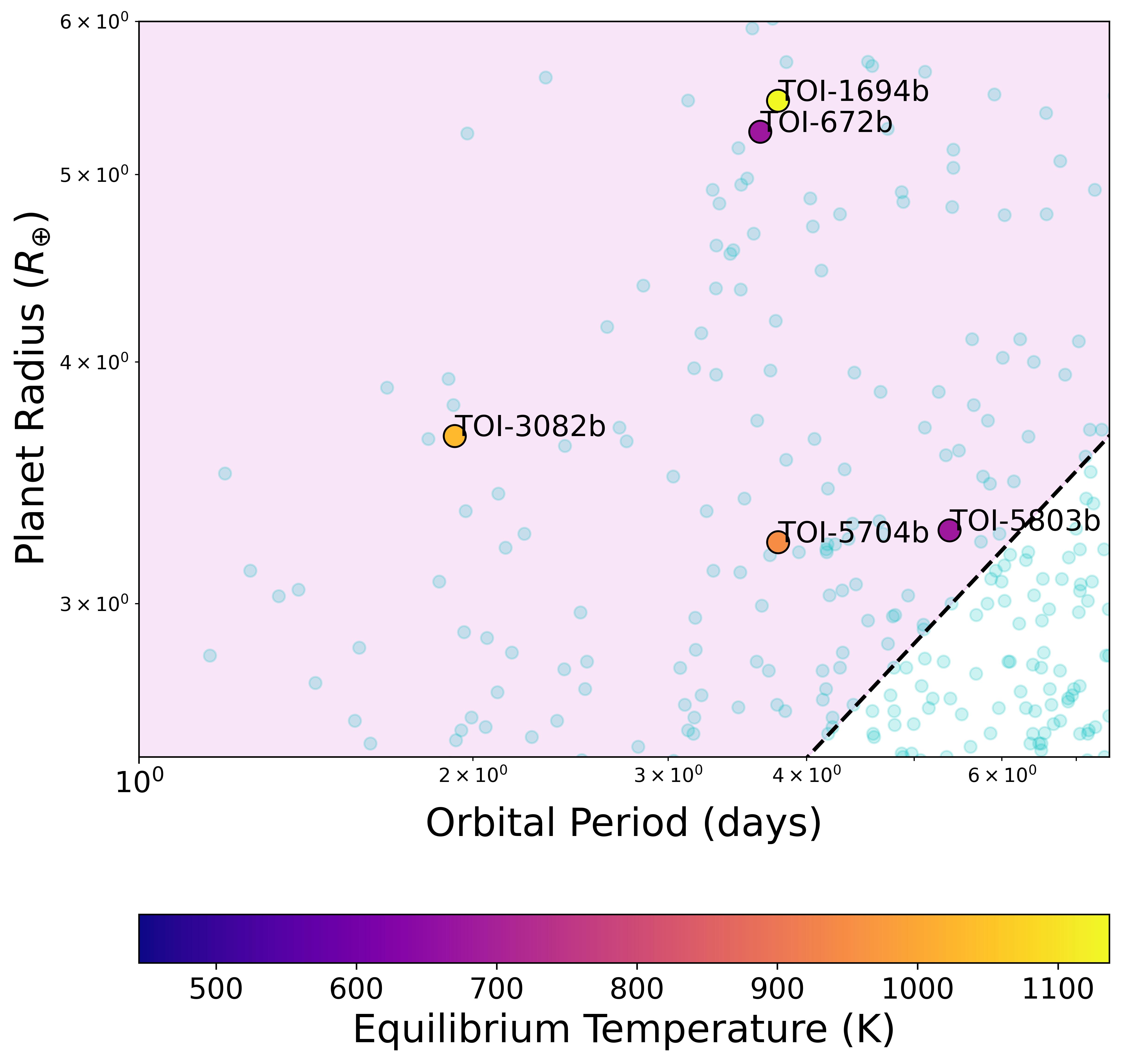}
    \includegraphics[scale = 0.41]{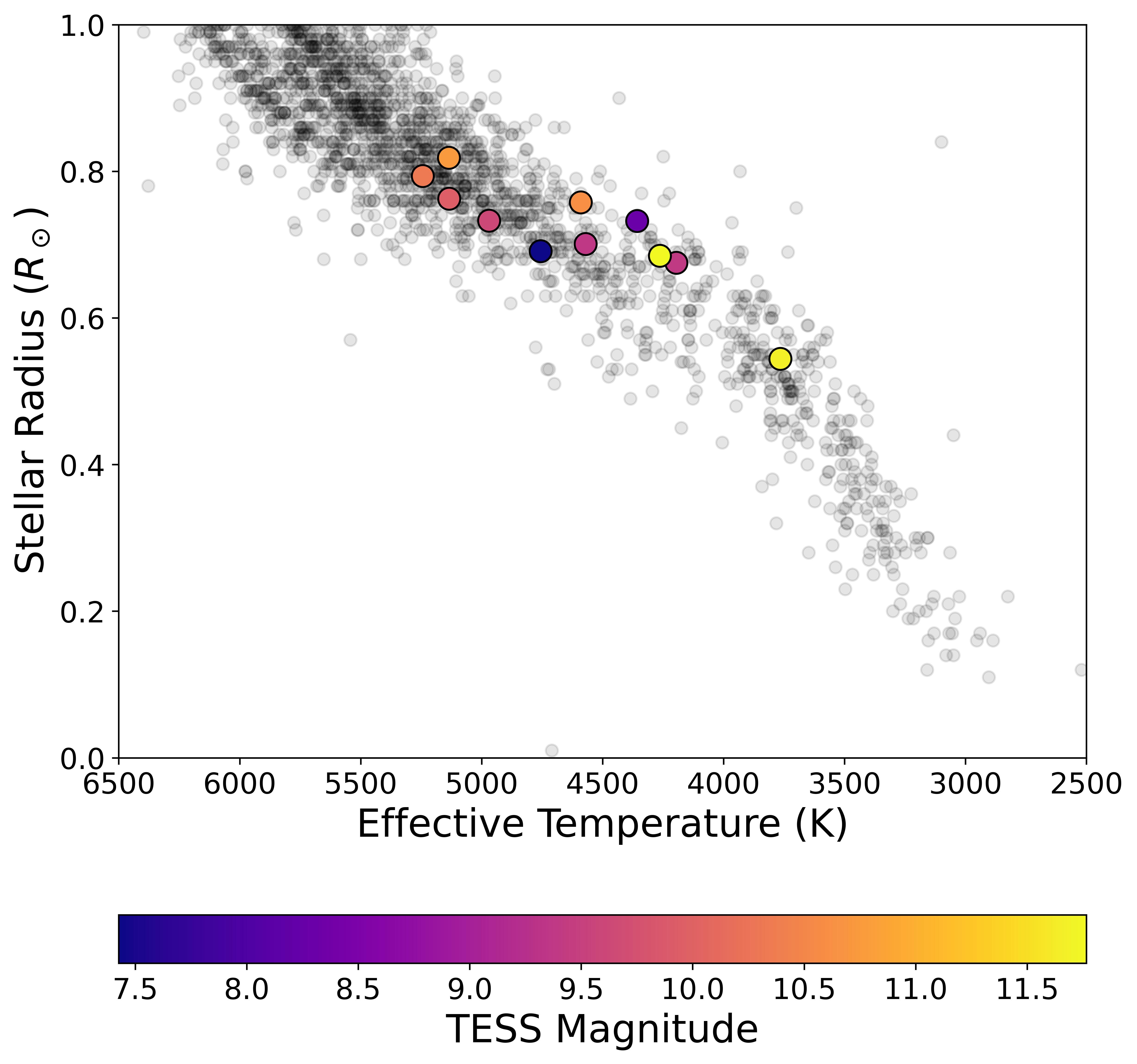}
    \caption{Properties of newly validated systems. Top panel: New planets and their comparison with previously known planets (with $<$ 100-day period), shaded region depicts the Hot Neptune Desert, figure in right panel is a zoomed-in version of the left figure with labels for the planets in the Hot Neptune Desert. Five of the planets lie in the Hot-Neptune Desert. Right panel: New host stars compared to hosts of known planets. Data for this plots were taken from NASA Exoplanet Archive.}
    \label{fig:planets}
\end{figure*}

\begin{figure*}
    \centering
    \includegraphics[scale = 0.34]{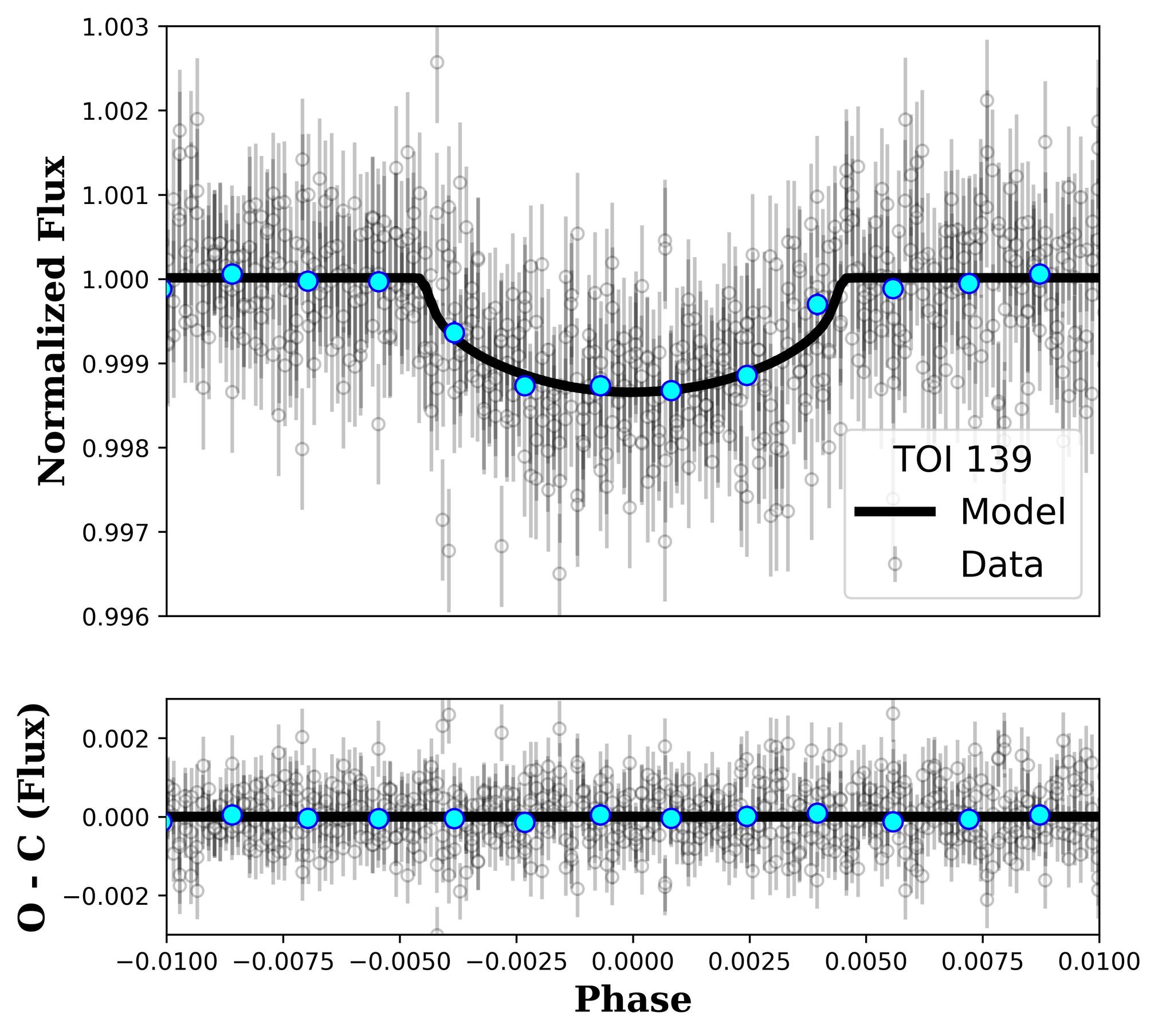}
    \includegraphics[scale = 0.34]{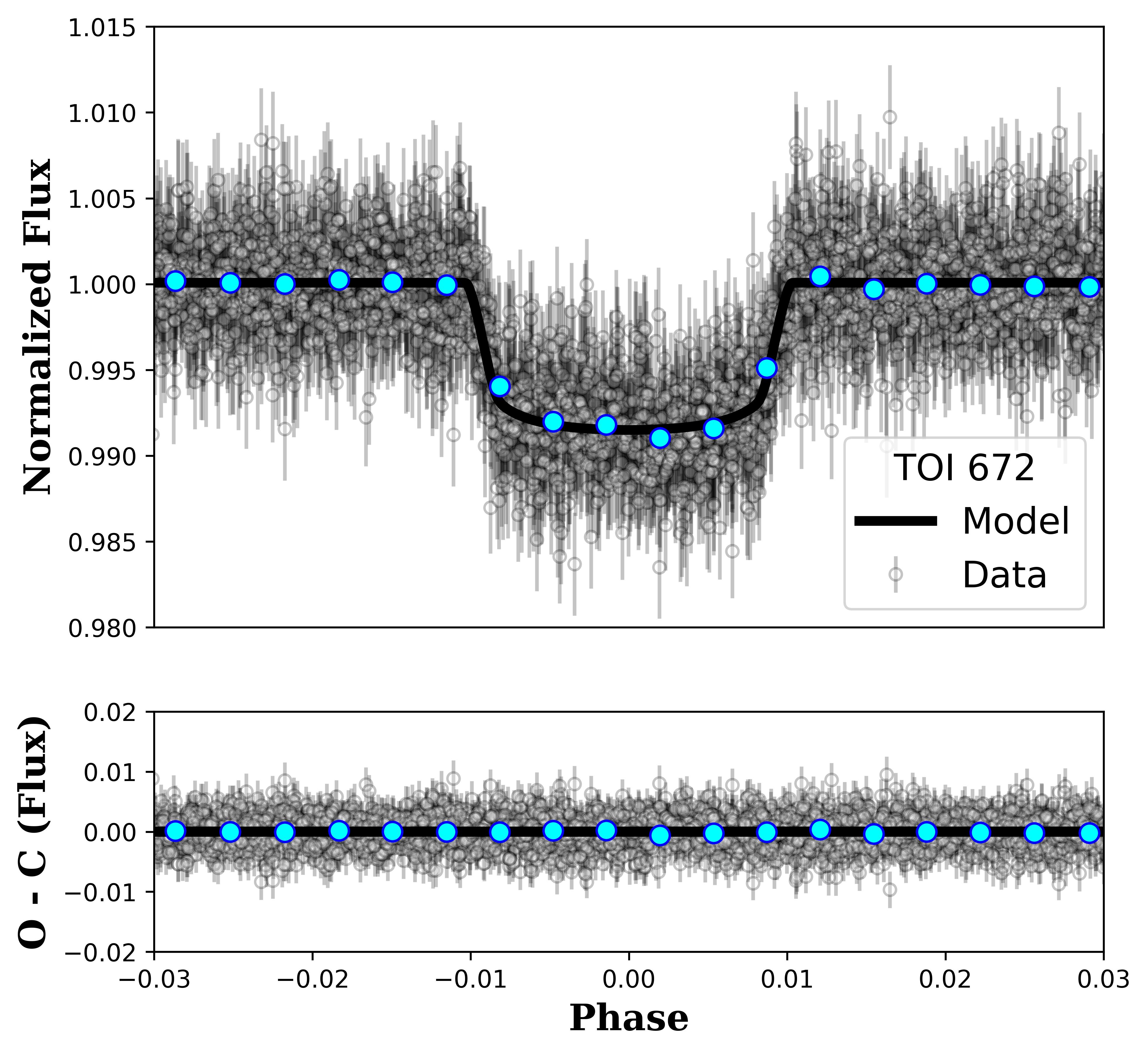}
    \includegraphics[scale = 0.34]{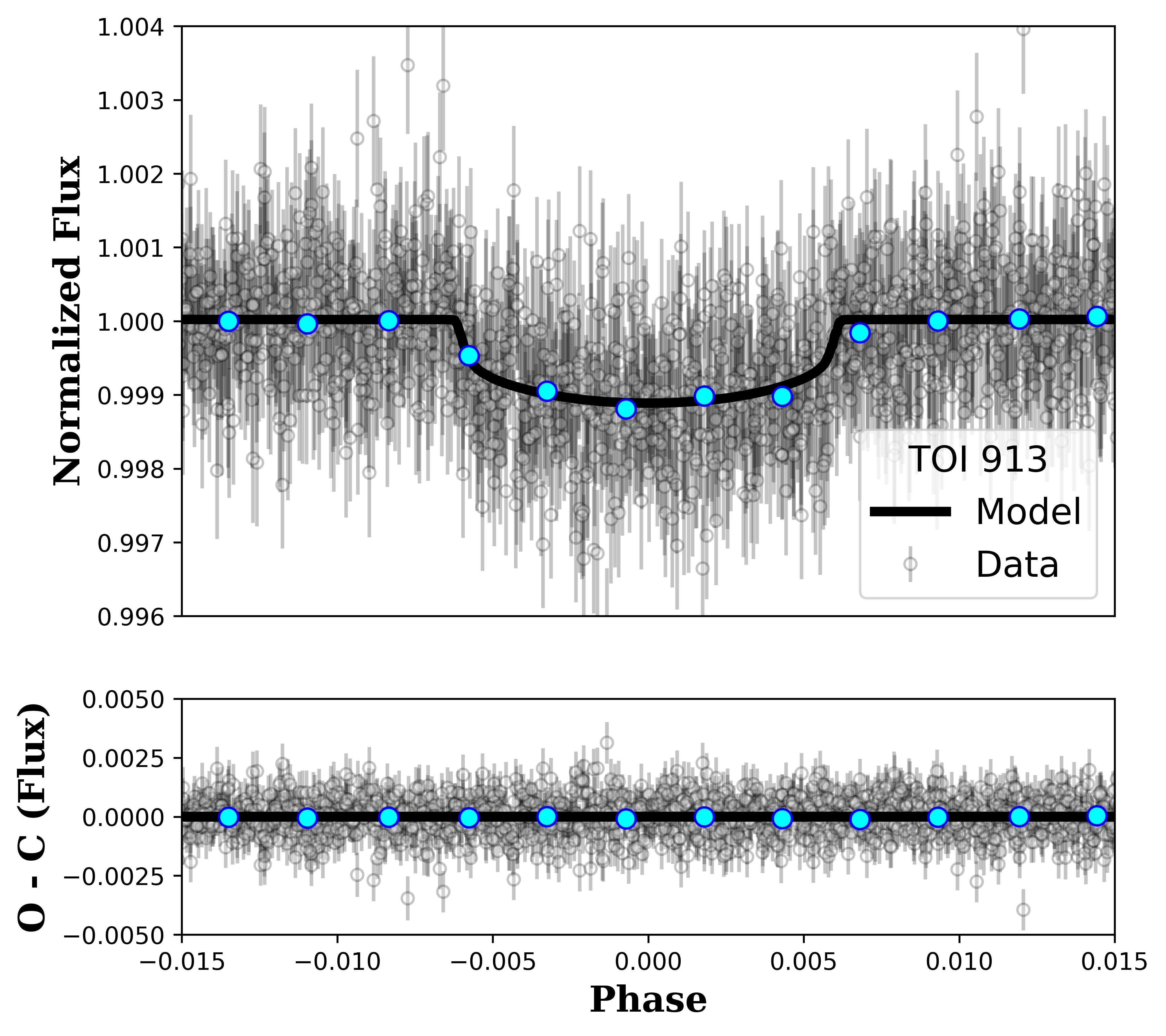}
    \includegraphics[scale = 0.34]{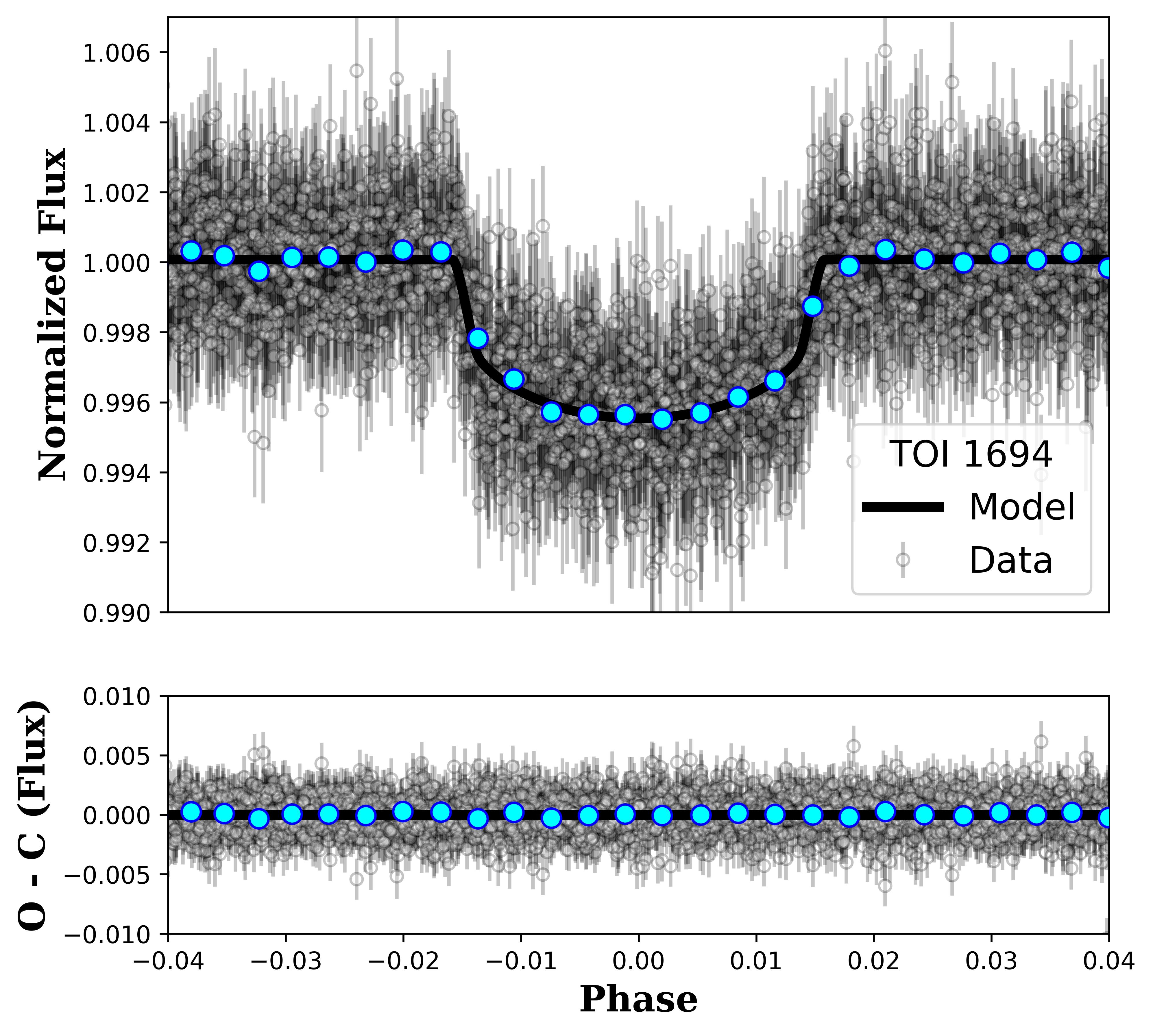}
    \includegraphics[scale = 0.34]{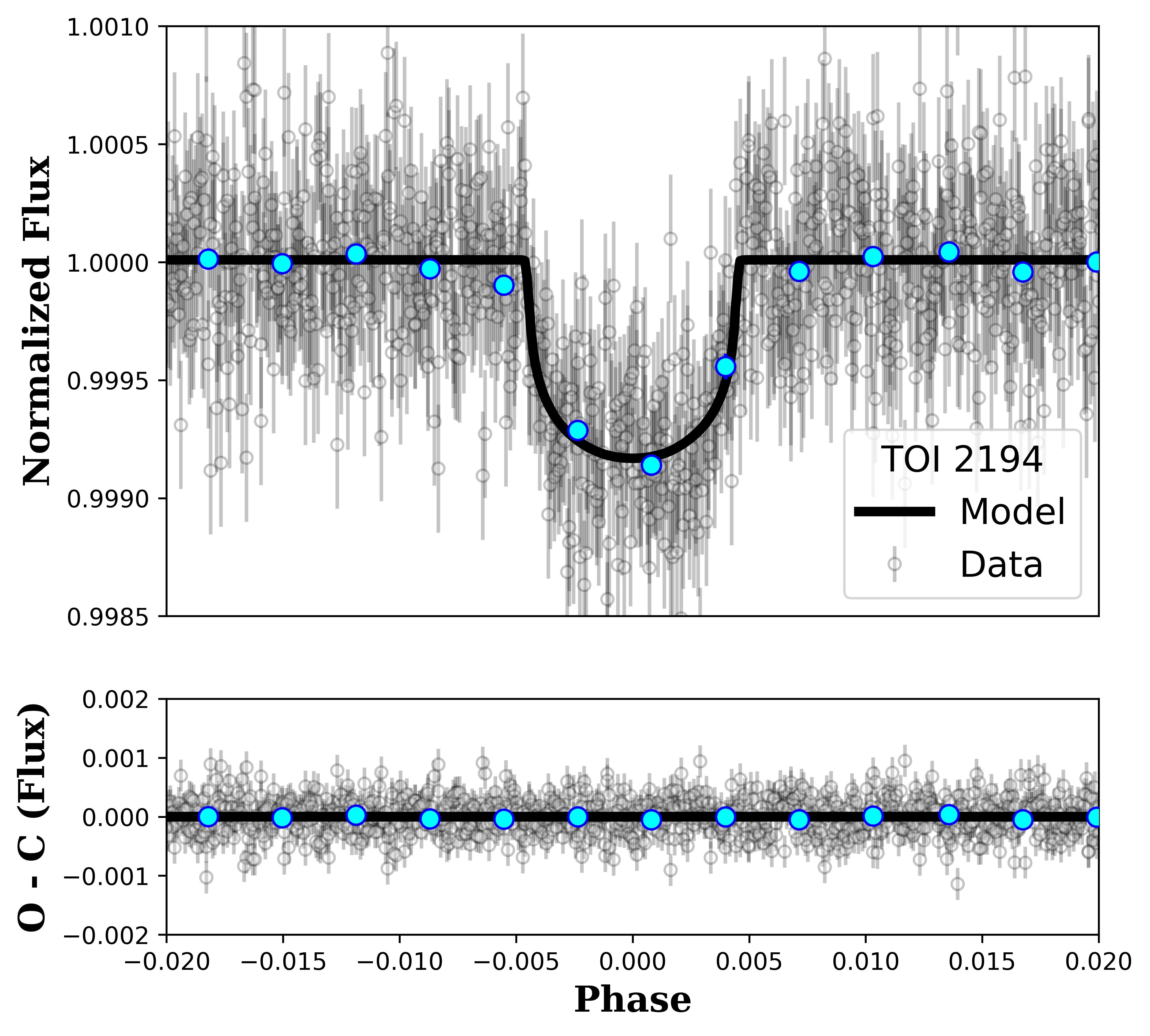}
    \includegraphics[scale = 0.34]{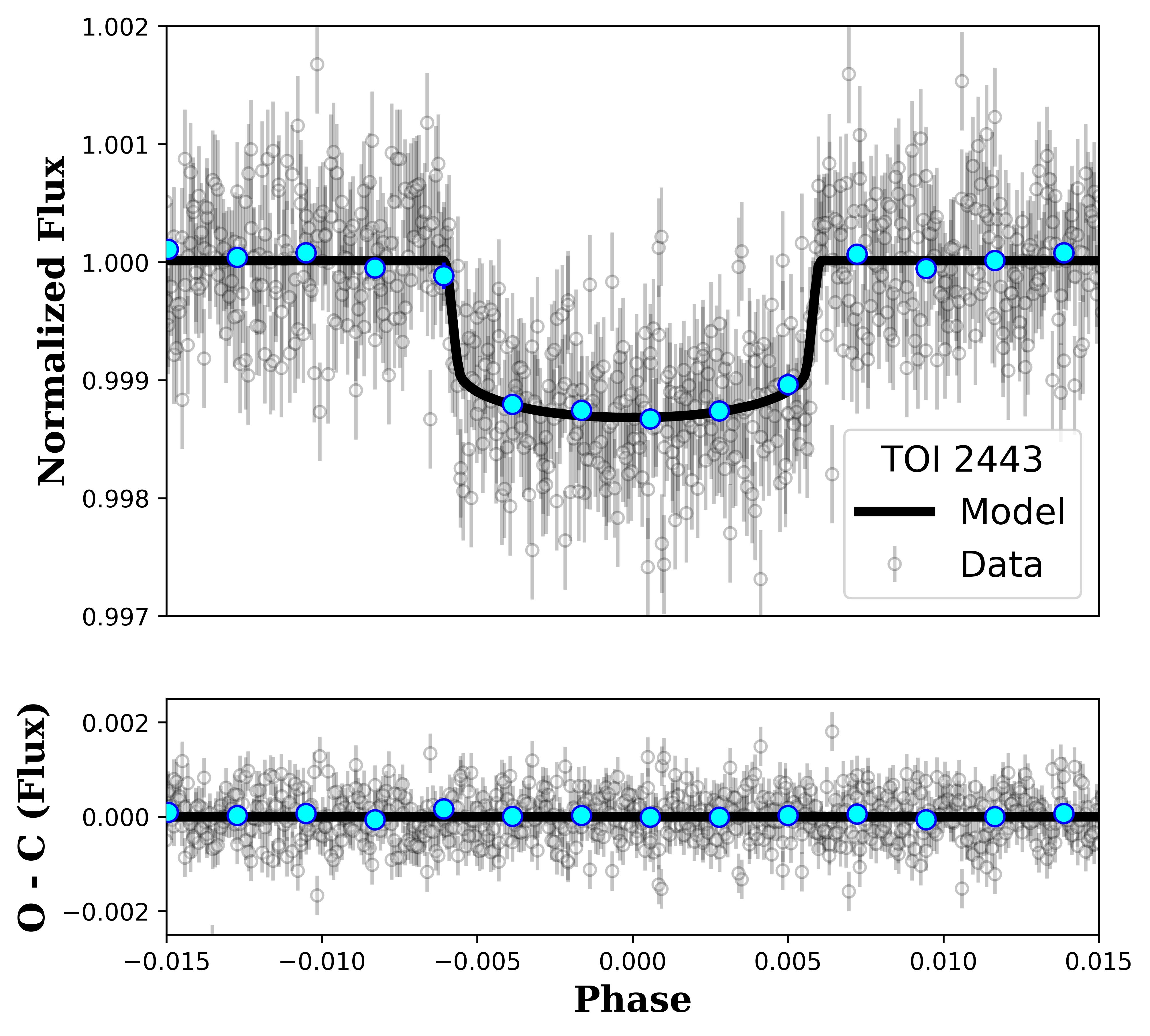}
    \includegraphics[scale = 0.34]{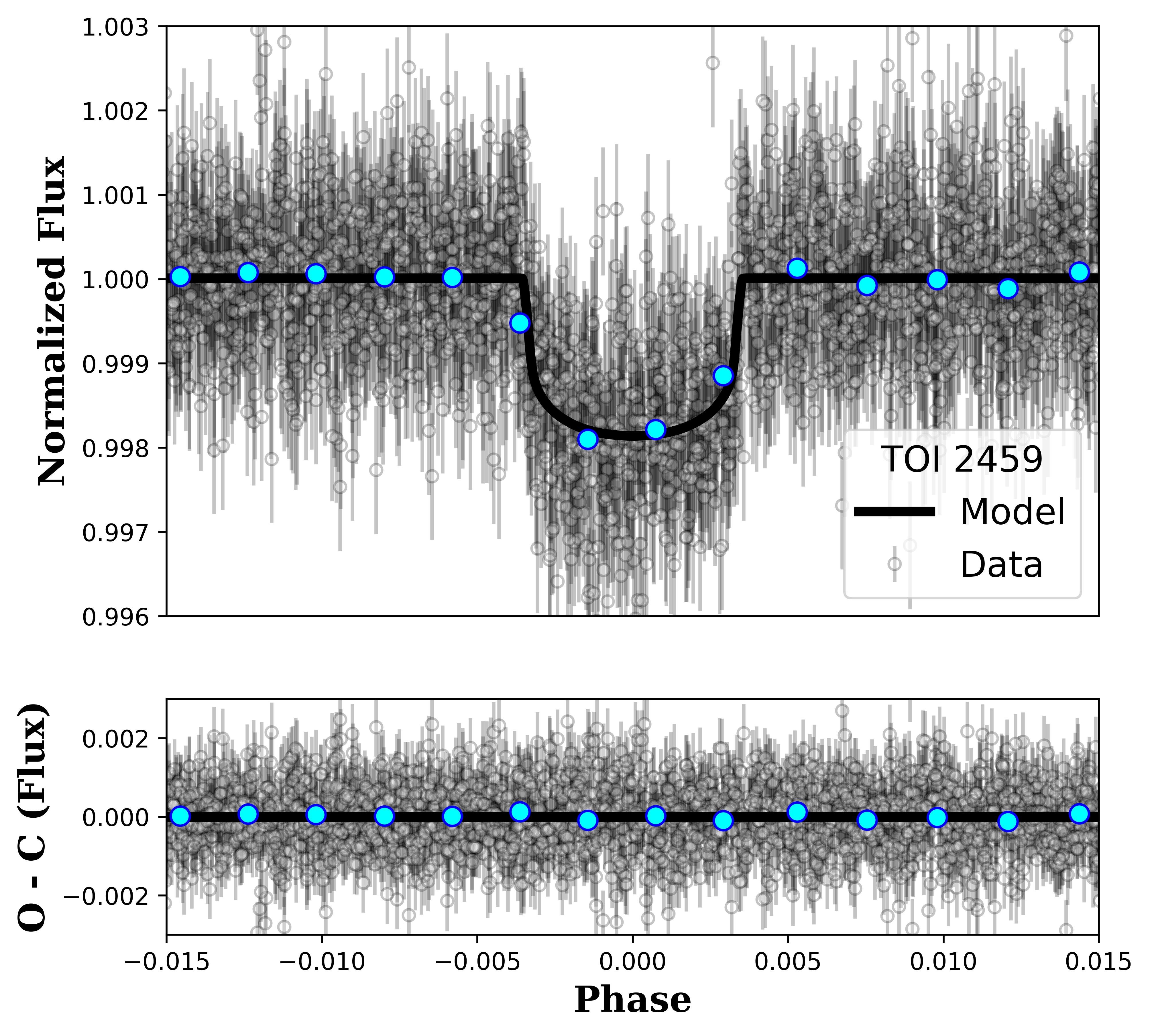}
    \includegraphics[scale = 0.34]{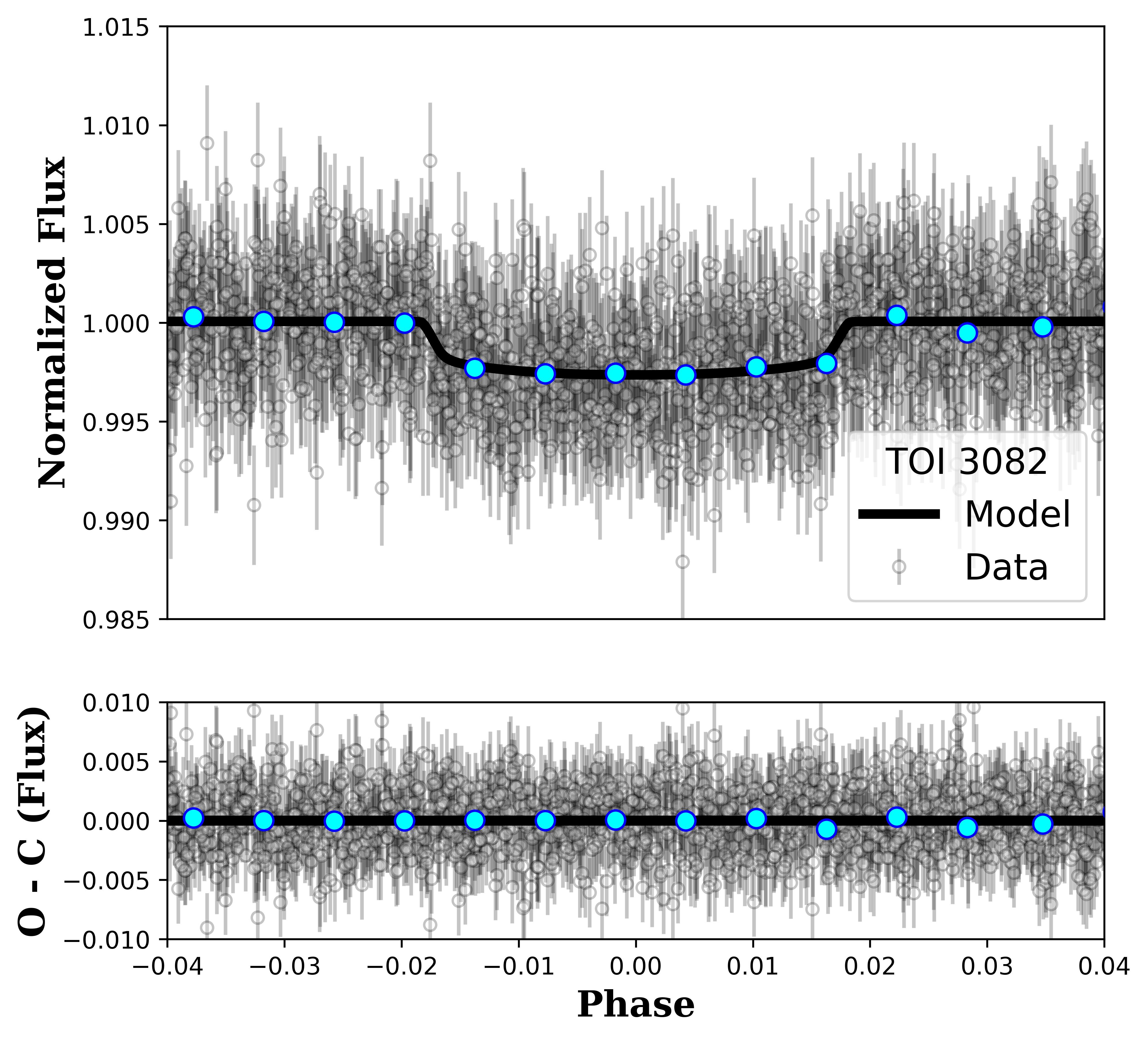}
    \includegraphics[scale = 0.34]{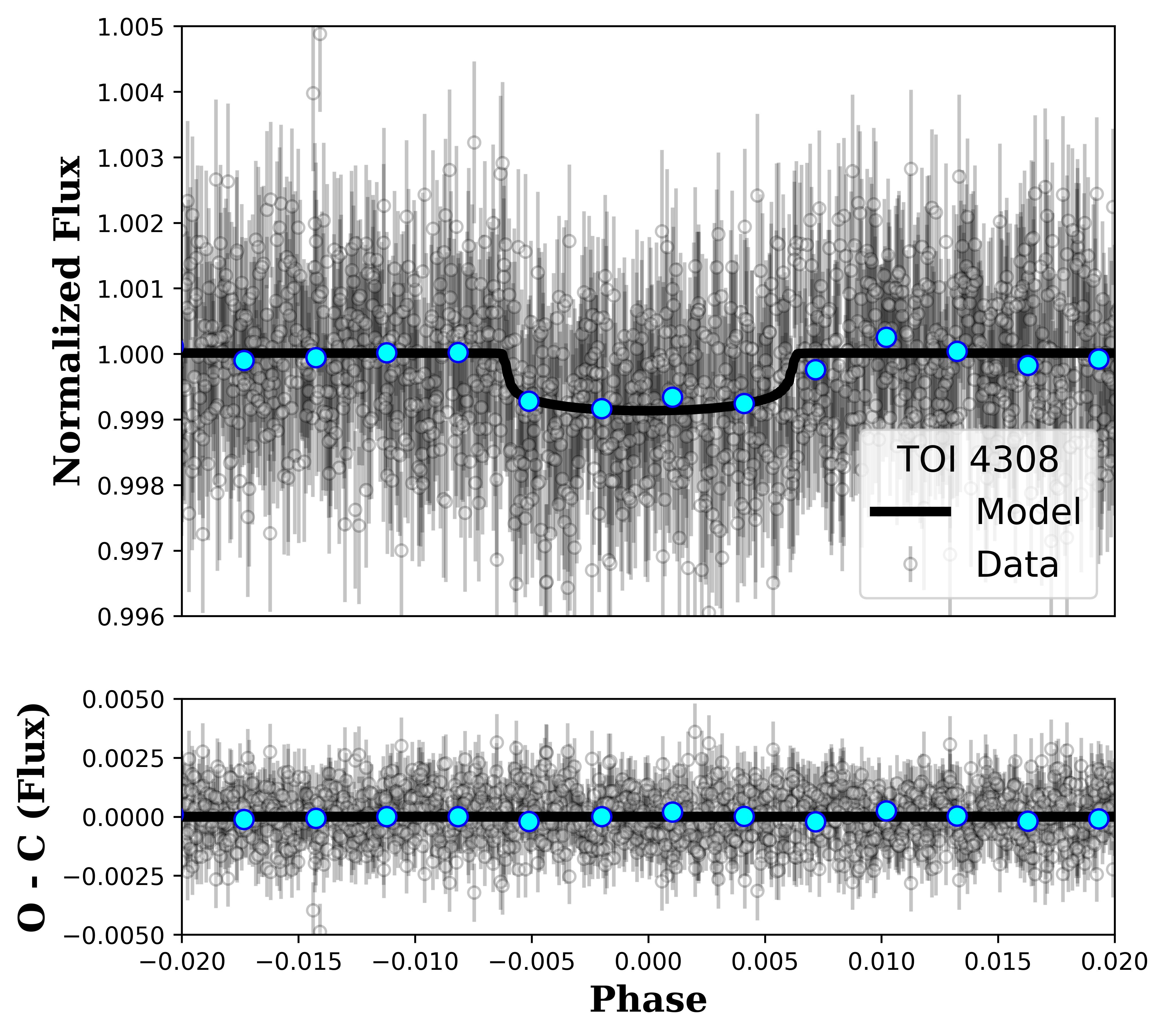}
    \includegraphics[scale = 0.34]{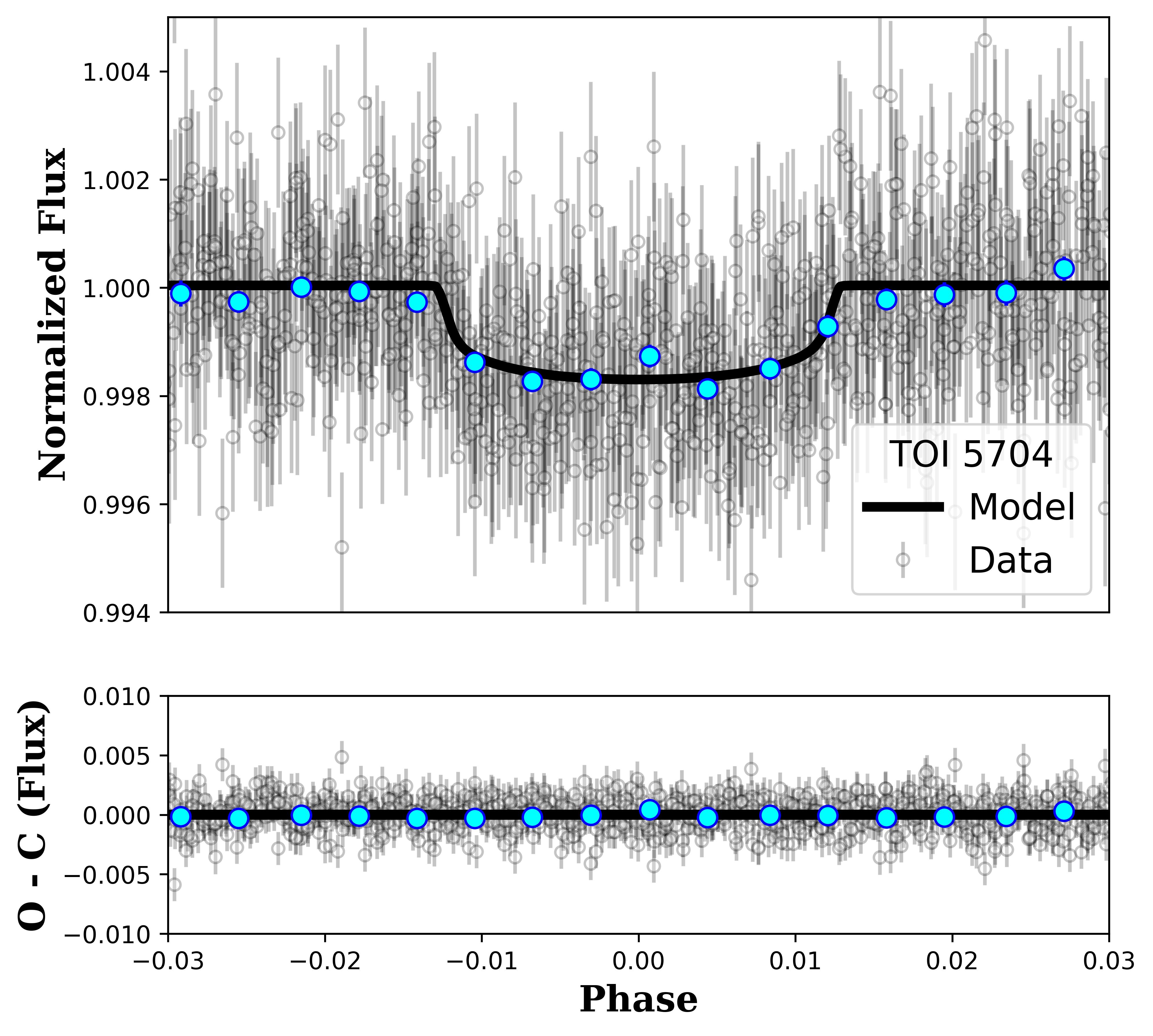}
    \includegraphics[scale = 0.34]{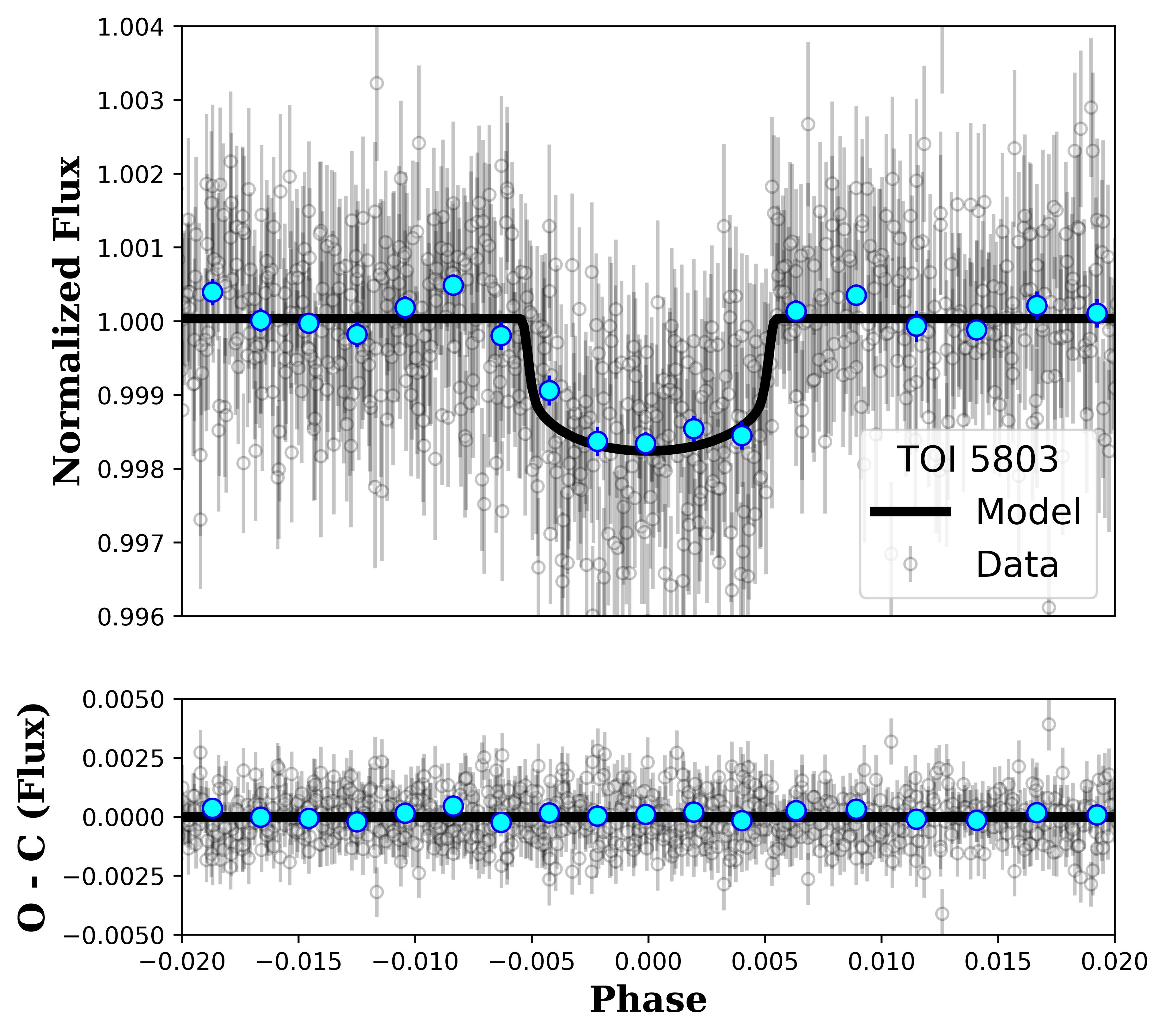}
    \caption{Phase-folded light curves of newly validated planetary systems. Black line shows the best-fit model, blue dots are binned observations.}
    \label{fig:LCs1}
\end{figure*}

We consider planetary candidates with high-resolution imaging showing no evidence of stellar companion and \texttt{TRICERATOPS} FPP of $< 1.5 \times 10^{-2}$ and NFPP of $ < 10^{-3} $ \citep{2021AJ....161...24G} to be statistically validated. \texttt{TRICERATOPS} undertakes the nearby stars (within 2.5' radius from the target) for which measured transit depths are non-zero  to calculate the NFPP. In Appendix \ref{A1}, we have compiled a list of such nearby stars that were considered by \texttt{TRICERATOPS} for calculating the NFPP for all of our validated planets. The tabulated probability clearly suggests that these stars are not contaminating our target star and thereby strongly suggesting that the transit source detected is originating solely from the target star. By examining the 24 candidates we validated 11 planetary systems. The properties of the new planets are shown in the top panel and the properties of new host stars are shown in the bottom panel of Figure \ref{fig:planets}. Our newly validated planets range in size from the super-Earth sized TOI-2194b (1.99 $R_{\earth}$) to the sub-Saturn sized TOI-672b (5.26 $R_{\earth}$) and TOI-1694b (5.46 $R_{\earth}$). The derived planetary and orbital parameters for all 11 planets are listed in Table \ref{tab:params}. Phase-folded transit light curves with the best-fit \texttt{Juliet} model are shown in Figure \ref{fig:LCs1}.  

\begin{figure*}
    \centering
    \includegraphics[scale = 0.42]{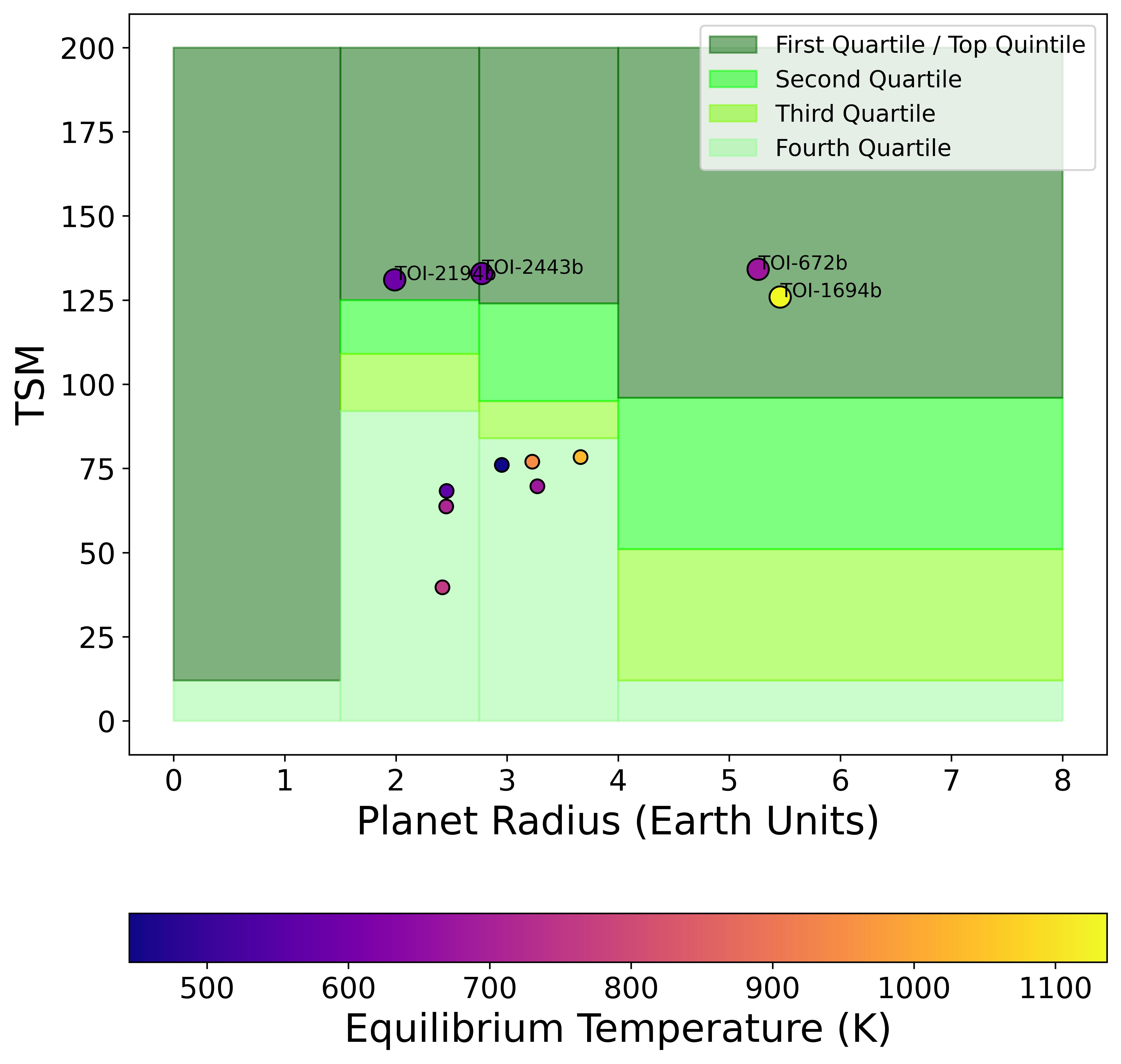}
    \includegraphics[scale = 0.42]{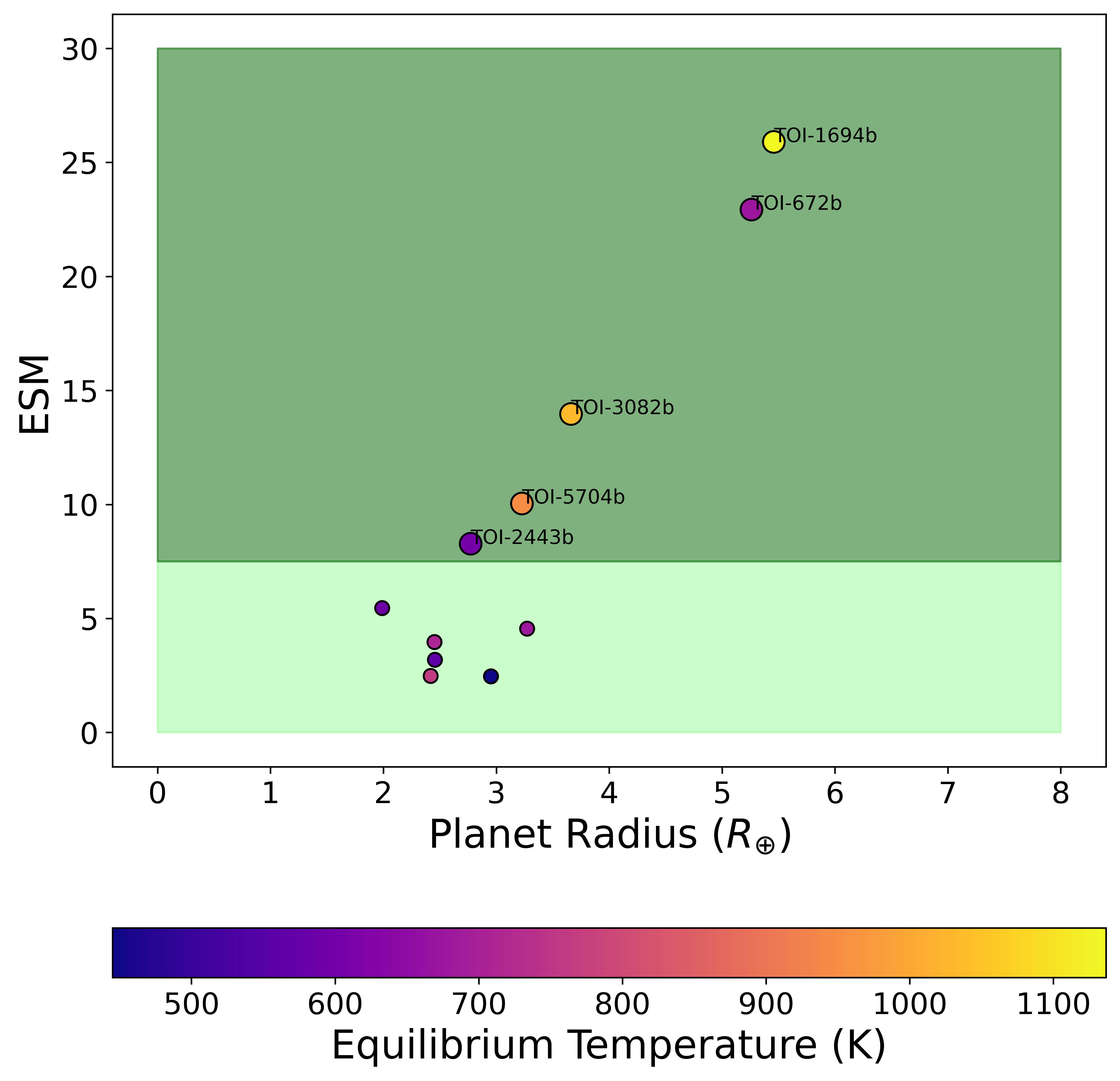}
    \caption{Transmission spectroscopy values (Left) and emission spectroscopy values (Right) for the newly validated planets, color-coded by their equilibrium temperature. In each case the shaded regions indicate areas of interest (dark green region) as identified by \citet{2018PASP..130k4401K}. Big dots represent the planets amenable for transmission or emission spectroscopy.}
    \label{fig:TESM}
\end{figure*}

In this section we will discuss some of the interesting features of new systems. By using an unbiased mass-radius empirical relationship \citep{2017ApJ...834...17C}, which was developed upon the probabilistic mass-radius relation condition on a sample of known exoplanets and late-type stars, we approximated the mass of the newly validated planets. It is to be noted here that these mass estimates should not be considered robust for characterization of the planets properties, we used these estimates to get better idea of these systems. Based on this mass estimates we also calculated the semi-amplitude of radial velocity that is induced on the host star by the orbiting planet. In order to facilitate the identification of the most optimal targets for atmospheric characterization among the TESS planet candidates, \cite{2018PASP..130k4401K} developed a method for calculating Transmission Spectroscopy Metrics (TSM) that is proportional to the expected transmission spectroscopy signal-to-noise (S/N), based on the strength of spectral features and the brightness of the host star, assuming cloud-free atmospheres and Emission Spectroscopy Metrics (ESM) that is proportional to the expected S/N of a JWST secondary eclipse detection at mid-IR wavelengths. The calculated TSM and ESM values for the candidates are tabulated in Table \ref{tab:estiparams}. This method allows for efficient prioritization of the most promising candidates for further study and characterization of their atmospheric properties. Distribution of TSM and ESM for planets is shown in Figure \ref{fig:TESM}. Using radial velocity mass measurements, it is recommended to quickly measure the original mass and follow up on targets that meet the suggested threshold values for these measurements.

\begin{table*}
    \centering
    \caption{Estimated parameters, Transmission and Emission Spectroscopy Metrics (TSM and ESM) for newly validated planetary systems.}
    \label{tab:estiparams}
    \begin{tabular}{c c c c c c c c}
    \hline
    Planet & TIC ID & $T_{eq}$ & $M_p$ & Density & K  & TSM & ESM \\
            &           &  [K]     &  [$M_{\earth}$] & [cgs] & m $s^{-1}$ & \\
    \hline
    \hline
    TOI-139b & 62483237 & 561.17 & 6.8 $\pm$ 3.0 & 2.45 & 2.4305 & 68.30  & 3.18 \\
    TOI-672b & 151825527 & 676.15 & 24.2 $\pm$ 10.7 & 0.91 & 15.1375 & 134.15 & 22.93 \\
    TOI-913b & 407126408 & 712.01 & 6.8 $\pm$ 2.9 & 2.46 & 2.1586 & 63.70  & 3.96 \\
    TOI-1694b & 396740648 & 1136.57 & 25.5 $\pm$ 11.9 & 0.87 & 11.8096 & 125.91 & 25.89 \\
    TOI-2194b & 271478281 & 590.88 & 4.9 $\pm$ 2.0 & 3.23 & 1.4536 & 131.02 & 5.45 \\
    TOI-2443b & 318753380 & 600.83 & 8.3 $\pm$ 3.6 & 2.09 & 2.7401 & 132.89 & 8.28 \\
    TOI-2459b & 192790476 & 445.01  & 9.1 $\pm$ 4.0 & 1.93 & 2.8544 & 76.04  & 2.46 \\
    TOI-3082b & 428699140 & 1032.78 & 13.2 $\pm$ 5.8 & 1.46 & 8.7974 & 78.37  & 13.96 \\
    TOI-4308b & 144193715 & 763.05 & 6.5 $\pm$ 2.7 & 2.50 & 2.1129 & 39.68  & 2.47 \\
    TOI-5704b & 148673433 & 949.07 & 9.49 $\pm$ 4.4 & 1.55 & 5.3273 & 76.99 & 10.04 \\
    TOI-5803b & 466382581 & 678.87 & 10.8 $\pm$ 4.8 & 1.69 & 4.3137 & 69.69  & 4.55 \\
    \hline
    \end{tabular}
\end{table*}

\begin{deluxetable*}{ccccccccc}
\renewcommand{\arraystretch}{1.5}
\tablecaption{Planetary and orbital parameters for the newly validated planetary systems using Juliet. \label{tab:params}} 
\tablehead{
\colhead{Planet} &  
\colhead{TIC ID} &
\colhead{Period} &
\colhead{Epoch Time} & 
\colhead{$R_p/R_s$} &
\colhead{$R_p$} & 
\colhead{b} &
\colhead{$a/R_s$} &
\colhead{i} \\
\colhead{} & 
\colhead{} & 
\colhead{[days]} &
\colhead{[BJD]} & 
\colhead{} &
\colhead{[$R_{\earth}$]} &
\colhead{} &
\colhead{} &
\colhead{[degree]}
}       
\startdata
TOI-139b & 62483237 & $11.070850^{+0.000024} _{-0.000030}$ & $2458334.8906^{+0.0010} _{-0.0010}$ & $0.0321^{+0.0028} _{-0.0016}$ & $2.4566^{+0.2122} _{-0.1245}$ & $0.395^{+0.254} _{-0.252}$ & $33.159^{+2.667} _{-5.336}$ & $89.32^{+0.45} _{-0.65}$ \\
TOI-672b & 151825527 & $3.633575^{+0.000001} _{-0.000001}$ & $2458546.4799^{+0.0002} _{-0.0002}$ & $0.0885^{+0.0014} _{-0.0017}$ & $5.2604^{+0.0827} _{-0.0985}$ & $0.424^{+0.108} _{-0.206}$ & $15.503^{+1.055} _{-0.934}$ & $88.43^{+0.82} _{-0.52}$ \\
TOI-913b & 407126408 & $11.098644^{+0.000587} _{-0.000581}$ & $2458625.2133^{+0.0024} _{-0.0023}$ & $0.0306^{+0.0016} _{-0.0013}$ & $2.4528^{+0.1269} _{-0.1009}$ & $0.387^{+0.248} _{-0.254}$ & $24.352^{+1.910} _{-3.825}$ & $89.10^{+0.61} _{-0.87}$ \\
TOI-1694b & 396740648 & $3.770179^{+0.000058} _{-0.000060}$ & $2458817.2662^{+0.0004} _{-0.0007}$ & $0.0610^{+0.0017} _{-0.0013}$ & $5.4585^{+0.4682} _{-0.7919}$ & $0.326^{+0.172} _{-0.198}$ & $10.206^{+0.468} _{-0.792}$ & $88.17^{+1.15} _{-1.19}$ \\
TOI-2194b & 271478281 & $15.337597^{+0.001585} _{-0.001616}$ & $2459037.3678^{+0.0013} _{-0.0011}$ & $0.0263^{+0.0017} _{-0.0009}$ & $1.9892^{+0.1313} _{-0.0668}$ & $0.412^{+0.288} _{-0.253}$ & $32.393^{+2.718} _{-6.864}$ & $89.27^{+0.47} _{-0.85}$ \\
TOI-2443b & 318753380 & $15.669494^{+0.000926} _{-0.001004}$ & $2459148.0988^{+0.0007} _{-0.0007}$ & $0.0347^{+0.0006} _{-0.0006}$ & $2.7731^{+0.0493} _{-0.0515}$ & $0.285^{+0.183} _{-0.175}$ & $26.293^{+0.952} _{-2.053}$ & $89.38^{+0.39} _{-0.48}$ \\
TOI-2459b & 192790476 & $19.104718^{+0.000023} _{-0.000024}$ & $2458452.3342^{+0.0007} _{-0.0007}$ & $0.0400^{+0.0012} _{-0.0009}$ & $2.9531^{+0.0916} _{-0.0658}$ & $0.321^{+0.242} _{-0.209}$ & $44.432^{+2.039} _{-5.404}$ & $89.59^{+0.28} _{-0.41}$ \\
TOI-3082b & 428699140 & $1.926907^{+0.000128} _{-0.000134}$ & $2459309.1199^{+0.0010} _{-0.0010}$ & $0.0489^{+0.0020} _{-0.0019}$ & $3.6621^{+0.1464} _{-0.1448}$ & $0.355^{+0.247} _{-0.223}$ & $8.519^{+0.600} _{-1.286}$ & $87.63^{+1.54} _{-2.39}$ \\
TOI-4308b & 144193715 & $9.151201^{+0.000036} _{-0.000037}$ & $2458333.4284^{+0.0026} _{-0.0029}$ & $0.0279^{+0.0014} _{-0.0015}$ & $2.4189^{+0.1195} _{-0.1333}$ & $0.384^{+0.259} _{-0.253}$ & $23.606^{+2.227} _{-3.906}$ & $89.07^{+0.64} _{-0.92}$ \\
TOI-5704b & 148673433 & $3.771116^{+0.0000115} _{-0.0000107}$ & $2459610.7568^{+0.0008} _{-0.0007}$ & $0.0389^{+0.0023} _{-0.0017}$ & $3.2274^{+0.1873} _{-0.1398}$ & $0.424^{+0.223} _{-0.272}$ & $11.695^{+1.048} _{-2.486}$ & $88.94^{+1.37} _{-2.29}$ \\
TOI-5803b & 466382581 & $5.383050^{+0.000207} _{-0.000200}$ & $2459802.7103^{+0.0004} _{-0.0005}$ & $0.0393^{+0.0015} _{-0.0014}$ & $3.2732^{+0.1251} _{-0.1194}$ & $0.349^{+0.223} _{-0.233}$ & $28.596^{+1.887} _{-3.423}$ & $89.30^{+0.48} _{-0.60}$ \\
\enddata
\end{deluxetable*}

\renewcommand{\arraystretch}{1}

\subsection{TOI-139}
TOI-139b is a sub-Neptune (2.4566 $R_{\earth}$) planet orbiting bright (Vmag = 10.55, Tmag = 9.36) star TOI-139 (0.70 $R_{\sun}$, 0.69 $M_{\sun}$), observed in TESS sectors 1 and 28. It orbits the star at a distance of 0.11 AU with an orbital period 11.07 days and having an equilibrium temperature 561.17 K. 'Alopeke and NIRC2 high resolution imaging showed no contaminating stellar companion and ground-based follow-up observations have ruled out NEBs in all nearby ($\approx 2'.5$) Gaia DR2 and TIC stars that are bright enough to have caused the TESS detection. The \texttt{TRICERATOPS} FPP and NFPP are listed in Table \ref{tab:FPPs} calculated using four available contrast curve files. These FPP values are consistent with the source of transit signal being on the target star. Using the \cite{2017ApJ...834...17C} mass-radius relationship we predicted the mass 6.8 $\pm$ 3.0 $M_{\earth}$. Based on this, the resultant RV semi-amplitude is 2.43 m  $s^{-1}$. TSM and ESM are estimated as 68.30 and 3.18 respectively which are below the recommended threshold of \cite{2018PASP..130k4401K}. So this target will not be favourable for either transmission or emission spectroscopy. 

\subsection{TOI-672}
TOI-672b is a sub-Saturn or super-Neptune (5.26 $R_{\earth}$) planet orbiting faint (Vmag = 13.57, Tmag = 11.67) star TOI-672 (0.54 $R_{\sun}$, 0.53 $M_{\sun}$), observed in TESS sectors 9, 10 and 36. It orbits the star at a distance of 0.039 AU with an orbital period 3.63 days and having an equilibrium temperature 676.15 K. Zorro high resolution imaging showed no contaminating stellar companion and ground-based follow-up observation verified the transit event occur within the target star follow-up aperture, and there is no strong filter dependent depth chromaticity, and there are no other obvious or Gaia DR2 or TIC stars contaminating the follow-up aperture that are bright enough to cause the TESS detection. The \texttt{TRICERATOPS} FPP and NFPP are listed in Table \ref{tab:FPPs} calculated using two available contrast curve files. These FPP values are consistent with the source of transit signal being on the target star. We can neglect the very low probability NFPP scenarios because ground-based observations have confirmed the transit event on the target. Using the \cite{2017ApJ...834...17C} mass-radius relationship we predicted the mass 24.2 $\pm$ 10.7 $M_{\earth}$. Based on this, the resultant RV semi-amplitude is 15.14 m  $s^{-1}$. TSM and ESM are estimated as 134.15 and 22.93 respectively which are above the recommended threshold of \cite{2018PASP..130k4401K}. So this target will be favourable for both transmission and emission spectroscopy.

\subsection{TOI-913}
TOI-913b is a sub-Neptune (2.45 $R_{\earth}$) planet orbiting bright (Vmag = 10.45, Tmag = 9.62) star TOI-913 (0.73 $R_{\sun}$, 0.82 $M_{\sun}$), observed in TESS sectors 12 and 13. It orbits the star at a distance of 0.083 AU with an orbital period 11.09 days and having an equilibrium temperature 712 K. Zorro high resolution imaging showed no contaminating stellar companion and the ground-based follow-up observations verified the transit event occur within the target star follow-up aperture and that there are no other obvious or Gaia DR2 stars contaminating the follow-up aperture that are bright enough to cause the TESS detection. The \texttt{TRICERATOPS} FPP and NFPP are listed in Table \ref{tab:FPPs} calculated using two available contrast curve files. These FPP values are consistent with the source of transit signal being on the target star. We can neglect the very low probability NFPP scenarios because ground-based observations have confirmed the transit event on the target. Using the \cite{2017ApJ...834...17C} mass-radius relationship we predicted the mass 6.8 $\pm$ 2.9 $M_{\earth}$. Based on this, the resultant RV semi-amplitude is 2.16 m  $s^{-1}$. TSM and ESM are estimated as 63.71 and 3.96 respectively which are below the recommended threshold of \cite{2018PASP..130k4401K}. So this target will not be favourable for either transmission or emission spectroscopy.

\subsection{TOI-1694}
TOI-1694b is a sub-Saturn or super-Neptune (5.46 $R_{\earth}$) planet orbiting bright (Vmag = 11.45, Tmag = 10.74) star TOI-1694 (0.82 $R_{\sun}$, 0.84 $M_{\sun}$), observed in TESS sectors 19 and 20. It orbits the star at a distance of 0.039 AU with an orbital period 3.77 days and having an equilibrium temperature 1136.57 K. 'Alopeke and NIRC2 high resolution imaging showed no contaminating stellar companion and the ground-based follow-up observations verified the transit event occur within the target star follow-up aperture, and there is no strong filter dependent depth chromaticity, and there are no other obvious or Gaia DR2 or TIC stars contaminating the follow-up aperture that are bright enough to cause the TESS detection. The \texttt{TRICERATOPS} FPP and NFPP are listed in Table \ref{tab:FPPs} calculated using three available contrast curve files. These FPP values are consistent with the source of transit signal being on the target star. Using the \cite{2017ApJ...834...17C} mass-radius relationship we predicted the mass 25.5 $\pm$ 11.9 $M_{\earth}$. Based on this, the resultant RV semi-amplitude is 11.81 m  $s^{-1}$. TSM and ESM are estimated as 125.91 and 25.89 respectively which are above the recommended threshold of \cite{2018PASP..130k4401K}. So this target will be favourable for both transmission and emission spectroscopy. This planet also has one Jupiter analog outer companion  TOI-1694c (M$\sin{i}$ = 1.05 M$_J$, P = 389 days) discovered by \citet{2023AJ....165...60V}. Our predicted mass 25.5 $\pm$ 11.9 $M_{\earth}$ for TOI-1694b using \citet{2017ApJ...834...17C} is also comparable to the true measured by \citet{2023AJ....165...60V}, that is 26.1 $\pm$ 2.2 $M_{\earth}$.

\subsection{TOI-2194}
TOI-2194b is a super-Earth (1.99 $R_{\earth}$) planet orbiting bright (Vmag = 8.42, Tmag = 7.42), metal-poor ([Fw/H] = -0.3720 $\pm$ 0.1) star TOI-2194 (0.69 $R_{\sun}$, 0.74 $M_{\sun}$), observed in TESS sector 27. It orbits the star at a distance of 0.10 AU with an orbital period 15.34 days and having an equilibrium temperature 590.88 K. HRCam high resolution imaging showed no contaminating stellar companion and the ground-based follow-up observations have ruled out NEBs in all nearby ($\approx 2'.5$) Gaia DR2 and TIC stars that are bright enough to have caused the TESS detection. The \texttt{TRICERATOPS} FPP and NFPP are listed in Table \ref{tab:FPPs} calculated using one available contrast curve file. These FPP values are consistent with the source of transit signal being on the target star. Using the \cite{2017ApJ...834...17C} mass-radius relationship we predicted the mass 4.9 $\pm$ 2.0 $M_{\earth}$. Based on this, the resultant RV semi-amplitude is 1.45 m  $s^{-1}$. TSM is estimated as 131.023 which is above the threshold of second quartile suggested by \cite{2018PASP..130k4401K}, that makes it a good target for transmission spectroscopy. In other hand ESM is 5.45 which is comparable but still below 7.5. So emission spectroscopy would be challenging.

\subsection{TOI-2443}
TOI-2443b is a sub-Neptune (2.77 $R_{\earth}$) planet orbiting bright (Vmag = 9.51, Tmag = 8.29) star TOI-2443 (0.73 $R_{\sun}$, 0.66 $M_{\sun}$), observed in TESS sector 31. It orbits the star at a distance of 0.089 AU with an orbital period 15.67 days and having an equilibrium temperature 600.83 K. This is the coolest planet validated in this project. 'Alopeke and PHARO high resolution imaging showed no contaminating stellar companion and the ground-based follow-up observations have ruled out NEBs in all nearby ($\approx 2'.5$) Gaia DR2 and TIC stars that are bright enough to have caused the TESS detection. The \texttt{TRICERATOPS} FPP and NFPP are listed in Table \ref{tab:FPPs} calculated using three available contrast curve files. These FPP values are consistent with the source of transit signal being on the target star. We can neglect the very low probability NFPP scenarios because ground-based observations have confirmed the transit event on the target. Using the \cite{2017ApJ...834...17C} mass-radius relationship we predicted the mass 8.3 $\pm$ 3.6 $M_{\earth}$. Based on this, the resultant RV semi-amplitude is 2.74 m  $s^{-1}$. TSM and ESM are estimated as 132.89 and 8.28 respectively which are above the recommended threshold of \cite{2018PASP..130k4401K}. So this target will be favourable for both transmission and emission spectroscopy.

\subsection{TOI-2459}
TOI-2459b is a sub-Neptune (2.95 $R_{\earth}$) planet orbiting bright (Vmag = 10.77, Tmag = 9.39) star TOI-2459 (0.67 $R_{\sun}$, 0.66 $M_{\sun}$), observed in TESS sectors 5, 6, 32 and 33. It orbits the star at a distance of 0.14 AU with an orbital period 19.10 days and having an equilibrium temperature 445 K.  high resolution imaging showed no contaminating stellar companion and the ground-based follow-up observations verified the transit event occur within the target star follow-up aperture and that there are no other obvious or Gaia DR2 stars contaminating the follow-up aperture that are bright enough to cause the TESS detection.. The \texttt{TRICERATOPS} FPP and NFPP are listed in Table \ref{tab:FPPs} calculated using one available contrast curve file. These FPP values are consistent with the source of transit signal being on the target star. We can neglect the very low probability NFPP scenarios because ground-based observations have confirmed the transit event on the target. Using the \cite{2017ApJ...834...17C} mass-radius relationship we predicted the mass 9.1 $\pm$ 4.0 $M_{\earth}$. Based on this, the resultant RV semi-amplitude is 2.85 m  $s^{-1}$. TSM and ESM are estimated as 76.04 and 2.46 respectively which are below the recommended threshold of \cite{2018PASP..130k4401K}. So this target will not be favourable for either transmission or emission spectroscopy.

\subsection{TOI-3082}
TOI-3082b is a Neptune-like (3.66 $R_{\earth}$) planet orbiting faint (Vmag = 12.93, Tmag = 11.77) star TOI-3082 (0.68 $R_{\sun}$, 0.66 $M_{\sun}$), observed in TESS sectors 37. It orbits the star at a distance of 0.027 AU with an orbital period 1.93 days and having an equilibrium temperature 1032.78 K. Ground-based follow-up observations verified the transit event occur within the target star follow-up aperture, and there is no strong filter dependent depth chromaticity, and there are no other obvious or Gaia DR2 or TIC stars contaminating the follow-up aperture that are bright enough to cause the TESS detection. The \texttt{TRICERATOPS} FPP and NFPP are listed in Table \ref{tab:FPPs} calculated without using any contrast curve file. These FPP values are consistent with the source of transit signal being on the target star. We can neglect the very low probability NFPP scenarios because ground-based observations have confirmed the transit event on the target. Using the \cite{2017ApJ...834...17C} mass-radius relationship we predicted the mass 13.2 $\pm$ 5.8 $M_{\earth}$. Based on this, the resultant RV semi-amplitude is 8.79 m  $s^{-1}$. TSM is estimated as 78.37 which is below the threshold set by \cite{2018PASP..130k4401K}. However ESM is predicted to be 13.37 above the 7.5, indicating that it is potentially a good target for emission spectroscopy.

\subsection{TOI-4308}
TOI-4308b is a sub-Neptune (2.42 $R_{\earth}$) planet orbiting bright (Vmag = 11.25, Tmag = 10.34) star TOI-4608 (0.79 $R_{\sun}$, 0.9 $M_{\sun}$), observed in TESS sector 1. It orbits the star at a distance of 0.087 AU with an orbital period 9.15 days and having an equilibrium temperature 763.05 K. HRCam high resolution imaging showed no contaminating stellar companion. The \texttt{TRICERATOPS} FPP and NFPP are listed in Table \ref{tab:FPPs} calculated using one available contrast curve file. These FPP values are consistent with the source of transit signal being on the target star. Using the \cite{2017ApJ...834...17C} mass-radius relationship we predicted the mass 6.5 $\pm$ 2.7 $M_{\earth}$. Based on this, the resultant RV semi-amplitude is 2.11 m  $s^{-1}$. TSM and ESM are estimated as 39.68 and 2.48 respectively which are below the recommended threshold of \cite{2018PASP..130k4401K}. So this target will not be favourable for either transmission or emission spectroscopy.

\subsection{TOI-5704}
TOI-5704b is a sub-Neptune (3.23 $R_{\earth}$) planet orbiting bright (Vmag = 11.529, Tmag = 10.6147) star TOI-5704 (0.76 $R_{\sun}$, 0.73 $M_{\sun}$), observed in TESS sectors 22 and 48. It orbits the star at a distance of 0.04 AU with an orbital period 3.77 days and having an equilibrium temperature 949.07 K. Ground-based follow-up observation found the transit event in  7$\arcsec$ target apertures that are contaminated with 1.5$\arcsec$ neighbor TIC 900281091. We calculated the probability of the signal originating from the contaminating star, which was found to be 1.79 $\times$ $10^{-10}$ (see Table \ref{tab:nearby_stars}), thus ruling out the possibility of contamination. The \texttt{TRICERATOPS} FPP and NFPP are listed in Table \ref{tab:FPPs} calculated without using any contrast curve file. These FPP values are consistent with the source of transit signal being on the target star. Using the \cite{2017ApJ...834...17C} mass-radius relationship we predicted the mass 9.49  $\pm$ 4.4 $M_{\earth}$. Based on this, the resultant RV semi-amplitude is 5.33 m  $s^{-1}$. TSM is estimated as 76.99 which is comparable but still below the threshold set by \cite{2018PASP..130k4401K}. However ESM is predicted to be 10.04 above the 7.5, indicating that it is potentially a good target for emission spectroscopy.

\subsection{TOI-5803}
TOI-5803b is a sub-Neptune (3.27 $R_{\earth}$) planet orbiting bright (Vmag = 10.65, Tmag = 9.94) star TOI-5803 (0.76 $R_{\sun}$, 0.87 $M_{\sun}$), observed in TESS sector 55. It orbits the star at a distance of 0.10 AU with an orbital period 5.38 days and having an equilibrium temperature 678.87 K. HRCam high resolution imaging showed no contaminating stellar companion. The \texttt{TRICERATOPS} FPP and NFPP are listed in Table \ref{tab:FPPs} calculated using one available contrast curve file. \texttt{TRICERATOPS} has observed one nearby star TIC 2025175669 at the separation of 6$\arcsec$.82 with $\Delta mag$ = 7, but the further calculation using \texttt{TRICERATOPS} found out that this target has non-zero transit depth so it eventually ruled out the possibility of any contamination. These FPP values are consistent with the source of transit signal being on the target star. Using the \cite{2017ApJ...834...17C} mass-radius relationship we predicted the mass 10.8  $\pm$ 4.8 $M_{\earth}$. Based on this, the resultant RV semi-amplitude is 4.31 m  $s^{-1}$. TSM and ESM are estimated as 69.70 and 4.55 respectively which are below the recommended threshold of \cite{2018PASP..130k4401K}. So this target will not be favourable for either transmission or emission spectroscopy.

\section{Likely Planets and Not Validated Candidates}
\label{Likely Planets and Not Validated Candidates}
In Table \ref{tab:FPPs}, we have listed the Likely Planets along with the Not Validated candidates. In the case of \texttt{TRICERATOPS}, the target would be classified as a "likely planet" if FPP is $<$ 0.5 and NFPP is $<$ 10$^{-03}$  and "Likely False Positive" if NFPP $>$ 10$^{-03}$ \citep{2021AJ....161...24G}. We have identified five likely planet targets that can be further followed up to establish their planetary nature. These targets are TOI-323, TOI-1180, TOI-2200, TOI-2408 and TOI-3913. And the targets remain unvalidated due to not passing the \texttt{TRICERATOPS} threshold are, TOI-493, TOI 815 (Noisy signal with 1 \% False Alarm Probability), TOI 1179 (\texttt{TRICERATOPS} has detected blended eclipsing binary), TOI 1732 (FPP for sector 47 passed the validation threshold but sector 20 showing FPP above the threshold), TOI-3568, TOI-3896, TOI-4090 and TOI-5584.

\section{Conclusions}
\label{Conclusion}
Using ground-based light curves, high resolution imaging, and the statistical validation tool TRICERATOPS, out of the 24 initial candidates selected for examination, 11 new TESS exoplanetary systems have been statistically validated. Among these recently validated planets, there are several intriguing targets that worthy for further investigation into their atmospheres. For example, based on the estimated Transmission Spectroscopy Metrics (TSM) values, TOI-2194b is considered a promising candidate for the investigation of its atmosphere via transmission spectroscopy. Similarly, TOI-3082b and TOI-5704b are considered to be optimal targets for investigating via emission spectroscopy, as per their estimated Emission Spectroscopy Metrics (ESM) values. Additionally, based on the TSM and ESM values, TOI-672b, TOI-1694b, and TOI-2443b are considered to be promising candidates for the investigation of their atmospheres via both transmission and emission spectroscopy. Furthermore, we have identified five potential planets that would benefit from further investigation through the use of radial velocity and high-resolution imaging techniques in order to establish their planetary nature with a high degree of certainty. These investigations would help to reveal more about the properties and behavior of these exoplanets and provide insights into the formation and evolution of planetary systems.

\section{Acknowledgements}
M.V.G. and I.A.S. acknowledge the support of Ministry of Science and Higher Education of the Russian Federation under the grant 075-15-2020-780 (N13.1902.21.0039). Funding for the TESS mission is provided by NASA's Science Mission Directorate. KAC acknowledges support from the TESS mission via subaward s3449 from MIT. Some of the observations in this paper made use of the High-Resolution Imaging instruments ‘Alopeke and Zorro and were obtained under Gemini LLP Proposal Number: GN/S-2021A-LP-105. ‘Alopeke and Zorro were funded by the NASA Exoplanet Exploration Program and built at the NASA Ames Research Center by Steve B. Howell, Nic Scott, Elliott P. Horch, and Emmett Quigley. Alopeke was mounted on the Gemini North telescope of the international Gemini Observatory, a program of NSF’s OIR Lab, which is managed by the Association of Universities for Research in Astronomy (AURA) under a cooperative agreement with the National Science Foundation. on behalf of the Gemini partnership: the National Science Foundation (United States), National Research Council (Canada), Agencia Nacional de Investigación y Desarrollo (Chile), Ministerio de Ciencia, Tecnología e Innovación (Argentina), Ministério da Ciência, Tecnologia, Inovações e Comunicações (Brazil), and Korea Astronomy and Space Science Institute (Republic of Korea). This work makes use of observations from the LCOGT network. Part of the LCOGT telescope time was granted by NOIRLab through the Mid-Scale Innovations Program (MSIP). MSIP is funded by NSF. This paper makes use of observations made with the MuSCAT2 instrument, developed by the Astrobiology Center, at TCS operated on the island of Tenerife by the IAC in the Spanish Observatorio del Teide. This paper is based on observations made with the MuSCAT3 instrument, developed by the Astrobiology Center and under financial supports by JSPS KAKENHI (JP18H05439) and JST PRESTO (JPMJPR1775), at Faulkes Telescope North on Maui, HI, operated by the Las Cumbres Observatory. This research has made use of the Exoplanet Follow-up Observation Program (ExoFOP; DOI:\dataset[10.26134/ExoFOP5]{http://dx.doi.org/10.26134/ExoFOP5}) website, which is operated by the California Institute of Technology, under contract with the National Aeronautics and Space Administration under the Exoplanet Exploration Program. This publication makes use of data products collected by the TESS mission and obtained from the MAST data archive at the Space Telescope Science Institute (STScI). The light curve and target pixel file data used in this paper can be found in \dataset[10.17909/t9-nmc8-f686]{http://dx.doi.org/10.17909/t9-nmc8-f686}. C. M. would like to gratefully acknowledge the entire Dragonfly Telephoto Array team, and Bob Abraham in particular, for allowing their telescope bright time to be put to use observing exoplanets. TRAPPIST-South is funded by the Belgian National Fund for Scientific Research (F.R.S.-FNRS) under grant PDR T.0120.21, with the participation of the Swiss National Science Fundation (SNF).M.G. is F.R.S-FNRS Research Director. E.J. is F.R.S-FNRS Senior Research Associate. The postdoctoral fellowship of KB is funded by F.R.S.-FNRS grant T.0109.20 and by the Francqui Foundation. This publication benefits from the support of the French Community of Belgium in the context of the FRIA Doctoral Grant awarded to MT. F.J.P. acknowledges financial support from the grant CEX2021-001131-S funded by MCIN/AEI/ 10.13039/501100011033. This research received funding from the European Research Council (ERC) under the European Union's Horizon 2020 research and innovation programme (grant agreement n$^\circ$ 803193/BEBOP), and from the Science and Technology Facilities Council (STFC; grant n$^\circ$ ST/S00193X/1). This work makes use of observations from the ASTEP telescope. ASTEP benefited from the support of the French and Italian polar agencies IPEV and PNRA in the framework of the Concordia station program, from INSU, ESA, the University of Birmingham and STFC. 

%% To help institutions obtain information on the effectiveness of their 
%% telescopes the AAS Journals has created a group of keywords for telescope 
%% facilities.
%
%% Following the acknowledgments section, use the following syntax and the
%% \facility{} or \facilities{} macros to list the keywords of facilities used 
%% in the research for the paper.  Each keyword is check against the master 
%% list during copy editing.  Individual instruments can be provided in 
%% parentheses, after the keyword, but they are not verified.

%%%\vspace{5mm}
%%%%%%\facilities{HST(STIS), Swift(XRT and UVOT), AAVSO, CTIO:1.3m,
%%%%CTIO:1.5m,CXO}

%% Similar to \facility{}, there is the optional \software command to allow 
%% authors a place to specify which programs were used during the creation of 
%% the manuscript. Authors should list each code and include either a
%% citation or url to the code inside ()s when available.

\software{\texttt{TLS} \citep{2019A&A...623A..39H}, 
\texttt{LATTE} \citep{2022ascl.soft05006E},
\texttt{TESS-Plots}\footnote{\url{https://github.com/mkunimoto/TESS-plots}},
\texttt{Lightkurve} \citep{2018ascl.soft12013L}, 
\texttt{Juliet} \citep{2018ascl.soft12016E} and 
\texttt{TRICERATOPS} \citep{2020ascl.soft02004G}.}

%% For this sample we use BibTeX plus aasjournals.bst to generate the
%% the bibliography. The sample631.bib file was populated from ADS. To
%% get the citations to show in the compiled file do the following:
%%
%% pdflatex sample631.tex
%% bibtext sample631
%% pdflatex sample631.tex
%% pdflatex sample631.tex

\bibliography{Reference.bib}{}
\bibliographystyle{aasjournal}

%% This command is needed to show the entire author+affiliation list when
%% the collaboration and author truncation commands are used.  It has to
%% go at the end of the manuscript.
%\allauthors

\appendix
\section{Nearby stars and their probability being a nearby planet or nearby eclipsing binary}
\label{A1}
In Table \ref{tab:nearby_stars}, we have compiled a list of nearby stars for which non-zero transit depths were measured using the \texttt{TRICERATOPS} tool. These stars were selected as targets to calculate the NFPP. The NFPP values were then presented in Table \ref{tab:FPPs}. Specifically, an NFPP value of 0.00 $\pm$ 0.00 suggests that there is no known nearby star for which a non-zero transit depth has been observed.
\begin{table}[h!]
    \centering
    \begin{tabular}{c c c c c c c}
    \hline
    Target & Nearby TICs & Separation & Transit Depth & \multicolumn{3}{c}{Probability} \\
           &             & [arcsec] & & NTP\footnote{Nearby Transiting Planet} & NEB\footnote{Nearby Eclipsing Binary} & NEBX2P\footnote{Nearby Eclipsing Binary with twice Period} \\
    \hline
    \hline
    TOI-139 & - & - & - & - & - & - \\
    \hline
    TOI-672 & 151825526 & 51.797 & 0.5419 & 0.00E+00 & 1.45E-46 & 5.58E-72 \\
    \hline
    TOI-913 & 407126405 & 9.538 & 0.0672 & 2.49E-69 & 1.80E-63 & 1.29E-57 \\
            & 407126402 & 46.165 & 0.1389 & 1.13E-69 & 3.67E-68 & 9.97E-67 \\
            & 407126397 & 58.015 & 0.3984 & 9.14E-70 & 5.68E-28 & 1.15E-26 \\
            & 407126407 & 63.915 & 0.7803 & 1.75E-70 & 2.48E-70 & 1.71E-70 \\
            & 407126425 & 99.975 & 0.5770 & 1.29E-70 & 1.06E-70 & 2.28E-71 \\
    \hline
    TOI-1694 & 396776943 & 52.393 & 0.8128 & 0.00E+00 & 9.63E-100 & 1.42E-132 \\
             & 396740632 & 79.123 & 0.7012 & 0.00E+00 & 8.27E-158 & 3.04E-157 \\
    \hline
    TOI-2194 & - & - & - & - & - & - \\
    \hline
    TOI-2443 & 318753384 & 54.738 & 0.5893 & 0.00E+00 & 5.58E-64 & 9.21E-89 \\
             & 318753383 & 62.408 & 0.1128 & 2.20E-200 & 1.54E-19 & 3.04E-35 \\
    \hline
    TOI-2459 & 192790481 & 18.900 & 0.1217 & 1.19E-21 & 8.85E-05 & 1.59E-09 \\
             & 192790483 & 51.229 & 0.0585 & 1.04E-54 & 1.80E-04 & 1.43E-09 \\
             & 192790473 & 72.392 & 0.1829 & 8.46E-41 & 2.26E-06 & 5.33E-10 \\
    \hline
    TOI-3082 & 428699131 & 21.134 & 0.0572 & 2.75E-59 & 4.76E-28 & 8.90E-29 \\
    \hline
    TOI-4308 & 144193716 & 44.501 & 0.6594 & 3.36E-16 & 5.48E-11 & 1.19E-10 \\
    \hline
    TOI-5704 & 900281091 & 2.587 & 0.1832 & 1.79E-17 & 5.24E-04 & 2.94E-05 \\
    \hline
    TOI-5803 & 466382573 & 64.612 & 0.6685 & 1.77E-58 & 5.15E-08 & 6.46E-11 \\
    \hline
    \end{tabular}
    \caption{All the nearby stars considered by \texttt{TRICERATOPS} to calculate NFPP.}
    \label{tab:nearby_stars}
\end{table}
%% Include this line if you are using the \added, \replaced, \deleted
%% commands to see a summary list of all changes at the end of the article.
%\listofchanges
% End of file `sample631.tex'.

\end{document}